\documentclass[longauth]{aa}  

\usepackage{graphicx}
\usepackage{txfonts}
\usepackage{hyperref}
\usepackage[rightcaption]{sidecap}
\usepackage{subfigure}
\usepackage{longtable}
\usepackage[utf8]{inputenc}
\usepackage[flushleft]{threeparttable}
\def\ms{\hbox{\,m\,s$^{-1}$}}         
\newcommand{\mearth}{M_\oplus}
\newcommand{\rearth}{R_{\rm \oplus}}

\newcommand{\rsun}{R_{\rm \odot}}

\begin{document}

   \title{A compact multi-planet system transiting HIP\,29442 (TOI-469) discovered by TESS and ESPRESSO\thanks{Based on Guaranteed Time Observations collected at the European Southern Observatory by the ESPRESSO Consortium under ESO programmes 1104.C-0350, 106.21M2.002, 106.21M2.003, 106.21M2.004, 106.21M2.007, and 108.2254.002.}}
   \subtitle{Radial velocities lead to the detection of transits with low signal-to-noise ratio}

   \author{M.\,Damasso \inst{1}
           \and J.\,Rodrigues \inst{2,3}
           \and A.~Castro-Gonz\'alez \inst{4}
           \and B.\,Lavie \inst{5}
           \and J.\,Davoult \inst{6,7}
           \and M.\,R.\,Zapatero Osorio \inst{8}
           \and J.\,Dou \inst{9}
           \and S.\,G.\,Sousa \inst{2}
           \and J.\,E.\,Owen \inst{10}
           \and P.\,Sossi \inst{11}
           \and V.\,Adibekyan \inst{2,3}
           \and H.\,Osborn \inst{6,7}
           \and Z.\,Leinhardt \inst{8}
           \and Y.\,Alibert \inst{6,7}
           \and C.\,Lovis \inst{5}
           \and E.\,Delgado\,Mena \inst{2}
           \and A.\,Sozzetti \inst{1}
           \and S.\,C.\,C.\,Barros \inst{2,3}
           \and D.\,Bossini \inst{2}
           \and C.\,Ziegler \inst{12}
           \and D.\,R.\,Ciardi \inst{13}
           \and E.~C.~Matthews \inst{14}
           \and P.\,J.\,Carter \inst{8}
           \and J.\,Lillo-Box \inst{4}
           \and A.\,Su\'arez Mascare\~no \inst{15,16}
           \and S.\,Cristiani \inst{17}
           \and F.\,Pepe \inst{4}
           \and R.\,Rebolo \inst{15,16,19}
           \and N.\,C.\,Santos \inst{2,3}
           \and C.~Allende~Prieto \inst{15,16}
           \and S.~Benatti \inst{20}
           \and F.~Bouchy \inst{4}
           \and C.\,Brice\~{n}o \inst{21}
           \and P.\,Di\,Marcantonio \inst{17}
           \and V.~D'Odorico \inst{17}
           \and X.~Dumusque \inst{4}
           \and J.~A.~Egger \inst{6,7}
           \and D.~Ehrenreich \inst{4}
           \and J.~Faria \inst{2,3}
           \and P.~Figueira \inst{5,2}
           \and R.~G\'enova~Santos \inst{15,16}
           \and E.\,J.\,Gonzales \inst{22}
           \and J.\,I.\,Gonz\'alez Hern\'andez \inst{15,16}
           \and N.\,Law \inst{23}
           \and G.~Lo~Curto \inst{24}
           \and A.\,W.\,Mann \inst{23}
           \and C.\,J.\,A.\,P.\,Martins \inst{2,25}
           \and A.~Mehner \inst{24}
           \and G.~Micela \inst{20}
           \and P.~Molaro \inst{17}
           \and N.~J.~Nunes \inst{26}
           \and E.~Palle \inst{15,16}
           \and E.~Poretti \inst{27}
           \and J.~E.~Schlieder \inst{28}
           \and S.~Udry \inst{4} 
          } 

    \institute{INAF - Osservatorio Astrofisico di Torino, Via Osservatorio 20, I-10025 Pino Torinese, Italy\\
              \email{mario.damasso@inaf.it}
               \and Instituto de Astrof\'isica e Ci\^encias do Espa\c{c}o, Universidade do Porto, CAUP, Rua das Estrelas, 4150-762 Porto, Portugal
               \and Departamento de F\'isica e Astronomia, Faculdade de Ci\^encias, Universidade do Porto, Rua do Campo Alegre, 4169-007 Porto, Portugal
               \and Centro de Astrobiolog\'{\i}a (CSIC-INTA), ESAC campus, E-28692 Villanueva de la Ca\~nada, Madrid, Spain
               \and D\'epartement d’Astronomie, Universit\'e de Gen\`eve, Chemin Pegasi 51, CH-1290 Versoix, Switzerland
               \and Center for Space and Habitability, University of Bern, Gesellschaftsstrasse 6, CH-3012 Bern, Switzerland 
               \and Physics Institute of University of Bern, Gesellschaftsstrasse\,6, CH-3012 Bern, Switzerland
               \and Centro de Astrobiolog\'{\i}a (CSIC-INTA), Carretera de Ajalvir km 4, E-28850 Torrej\'on de Ardoz, Madrid, Spain
               \and School of Physics, H.H. Wills Physics Laboratory, University of Bristol, Tyndall Avenue, Bristol BS8 1TL, UK 
               \and Astrophysics Group, Department of Physics, Imperial College London, Prince Consort Rd, London, SW7 2AZ, UK
               \and Institute of Geochemistry and Petrology, ETH Zürich, CH-8092, Zürich, Switzerland           
               \and Department of Physics, Engineering and Astronomy, Stephen F. Austin State University, 1936 North St, Nacogdoches, TX 75962, USA
               \and NASA Exoplanet Science Institute-Caltech/IPAC, Pasadena, CA 91125, USA
               \and Max Planck Institute for Astronomy, Heidelberg, Germany   
               \and Instituto de Astrofisica de Canarias,  Via Lactea, E-38200 La Laguna, Tenerife, Spain
               \and Universidad de La Laguna, Departamento de Astrof\'isica, E- 38206 La Laguna, Tenerife, Spain
               \and INAF -- Osservatorio Astronomico di Trieste, Via Tiepolo 11, I-34143 Trieste, Italy  
               \and Institute for Fundamental Physics of the Universe, IFPU, Via Beirut 2, 34151 Grignano, Trieste, Italy
               \and Consejo Superior de Investigaciones Cient\'ificas, E-28006 Madrid, Spain 
               \and INAF – Osservatorio Astronomico di Palermo, Piazza del Parlamento 1, 90134 Palermo, Italy
               \and Cerro Tololo Inter-American Observatory/NSF’s NOIRLab, Casilla 603, La Serena, Chile
               \and Department of Astronomy and Astrophysics, University of California, Santa Cruz, CA 95064, USA
               \and Department of Physics and Astronomy, The University of North Carolina at Chapel Hill, Chapel Hill, NC 27599-3255, USA
               \and ESO, European Southern Observatory, Alonso de Cordova 3107, Vitacura, Santiago
               \and Centro de Astrof\'{\i}sica da Universidade do Porto, Rua das Estrelas,
                4150-762 Porto, Portugal
               \and Instituto de Astrof\'isica e Ciências do Espa\c{c}o, Faculdade de Ci\^encias da Universidade de Lisboa, Campo Grande, PT1749-016 Lisboa, Portugal
               \and INAF – Osservatorio Astronomico di Brera, Via Bianchi 46, 23807 Merate, Italy
               \and Exoplanets and Stellar Astrophysics Laboratory, NASA Goddard Space Flight Center, 8800 Greenbelt Road, Greenbelt, MD 20771, USA
             }

   \date{Received; accepted}

  \abstract
   {One of the goals of the ESPRESSO Guaranteed Time Observations (GTO) consortium is the precise characterisation of a selected sample of planetary systems discovered by TESS. One such a target is the K0V star HIP\,29442 (TOI-469), already known to host a validated sub-Neptune companion TOI-469.01, that we followed-up with ESPRESSO. } 
  {We aim to verify the planetary nature of TOI-469.01 by obtaining precise mass, radius and ephemeris, and constraining its bulk physical structure and composition. }
   {Following a Bayesian approach, we modelled radial velocity and photometric time series to measure the dynamical mass, radius, and ephemeris, and to characterise the internal structure and composition of TOI-469.01.}
   {We confirmed the planetary nature of TOI-469.01 (now renamed HIP\,29442\,$b$), and thanks to the ESPRESSO radial velocities we discovered two additional close-in companions. Through an in-depth analysis of the TESS light curve, we could also detect their low signal-to-noise transit signals. We characterised the additional companions, and conclude that HIP\,29442 is a compact multi-planet system. The three planets have orbital periods $P_{\rm orb,\,b}=13.63083\pm0.00003$, $P_{\rm orb,\, c}=3.53796\pm0.00003$, and $P_{\rm orb,\,d}=6.42975^{+0.00009}_{-0.00010}$ days, and we measured their masses with high precision: $m_{\rm p,\,b}=9.6\pm0.8~\mearth$, $m_{\rm p,\,c}=4.5\pm0.3~\mearth$, and $m_{\rm p,\,d}=5.1\pm0.4~\mearth$. We measured radii and bulk densities of all the planets (the 3$\sigma$ confidence intervals are shown in parenthesis): $R_{\rm p,\,b}=3.48^{+0.07\,(+0.19)}_{-0.08\,(-0.28)} ~\rearth$ and $\rho_{\rm p,\,b}=1.3\pm0.2\,(0.3)\, g~cm^{-3}$; $R_{\rm p,\,c}=1.58^{+0.10\,(+0.30)}_{-0.11\,(-0.34)}~\rearth$ and $\rho_{\rm p,\,c}=6.3^{+1.7\,(+6.0)}_{-1.3\,(-2.7)}\, g~cm^{-3}$; $R_{\rm p,\,d}=1.37\pm0.11^{(+0.32)}_{(-0.43)}~\rearth$ and $\rho_{\rm p,\,d}=11.0^{+3.4\,(+21.0)}_{-2.4\,(-6.3)}\, g~cm^{-3}$. Due to noisy light curves, we used the more conservative 3$\sigma$ confidence intervals for the radii as input to the interior structure modelling. We find that HIP\,29442\,$b$ appears as a typical sub-Neptune, likely surrounded by a gas layer of pure H-He with a mass of $0.27^{+0.24}_{-0.17} \mearth$ and a thickness of $1.4\pm0.5 \rearth$. For the innermost companions HIP\,29442\,$c$ HIP\,29442\,$d$, the model supports an Earth-like composition.}
   {The compact multi-planet system orbiting HIP\,29442 offers the opportunity to study in one shot time planets straddling the gap in the observed radius distribution of close-in, small-size exoplanets. High-precision photometric follow-up is required to get more accurate and precise radius measurements, especially for planets $c$ and $d$. This, together with our determined high-precision masses, will provide accurate and precise planets' bulk structure, and enable an accurate investigation of the system's evolution. }

   \keywords{Stars: individual: HIP29442; Stars: individual: TOI-469; Planetary systems; Techniques: photometric; Techniques: radial velocities; Planets and satellites: interior; Planets and satellites: formation }
   
\authorrunning{Damasso et al.}
\titlerunning{A compact three-planet system around HIP\,29442}
   \maketitle
%
\section{Introduction} \label{sec:intro}

After starting operations on October 2018, the ESPRESSO high-resolution spectrograph \citep{pepe21} of ESO's Very Large Telescope (VLT) demonstrated its unprecedented capabilities to reach radial velocity (RV) precision at a level of a few tens cm s$^{-1}$, and to guarantee stability over several months. ESPRESSO observations allowed to precisely measure the mean densities of transiting exoplanets detected by TESS \citep{ricker2016}, and to well constrain their fundamental physical properties, as demonstrated by the spectroscopic follow-up of several planetary systems (e.g. \citealt{Demangeon2021,Leleu2021,sozzetti2021,vaneylen2021MNRAS.507.2154V,barros2022,lavie2023A&A...673A..69L}).   

The radial velocity follow-up of known transiting planets hold surprises in some cases. For instance, ESPRESSO allowed the discovery of additional candidate planets in multi-planet systems (e.g. \citealt{lillobox2020,sozzetti2021,barros2022}), and in the case of LTT 1445A (TOI-455) the spectroscopic detection of LTT 1445A\,$c$ triggered a thorough analysis of the TESS photometry, which revealed previously undetected transits for this planet \citep{lavie2023A&A...673A..69L}. Undetected transits in systems with known sub-Neptunes/mini-Neptunes (here defined as planets with a radius $2\lesssim R \lesssim4~R_\oplus$) are particularly interesting, because such shallow transits could be likely produced by smaller-size Earths or super-Earths (with radii $1\lesssim R\lesssim2~R_\oplus$). Indeed, such a system would host planets residing in different locations on the well-known bi-modal radius distribution of small-size ($R<4 R_\oplus$, i.e. from Earth- to Neptune-size), close-in planets that emerged from the population of transiting planets discovered by \textit{Kepler} \citep{Fulton_2017}. The observed distribution shows two distinct peaks at $\sim1.3$ and $\sim$2.4 $R_\oplus$, with a scarcity of planets found between $\sim1.5$--2 $R_\oplus$ (an interval thus known as the ``radius gap''), suggesting the existence of two well-defined families of planets, usually identified as super-Earths and sub-Neptunes, respectively. A mechanism able to predict the existence of the gap for close-in planets is the atmospheric mass-loss due to photo-evaporation driven by the high-energy irradiation of the hosts (e.g. \citealt{Owen2013ApJ...775..105O,jin2014ApJ...795...65J,owen2017ApJ...847...29O}), but whether this is the only or dominant process at play to explain the transition from super-Earths to sub-Neptunes is still debated (e.g. core-powered atmospheric mass loss models have also been proposed, such as \citealt{ginzburg2018MNRAS.476..759G,gupta2019MNRAS.487...24G}). 

A first and critical step in understanding the processes that determine the current location of a planet within the radius distribution is an accurate and precise determination of its physical properties, such as mass and bulk density, that allow us to determine its average composition, and whether a planet is surrounded or not by a significant gaseous envelope. Multi-planet systems where transiting close-in planets cross over the radius distribution offer the fascinating opportunity to investigate planets' evolution within the same environment, and to constrain the models of the radius valley over some of the system parameters. Indeed, such planets have formed from the same protoplanetary disk, and their evolution has been taking place under the influence of the same stellar high-energy irradiation field. Examples of systems which host planets that straddle the radius valley are {\it Kepler}-36 \citep{Carter2012,Owen2016}, K2-3 \citep{damasso2018A&A...615A..69D,diamond2022AJ....164..172D}, K2-36 \citep{damasso2019A&A...624A..38D}, L231-32 \citealt{Gunther2019,VanEylen2021}, LTT 3780 \citep{cloutier2020AJ....160....3C}, TOI-1266 \citep{stefansson2020AJ....160..259S}, K2-32 and K2-233 \citep{lillobox2020A&A...640A..48L}, and TOI-1468 \citep{chaturvedi2022A&A...666A.155C} just to mention some. They represent very interesting targets to explore the role and relevance of evolutionary 
processes, such as the aforementioned photo-evaporation and core-powered mass loss, in shaping the system, and to also test alternative hypotheses, such as those which connect the emergence of the radius gap with the location of small-size planets within the system at the time of birth. Such models attribute the existence of the radius valley primarily to the differences in the bulk composition of the planets, identifying the super-Earths as rocky planets and sub-Neptunes as mostly water-ice-rich worlds \citep[e.g.][]{zeng2019PNAS..116.9723Z,izidoro2022ApJ...939L..19I,luque2022Sci...377.1211L}, with the first forming within the snow line, while the latter outside it, later migrating inward. Moreover, the cases of the planets Kepler-138\,$d$ \citep{piaulet2023NatAs...7..206P} and TOI-244\,b \citep{2023A&A...675A..52C}, which are low-density super-Earths skimming the radius gap ($R_{\rm p}=1.51\pm0.04 \rearth$ and $R_{\rm p}=1.52\pm0.12 \rearth$, respectively), with a composition consistent with that of a volatile-rich water world, shows that we should not expect all super-Earth-sized planets to have an Earth-like compositions.

That is the scenery where our story takes place. In this work, we present the detection and a characterisation of a packed multi-planet system, with close-in planets that occupy different locations in the bi-modal radius distribution. They orbit the star HIP~29442 (TIC 33692729, TOI-469, where TIC and TOI stand for TESS Input Catalogue and TESS Object of Interest, respectively), and have been detected using data from TESS and ESPRESSO, which are described in Sect. \ref{sec:descriptiondata}. This K-type main sequence star (characterised in Sect. \ref{sec:stellarparameters} and \ref{sec:staractivity}), was originally included in the Exoplanet Follow-up Observing Program (ExoFOP) database\footnote{\url{https://exofop.ipac.caltech.edu/tess/}} as it hosts a planet candidate TOI-469.01 (orbital period 13.6 days; radius 3.3 $\rearth$), later validated by \cite{Giacalone2021}. Recently, a mass measurement has been obtained by \cite{akana2023arXiv230616587A} thanks to Keck-HIRES RVs ($5.8\pm2.4$ $\mearth$). This short-period sub-Neptune sized planet was deemed interesting for follow-up characterisation with ESPRESSO within the Guaranteed Time Observations (GTO) Consortium. Thanks to ESPRESSO RVs, we confirmed the planetary nature of TOI-469.01, and detected the presence of two additional periodic signals. Through a detailed analysis of the TESS data, we discovered transit signals with periods that correspond to those detected in the RV time series. We describe the system's detection and analysis in Sect. \ref{sec:rvlcanalysis}. The newly detected pair of signals are due to small-size planets that orbit their host with periods shorter than TOI-469.01. Thus, HIP~29442 turns out to be a compact multi-planet system. 
The results of a first internal structure modelling of the three planets is presented in Sect. \ref{sec:internalstructure}. We draw conclusions in Sect. \ref{sec:conclusion}, exploring the role of photo-evaporation in the system's evolution, based on the current properties of the planets.


\section{Materials and methods }
\label{sec:descriptiondata}
\subsection{TESS light curve}
\label{sec:tess_lc}
HIP\,29442 was observed by TESS during Sector 6 (hereafter S6; 11 December 2018 - 7 January 2019) and Sector 33 (hereafter S33; 17 December 2020 - 13 January 2021) on Camera 2, in short-cadence mode ($t_{\rm exp} =$ 2 min). The observations were processed using the Science Processing Operations Center (SPOC) pipeline \citep{SPOC2016}. For our analysis, we used the Presearch Data Conditioning Simple Aperture Photometry (PDCSAP) flux \citep{PDCSAP2012, PDCSAP2014, PDCSAP2012II} which has common trends removed, and it is corrected for crowding from known nearby stars. Observations flagged by the pipeline as low-quality were removed. Fig. \ref{fig:tpfplotter} shows the TESS Target Pixel Files (TPF) centered on HIP\,29442, with sources cross-matched with the Gaia DR3 catalogue overplotted \citep{GAIAEDR3}. There are no bright sources falling within the TESS aperture which can significantly contaminate the light curve, and dilute the transits, or additional widely separated companions with similar parallaxes and proper motions \citep{mugrauer2020,mugrauer2021} identified by Gaia. Moreover, HIP\,29442 has a Gaia DR3 Renormalised Unit Weight Error (RUWE) of 1.11, consistent with the single star model. The possibility of contamination by closer stars (within 3$\arcsec$ from HIP\,29442) has been also investigated in more detail through speckle and adaptive optics imaging, as discussed in Sect. \ref{sec:imaging}.

\subsection{Imaging observations} \label{sec:imaging}
\begin{figure}
    \centering
    \includegraphics[width=0.8\linewidth]{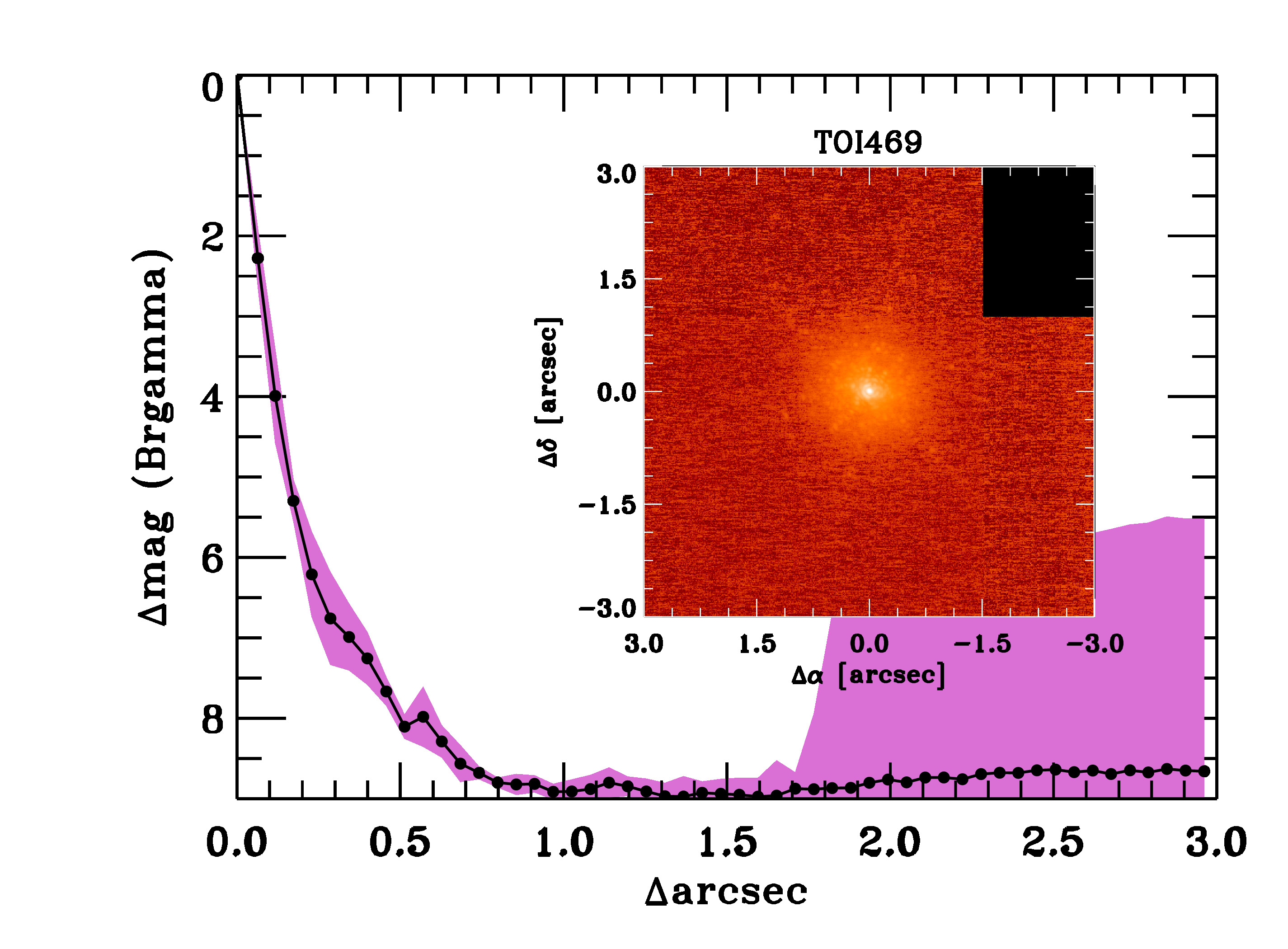}\\
    \includegraphics[width=0.7\linewidth]{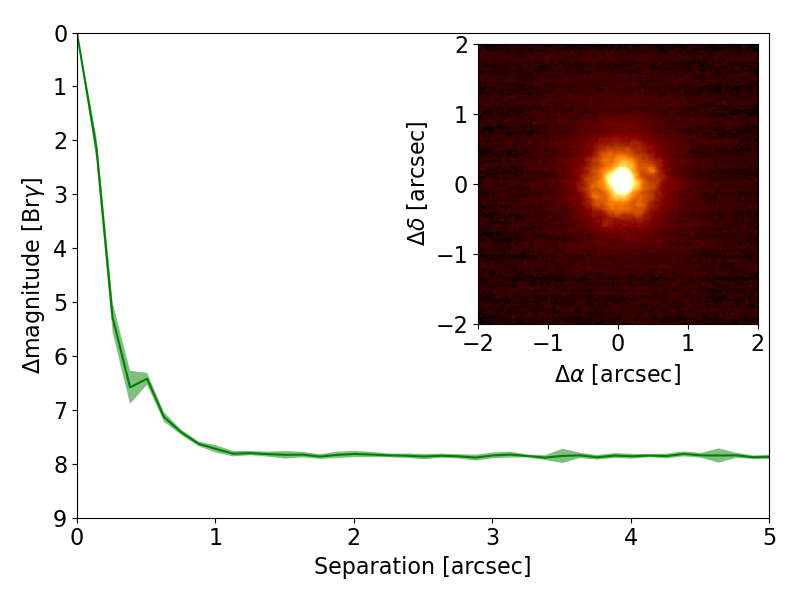}\\
    \includegraphics[width=0.7\linewidth]{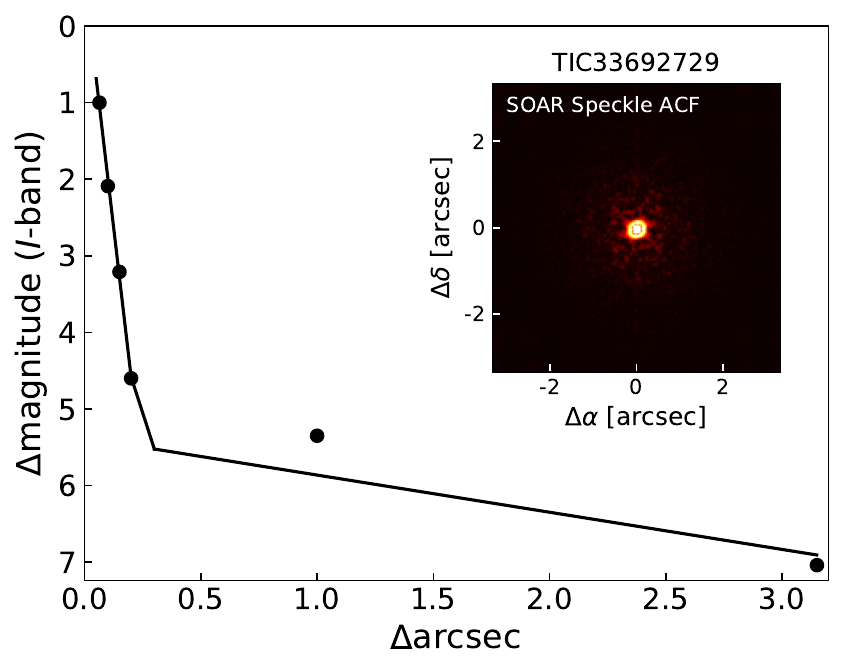}
    \caption{Imaging observations. \textit{Upper panel.} Keck NIR AO imaging and sensitivity curves for HIP~29442 taken in the Br-$\gamma$ filter on 25 March 2019. The image reaches a contrast of $\sim 8$ magnitudes fainter than the host star within 0.5$\arcsec$. \textit{Middle panel.} Gemini/NIRI AO imaging of HIP~29442. Observation taken on 5 April 2019 in the Br-$\gamma$ filter. \textit{Lower panel.} Speckle imaging of HIP~29442 taken with the SOAR telescope on 17 March 2019. No nearby stars were detected within 3$\arcsec$ of the target. 
    \textit{Insets:} Image of the central portion of the data, centred on the star. }
    \label{fig:aospeckle}
\end{figure}

As part of standard process for validating transiting exoplanets, high-angular resolution imaging is needed to search for nearby sources that can contaminate the TESS photometry, resulting in an underestimated planetary radius, or be the source of astrophysical false positives, such as background eclipsing binaries \citep[e.g.][]{lillobox2014A&A...566A.103L,ciardi2015}. 
HIP\,29442 was observed with infrared high-resolution adaptive optics (AO) imaging at Keck Observatory with the NIRC2 instrument on Keck-II behind the natural guide star AO system \citep{wizinowich2000,schlieder2021FrASS...8...63S}. The observations were made on 25 Mar 2019 in the standard 3-point dither pattern that is used with NIRC2 to avoid the left lower quadrant of the detector which is typically noisier than the other three quadrants. The dither pattern step size was $3\arcsec$ and was repeated twice, with each dither offset from the previous dither by $0.5\arcsec$. The camera was in the narrow-angle mode with a full field of view of $\sim10\arcsec$ and a pixel scale of approximately $0.0099442\arcsec$ per pixel. The observations were made in the narrow-band Br-$\gamma$ filter ($\lambda_{o} = 2.1686; \Delta\lambda = 0.0326\mu$m) with an integration time of 30 seconds with one coadd per frame, for a total of 270 seconds on target. The final resolution of the combined dithers was determined from the full-width half-maximum of the point spread function 0.057$\arcsec$.  
The sensitivities of the final combined AO image were determined by injecting simulated sources azimuthally around the primary target every $20^\circ $ at separations of integer multiples of the central source's full width at half maximum (FWHM) \citep{furlan2017}. The brightness of each injected source was scaled until standard aperture photometry detected it with $5\sigma$ significance. The resulting brightness of the injected sources relative to the target set the contrast limits at that injection location. The final $5\sigma$ limit at each separation was determined from the average of all of the determined limits at that separation and the uncertainty on the limit was set by the rms dispersion of the azimuthal slices at a given radial distance. The final sensitivity curve for the Keck data is shown in the upper panel of Fig. \ref{fig:aospeckle}). We reached a contrast upper limit of $\sim$8 mag at 0.5$\arcsec$ from the target star. 

We vet for visual companions also with Gemini/NIRI AO imaging. We collected 9 science images, each with integration time 1.5s, on 5 April 2019, and offset the telescope by $\sim2\arcsec$ in a grid-like dither pattern between each image. These science images were reduced following standard data reduction practices: we removed bad pixels, applied flat-field and sky background corrections (using a sky background reconstructed from the dithered science images themselves), aligned the images based on fitting the stellar position, and then co-added the data. No candidates were identified within the field of view, and we reach a contrast of 7.85 magnitudes in the background limited regime, and are sensitive to companions within 5 mag of the host beyond 250 mas. The contrast, and a thumbnail image of the target, are shown in the middle panel of Fig. \ref{fig:aospeckle}.

We also used speckle imaging to search for stellar companions to HIP\,29442 with the 4.1-m Southern Astrophysical Research (SOAR) telescope \citep{tokovinin2018PASP..130c5002T}. We observed the star on 17 March 2019 in Cousins I-band, which is a visible bandpass similar as TESS. This observation was sensitive to a 5.5-magnitude fainter star at an angular distance of 1$\arcsec$ from the target. More details of the observations within the SOAR TESS survey are available in \cite{ziegler2020AJ....159...19Z}. The 5$\sigma$ detection sensitivity and speckle auto-correlation functions from the observations are shown in the bottom panel of Fig. \ref{fig:aospeckle}. No nearby stars were detected within 3$\arcsec$ of HIP\,29442 in the SOAR observations. No additional stellar companions were detected in agreement with the AO observations.

\subsection{Spectroscopic follow-up with ESPRESSO }
We observed HIP\,29442 with the ESPRESSO spectrograph installed at the Very Large Telescope
(VLT) in the Paranal Observatory in Chile \citep{pepe21} between 9 October 2019 and 24 March 2022 (time span of 897 d). We collected 83 spectra in high resolution (HR), single telescope mode, with an exposure time of 900 s, which were reduced with the data reduction software (DRS) v.3.0.0, using a G9 mask. The observations have been executed with fibre B illuminated by a Fabry-P\'erot light as a spectral reference to measure the instrument’s internal drift \citep{pepe21}. The RVs, bisector inverse slope (BIS), and FWHM were calculated from the cross-correlation function (CCF). The RVs have a median photon noise precision of 0.31 \ms and an rms of 3.53 \ms, and are listed in Table \ref{table:dataespresso}. Due to the COVID-19 pandemic, ESPRESSO was in downtime for nine months, between 24 March and 24 December 2020. Therefore, in our modelling we considered the possible presence of a relative offset between data obtained before and after the interruption of the observations (Section \ref{sec:rvlcanalysis}). 


\section{Fundamental stellar parameters}
\label{sec:stellarparameters}

We compiled a list of astrometric, photometric, and spectroscopic parameters of HIP\,29442 (Table \ref{tab:stellarparam}). The stellar atmospheric parameters ($T_{\mathrm{eff}}$, $\log g$, microturbulence, [Fe/H]) were derived from a combined ESPRESSO spectrum  using ARES+MOOG, following the same methodology described in \cite{Sousa-21, Sousa-14, Santos-13}. The latest version of ARES \footnote{The latest version of the code \textsc{ARES v2} can be downloaded at \url{https://github.com/sousasag/ARES}} \citep{Sousa-07, Sousa-15} was used to consistently measure the equivalent widths (EW) of selected iron lines based on the line list presented in \cite{Sousa-08}. We used a minimisation process to find ionisation and excitation equilibrium and converge to the best set of spectroscopic parameters. This process makes use of a grid of Kurucz model atmospheres \citep{Kurucz-93} and the radiative transfer code MOOG \citep{Sneden-73}. The trigonometric surface gravity was also derived using eDR3 Gaia data following the same methodology as described in \citet[][]{Sousa-21}. The age, mass and radius were determined using the Bayesian code \texttt{PARAM}\footnote{\url{http://stev.oapd.inaf.it/cgi-bin/param}} \citep{PARAM,PARAM2,PARAM3} v1.4. The derived values are 11.7$^{+3.7}_{-3.8}$ Gyr ($>$4.07 Gyr at 95$\%$ of confidence level; note that here we report the 68$\%$ confidence interval derived from the posterior distribution, without constraining its upper limit to be 13.78 Gyr), $M_\star=0.88^{+0.04}_{-0.03}$ M$_\odot$, and $R=0.96^{+0.04}_{-0.03}$ R$_\odot$, respectively.
The code followed a grid-based approach where a well-sampled grid of stellar evolutionary tracks is matched to the observed quantities $T_{\rm eff}$, [Fe/H], and luminosity. The derivation of the errors on the mass and radius are calculated from the PDF of the posteriors, in which we take the median as the central value and the 16th and 84th percentiles (1$\sigma$ confidence interval). The grid of stellar evolutionary isochrones is taken from the code \texttt{PARSEC}\footnote{\url{http://stev.oapd.inaf.it/cgi-bin/cmd}} v2.1 \citep{Bressan2012}. Mass and radius derived with PARAM are the values adopted in our work.
The luminosity is computed by converting the bolometric magnitude, calculated from the $K_s$ 2MASS magnitude \citep{Skrutskie06} corrected for the corresponding bolometric correction and for the distance. 
The bolometric correction was estimated through the online tool YBC\footnote{\url{http://stev.oapd.inaf.it/YBC/index.html}} \citep[PARSEC Bolometric Correction;][]{Chen2019} with the input quantities $T_\mathrm{eff}$, [Fe/H], $\log{g}$). The distance is calculated by the inverse of the Gaia eDR3 parallax \citep{GAIAEDR3}. 
The interstellar absorption has been neglected, assuming that our target has low reddening due to its proximity to the Sun, and taking into account that the contribution of the reddening in the band ${K_s}$ is generally lower or comparable to the photometric error. We derived a bolometric luminosity $L=0.70\pm 0.03$ L$_\odot$. 

We calculated a model-independent bolometric luminosity by fitting the spectral energy distribution (SED) of TOI-469, taking advantage of the precise photometry available in several bands between $\sim0.3$--25 $\mu$m, and the Gaia EDR3 parallax (Fig. \ref{fig:sed}). We find $L=0.695\pm0.011$ L$_\odot$. The best-fit black-body curve for $T_{\rm eff}=5289\pm69$ K shows that there is no infrared-flux excesses from the star up to 25 $\mu$m. From the SED-derived bolometric luminosity and $T_{\rm eff}$, we determined the stellar radius by applying the Stefan-Boltzmann equation. We get $R=0.99^{+0.04}_{-0.03}$ R$_{\odot}$, in agreement with the radius calculated using stellar evolutionary models.

We also performed a chemical abundance analysis of HIP\,29442, using the combined ESPRESSO spectrum. Under the assumption of local thermodynamic equilibrium, the classical curve-of-growth analysis technique was used \cite[e.g.][]{nissen2015A&A...579A..52N,casali2020A&A...643A..12C,dasilva2012A&A...542A..84D,bensby2014A&A...562A..71B}. The stellar abundances of the elements were also determined using the same methods and models used to determine the stellar parameters. For the derivation of chemical abundances of the refractory elements we closely followed the methods described in e.g. \cite{Adibekyan-12, Adibekyan-15}. Although the EWs of the spectral lines were measured automatically with \textsc{ARES}, we undertook a detailed visual review of the EWs for the elements Na, Al, and Mg with only two or three lines available. Abundances of the volatile elements, C and O, were derived following the method of \cite{delgado2021, Bertrandelis-15}. Since the two spectral lines of oxygen are usually weak, the EWs of these lines were manually measured with the task \texttt{splot} in IRAF. In addition, a careful inspection of the individual spectra were done to remove the spectra contaminated by telluric lines or the oxygen airglow at $6300\AA$. The abundance of oxygen from different line indicators tend to show dissimilar values. In particular, for relatively cool stars (as it is the case of HIP\,29442), the oxygen line at $6300~\AA$ tends to provide lower abundances that usually translate into larger C/O ratios \citep{delgado2021}. Therefore, we only use the oxygen abundance from the \ion{O}{I} line at $6158\AA$ that is not dependent in $ T_{\rm eff}$. All the [X/H] ratios are obtained by doing a differential analysis with respect to a high signal-to-noise ratio (S/N) solar (Vesta) spectrum. The results are shown in Table \ref{tab:stellarparam}. Overall, the chemical pattern of HIP\,29442 resembles that of a typical Galactic thin disk star in the solar neighbourhood \citep[e.g.][]{Adibekyan-12}. 


We computed the Galactic space velocity ($U,V,W$) of HIP 29442 based on its coordinates, proper motion, parallax, and systemic RV extracted from Gaia DR3. Considering the solar peculiar motion from \citet{2003A&A...409..523R}, we obtain $U$ = 87.10 $\pm$ 0.10 $\rm km/s$, $V$ = \mbox{-2.71} $\pm$ 0.10 $\rm km/s$, and $W$ = \mbox{-16.75} $\pm$ 0.04 $\rm km/s$ with respect to the local standard of rest. We used these velocities to estimate the probability that HIP 29442 belongs to the thin disk (D), the thick disk (TD), and the halo (H), based on the kinematic approach from \citet{2003A&A...410..527B} and the kinematic characteristics for the stellar components in the Solar neighbourhood from \citet{2003A&A...409..523R}. We obtain D = 96.376$\%$, TD = 3.610$\%$, and H = 0.014$\%$. Thus, it is very likely that HIP 29442 is a member of the Galactic thin-disk population, which is a conclusion in agreement with its chemical pattern.

\begin{table}
    \centering
     \tiny
     \caption{Stellar properties of HIP~29442 (HD~42813; TOI-469).}
    \begin{tabular}{lcc}
    \hline
    \noalign{\smallskip}
    Parameter     &  Value & Refs. \\
    \noalign{\smallskip}
    \hline
    \noalign{\smallskip}
    \noalign{\smallskip}
    \noalign{\smallskip}
    $\alpha$ (J2000) & 06h 12m 13.97s & [1]  \\
    \noalign{\smallskip}
    $\delta$ (J2000) & $-14^\circ$ 39$^{\prime}$ 00.06$^{\prime\prime}$ & [1]  \\
    \noalign{\smallskip}
    $\mu_\alpha \cdot \cos \delta$ [mas yr$^{-1}$] & $-79.132\pm0.013$ & [1]  \\
    \noalign{\smallskip}
    $\mu_\delta$ (mas yr$^{-1}$) & $162.696\pm0.013$ & [1]  \\
    \noalign{\smallskip}
    $\varpi$ (mas) & $14.7065\pm0.0159$ & [1]   \\
    \noalign{\smallskip}
    $d$ [pc] & 67.907$\pm$0.075 & [2] \\
    \noalign{\smallskip}
    \noalign{\smallskip}
    $B$ & $10.25\pm0.03$ & [3]\\
    \noalign{\smallskip}
    $V$ &  $9.67$ & [3]\\
    \noalign{\smallskip}
    $G$ &  $9.282\pm0.003$ & [1] \\
    \noalign{\smallskip}
    $J$  & $8.056\pm0.029$ & [4] \\
    \noalign{\smallskip}
    $H$   & $7.718\pm0.049$ & [4] \\
    \noalign{\smallskip}
    $K_s$ &  $7.587\pm0.024$ &  [4]\\
    \noalign{\smallskip}
    \noalign{\smallskip}
    $T_{\rm eff}~(K)$ 	& 5289 $\pm$ 69 & [5]  \\
    \noalign{\smallskip}
    $v\sin i_{\star}$ (km $s^{-1}$) & $1.94 \pm 0.45$ & [5] \\
    \noalign{\smallskip}
    $\rm log\,g\, (cgs)\tablefootmark{a}$ & 4.24 $\pm$ 0.13 & [5]   \\
    \noalign{\smallskip}
    $\rm log\,g_{trig.}\, (cgs)\tablefootmark{b}$ & 4.39 $\pm$ 0.03 & [5]  \\
    \noalign{\smallskip}
    $\rm \xi_{\rm t}~(km~s^{-1})$  & 0.74 $\pm$ 0.06 & [5]  \\
    \noalign{\smallskip}
    $<\log R^{\prime}_{\rm HK}>$ (dex) & -5.25 & [5] \\
    \noalign{\smallskip}
    $P_{\rm rot,\,\star}$ (days) & $40.0^{+2.7}_{-2.4}$ & [5]$^*$ \\    
    \noalign{\smallskip} 
    $\rm R_{\star}\, (R_{\odot})$ & $0.993^{+0.035}_{-0.033}$ & [5]$^{**}$ \\
     & $0.96^{+0.04}_{-0.03}$ & [5]$^{***}$  \\
    \noalign{\smallskip}
    $\rm M_{\star}\, (M_{\odot})$ & $0.88^{+0.04}_{-0.03}$ & [5] \\
    \noalign{\smallskip}
    $\rho_{\star}$ ($\rho_{\odot})$ & $1.01\pm0.07$ & [5] \\
    \noalign{\smallskip}
    Bolom. luminosity, $L_{\star}$ ($L_{\odot}$) & $0.695\pm0.011$ & [5]** \\
    \noalign{\smallskip}
    & $0.70\pm 0.03$ & [5]***\\
    \noalign{\smallskip}
    Age\tablefootmark{c} (Gyr) & 11.7$^{+3.7}_{-3.8}$ & [5]*** \\
    & ($>4.07$ at 95$\%$ confidence) & \\
    \noalign{\smallskip}
    $\rm [Fe/H]$ (dex) & 0.24 $\pm$ 0.05 & [5]	\\
    \noalign{\smallskip}
    $\rm [NaI/H]$ (dex) & 0.22 $\pm$ 0.05 & [5] \\
    \noalign{\smallskip}
    $\rm [MgI/H]$ (dex) & 0.26 $\pm$ 0.04 & [5] \\
    \noalign{\smallskip}
    $\rm [AlI/H]$ (dex) & 0.21 $\pm$ 0.07 & [5] \\
    \noalign{\smallskip}
    $\rm [SiI/H]$ (dex) & 0.21 $\pm$ 0.04 & [5] \\
    \noalign{\smallskip}
    $\rm [TiI/H]$ (dex) & 0.25 $\pm$ 0.07 & [5] \\
    \noalign{\smallskip}
    $\rm [NiI/H]$ (dex) & 0.22 $\pm$ 0.05 & [5] \\
    \noalign{\smallskip}
    $\rm [C/H]$ (dex) & 0.15 $\pm$ 0.05 & [5] \\
    \noalign{\smallskip}
    $\rm [O/H]$ (dex), at 6158~\AA & 0.26 $\pm$ 0.06 & [5] \\
    \noalign{\smallskip}
    ($U$,$V$,$W$) (km s$^{-1}$) & ($86.30\pm0.11$, $3.23\pm0.10$ & [5]  \\  
    & $-15.40\pm0.04$) & \\
    \noalign{\smallskip}
    \hline
    \end{tabular}
        \label{tab:stellarparam}
    \tablefoot{
    \tablefoottext{a}{From spectral analysis.}
    \tablefoottext{b}{Trigonometric surface gravity using Gaia DR3 data.}
    \tablefoottext{c}{Here we report the 68\% confidence interval derived from the posterior distribution, without constraining the age upper limit to be 13.78 Gyr}}
    \tablebib{[1] \cite{gaiadr32023A&A...674A...1G,gaia2016A&A...595A...1G}; [2] \cite{bailer2021AJ....161..147B}; [3] \cite{2012yCat.1322....0Z}; [4] \cite{2003yCat.2246....0C}; [5] This work; [5]$^*$ Derived from a GP regression of RVs; [5]$^{**}$ From Stefan-Boltzmann equation (adopted value); [5]$^{***}$ PARAM 1.5 optimisation code. }
\end{table}


\section{Stellar activity characterisation}
\label{sec:staractivity}
\begin{figure*}
    \centering
    \includegraphics[scale=0.5]{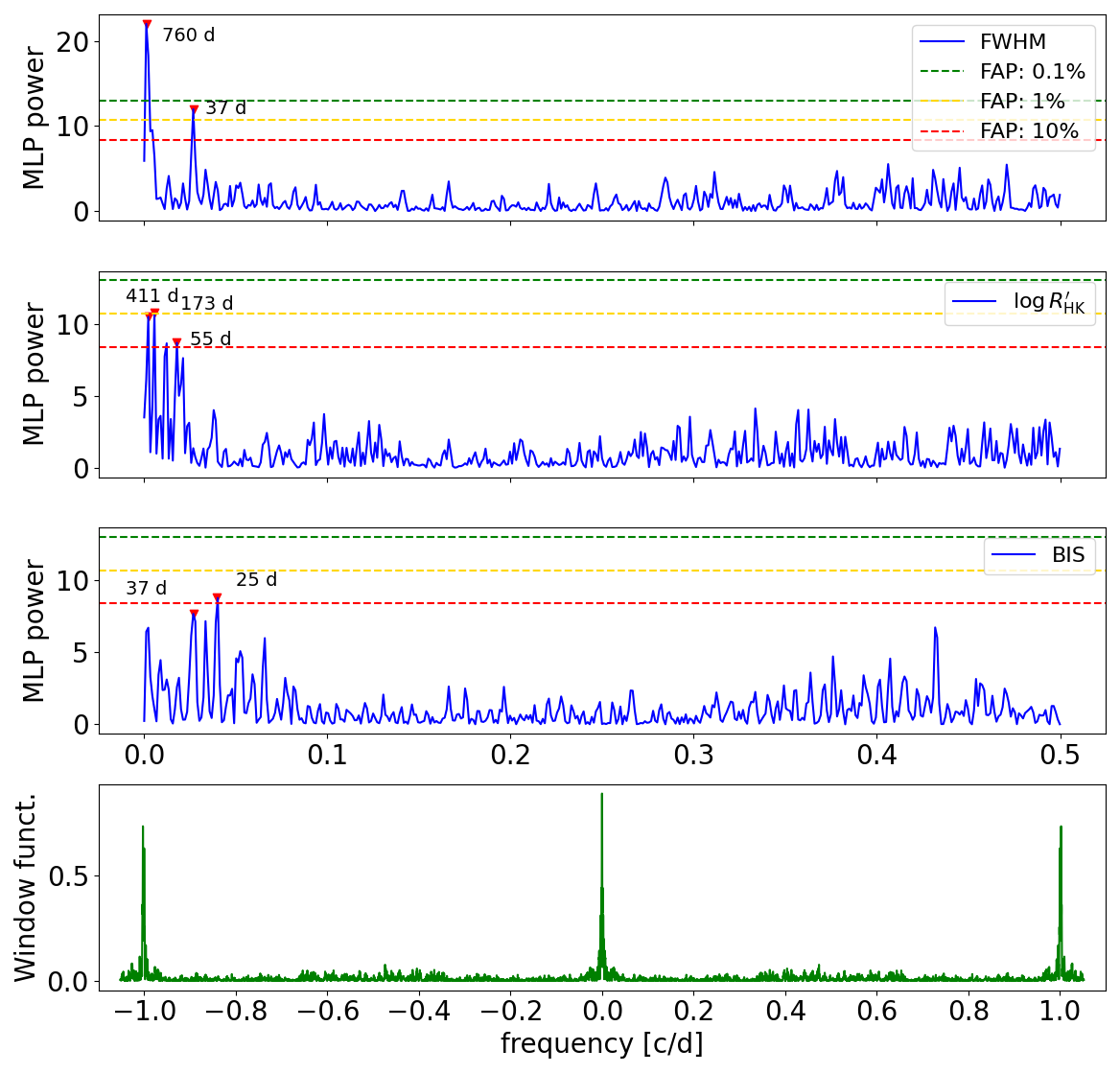}
    \caption{Maximum likelihood periodograms of spectroscopic activity diagnostics of HIP\,29442. The window function of the data is shown in the lower panel. The periodograms include levels of analytical FAP.}
    \label{fig:mlpactdiagnostics}
\end{figure*}

HIP\,29442 appears to be a chromospherically quiet star over the timespan of the observations, with the $\log R^{\prime}_{\rm HK}$ index, calculated by the DRS v.3.0.0 pipeline using the Calcium (Ca) H$\&$K lines, first deriving the well-known Mount Wilson chromospheric S-index, then converting it to $\log R^{\prime}_{\rm HK}$ following \cite{noyes1984ApJ...279..763N}. The $\log R^{\prime}_{\rm HK}$ index having an average of -5.25 dex and rms of 0.12 dex. We investigated the frequency content of the $\log R^{\prime}_{\rm HK}$ index, and of the spectral FWHM and BIS calculated from the CCF. The analysis consisted of the calculation of the maximum likelihood periodograms (MLP; \citealt{zech2019}), using the implementation of the MLP code included in the \textsc{Exo-Striker} packagThe $\log R^{\prime}_{\rm HK}$ e\footnote{see \url{https://ascl.net/1906.004}} (version 0.65). Differently from the Generalised Lomb-Scargle periodogram (GLS; \citealt{zech2009}), the y-axis of the MLP represents the difference ${\rm \Delta}(\ln\,L)$ between the logarithm of the likelihood function corresponding to the best-fit sine function for each tested frequency, and that of a constant function. The levels of false alarm probability (FAP) are calculated analytically. The MLP algorithm includes zero points as free parameters in case of dataset from different instruments, as well as instrumental uncorrelated jitter terms. In our case, we treated the data collected pre- and post-COVID interruption as two independent datasets.
The MLPs are shown in Fig. \ref{fig:mlpactdiagnostics}. The periodogram of the FWHM shows a significant peak at 760 days, which is close to the time span of the dataset, and a second peak at 37 days, which is also detected in the MLPs of the RV residuals (see Sect. \ref{sec:rvfreqanalysis}). The origin of the first and lower frequency signal is unclear. The periodograms of the $\log R^{\prime}_{\rm HK}$ and BIS activity diagnostics do not show significant peaks around 37 days. That of $\log R^{\prime}_{\rm HK}$ shows 1\% significant peaks at 173 and 411 days, the latter also appearing with low significance in the periodogram of the RV residuals.     
We interpret the 37-d signal as the signature of the stellar rotation period. Using our derived values of the projected rotation velocity $v \sin i_{\star}$ ($1.94 \pm 0.45$ km/s) and stellar radius $R_{\star}$ ($0.96^{+0.04}_{-0.03} \rsun$), we can assess the maximum stellar rotation period. We find $P_{\rm rot,\, \star max}=24.9^{+7.5}_{-4.7}$ days. This value is compatible within $2\sigma$ with the periods detected in the activity diagnostics and RVs. This inconsistency between $P_{\rm rot,\, \star max}$ and the period detected in the data can be justified taking into account that, since HIP\,29442 is a slow rotator, it is difficult to measure a value of $v \sin i_{\star}$ lower than that we determined (i.e. reach a sensitivity to longer $P_{\rm rot,\, \star max}$), because we are limited by the width of the instrument profile. 

We note that there is no evidence for a clear rotational modulation in the TESS light curve. We searched for a possible rotation modulation within the ASAS-SN public data\footnote{\url{https://asas-sn.osu.edu/}} \citep{shappee2014ApJ...788...48S,kochanek2017PASP..129j4502K}. HIP\,29442 has been observed with six cameras since 2013 (five facilities used a g-band filter, one employed a V-band filter), making a total of 1313 photometric data points to date. We shifted the photometric aperture on a yearly basis to correct for the target proper motion, in order to ensure a proper centring of the star, hence avoiding flux loss from the aperture. Data from the six cameras are shown in Fig. \ref{fig:asas}. Observations in V-band have been collected up to 2019. Observations in g-band appear quite different from camera to camera. Data from camera bE are particularly affected by large systematics, and there is a clear jump visible in the last part of the data that we cannot explain. Therefore, we discarded data from camera bE. We applied a 3$\sigma$ clipping to the data, performed a frequency analysis to the photometric time series, exploring frequencies up to 1/100 c/d, in order to be sensitive to a rotational modulation. Given the limited time coverage of the data, and that the g-band time series show not consistent long-term trends, albeit they cover a similar time span, we cannot draw conclusions about the presence of real long-term photometric modulations. We have calculated Lomb-Scargle periodograms, without finding significant peaks, concluding that ASAS-SN data do not allow us to detect the stellar rotation period.

\section{RV and light curve analysis}
\label{sec:rvlcanalysis}
\subsection{Frequency content analysis of the RVs}
\label{sec:rvfreqanalysis}

\begin{figure}
    \centering
    \includegraphics[width=\linewidth]{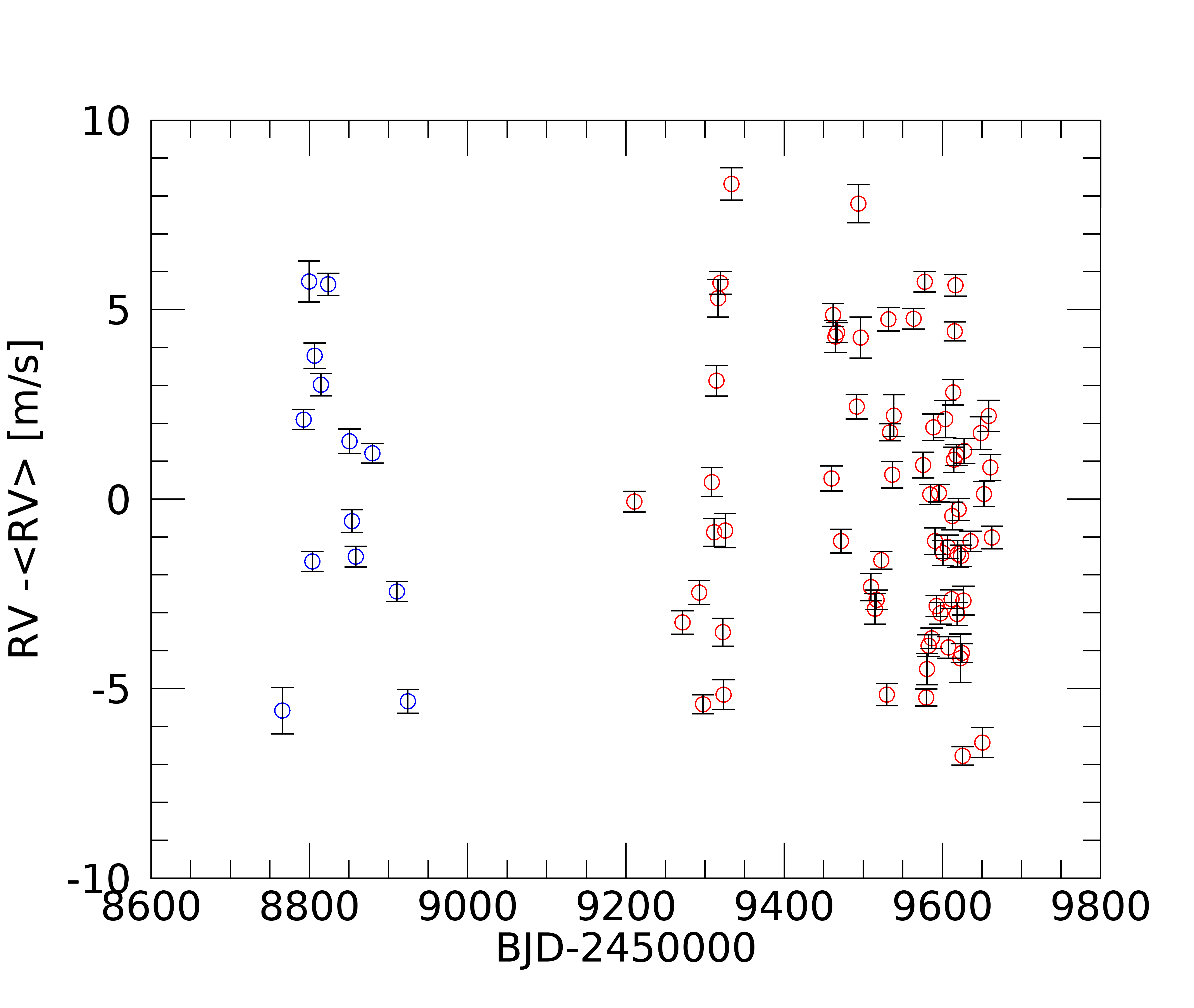}
    \caption{ESPRESSO RVs. Blue circles: pre-COVID dataset; red circles: post-COVID dataset. The mean value of the full dataset has been subtracted from the original data. No offset has been subtracted from the two sub-samples. The error bars represent the formal RV uncertainties derived by the DRS pipeline. }
    \label{fig:rv}
\end{figure}

\begin{figure*}
    \centering
    \includegraphics[scale=0.5]{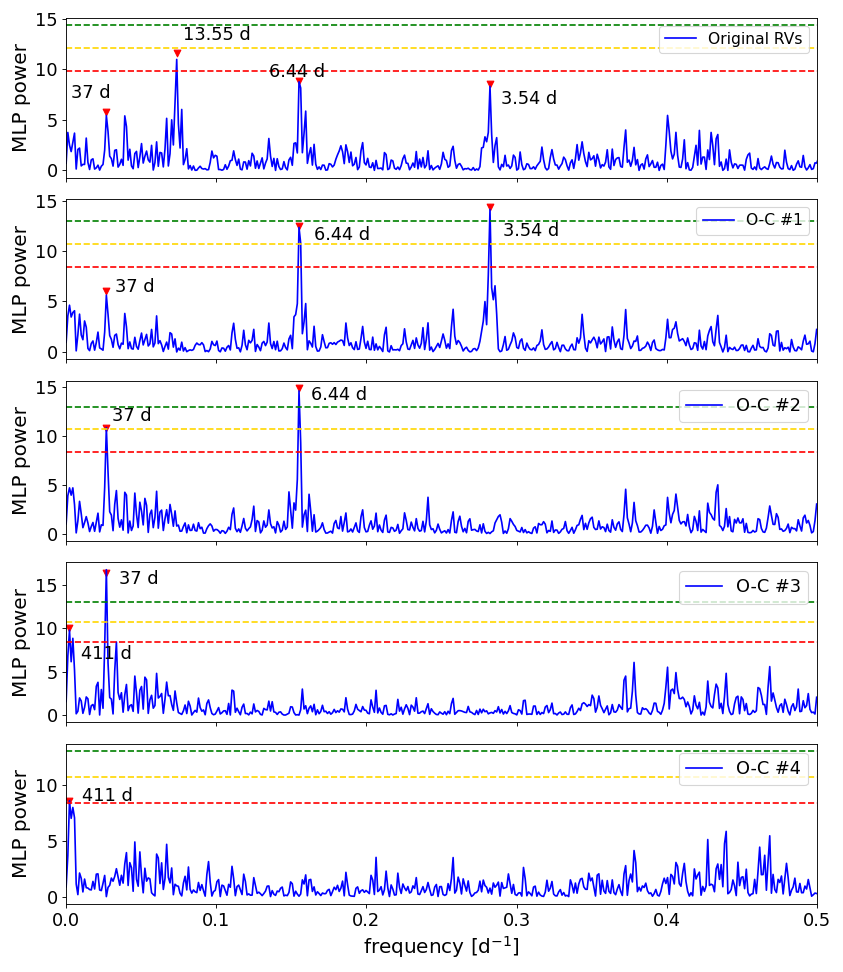}
    \caption{Maximum likelihood periodograms of the original RVs (upper panel), and of the RV residuals O-C after subtracting recursively the best-fit sinusoidal signal with the highest peak frequency. The periodograms include levels of analytical FAP. The window function of the data is the same as in Fig. \ref{fig:mlpactdiagnostics}. }
    \label{fig:rvmlpwf}
\end{figure*}

The time series of the ESPRESSO RVs is shown in Fig. \ref{fig:rv}. Using MLPs, we determined the frequency content of the original RV dataset and of the residuals, after subtracting the best-fit sinusoidal signal with the highest peak frequency recursively (Fig. \ref{fig:rvmlpwf}). The MLP of the original data shows the main peak very well compatible with the orbital period of TOI-469.01 ($\sim$13.55 days), and two additional peaks at 3.54 and 6.44 days. They are still present in the MLP after a first pre-whitening (i.e. after removing the best-fit sinusoid with a period corresponding to the peak with the highest power in the MLP) and with higher significance (second panel of Fig. \ref{fig:rvmlpwf}). A further pre-whitening increases the significance of the 6.44-day period, and also that of a 37-day signal, that was detected in the previous MLPs with low power. A peak at 37 days is also present in the periodograms of the FWHM with a FAP lower than 1$\%$ (Sect. \ref{sec:staractivity}), and we attribute the corresponding signal to stellar magnetic activity, i.e. likely being the rotational frequency of HIP\,29442. The MLP of the RV residuals, labelled O-C $\#$3 and $\#$4 in Fig. \ref{fig:rvmlpwf}, show a peak at 411 days, that was not significant in the previous periodograms. This signal has a counterpart in the periodogram of the $\log R^{\prime}_{\rm HK}$ (\ref{sec:staractivity}), and could be induced by some form of stellar activity. 

The preliminary results provided by the MLP show that the signal at $\sim 13.55$ days has a semi-amplitude of $\sim 2.6\pm0.5 \ms$, which allows us to label TOI-469.01 as a planet (minimum mass $m\sin i \cong 9 \mearth$), thoroughly characterised in Sect. \ref{sec:rvfit}. Therefore, hereafter we refer to TOI-469.01 as HIP\,29442\,$b$. Following the detection in the MLP of the lower-amplitude signals at 3.54 and 6.44 days ($K=2.2\pm0.4$ and $1.9\pm0.3$ $\ms$, respectively, both corresponding to $\sim$5 $\mearth$), we performed an in-depth analysis of the TESS light curve to search for transits at such periods that remained undetected and have not been reported in the ExoFOP website. 

\subsection{Search for additional transits in the TESS light curve}
\label{sec:newtesstransits}

The discovery of two additional signals in the RVs triggered a careful analysis of the light curve in order to reveal shallow transits not detected in previous analyses. In the beginning, we used a first version of the detrended light curve. We modelled the variability with a Gaussian process (GP) regression, using a covariance matrix obtained by combining a SHOTerm and RotationTerm kernel as implemented in \texttt{Celerite2} \citep{celerite1, celerite2}. The SHOTerm represents a stochastically driven, damped harmonic oscillator, with a power spectrum of that can be written as

\begin{equation} \label{eq:sho}
    S(w) = \sqrt{\frac{2}{\pi}} \frac{S_{0} w_{0}^{4}}{(w^{2}-w_{0}^{2})^{2} + 2w^{2} w_{0}^{2}},
\end{equation}

\noindent where $S_{0}$ is the amplitude, and $w_{0}$ is the angular frequency corresponding to the break point in the power spectral density of the kernel. A mixture of two SHO terms can be used to model stellar rotation. The model was trained on a light curve consisting of 44-min binned PDCSAP data to avoid fitting short-term variations. The parameters were optimised to find the maximum a posteriori parameters values, then the model was sampled using the \texttt{PyMC3-Extras} \citep{exoplanet:pymc3} library and the \texttt{exoplanet} package \citep{exoplanet:joss}, using two chains with 2500 tuning steps and another 2500 steps each. The corresponding trend was then interpolated and removed from the original light curve. We also flag and remove observations more than five times the standard deviation above the mean of the corrected flux. A similar approach is used for observations more than five times the standard deviation under the mean corrected flux, with the difference that each observation $o_i$ whose two neighbours $o_{i-1},o_{i+1}$ were considered outliers as well were considered of probable physical origin, therefore not removed from the light curve. As this is a blind analysis to search for additional transits, the model does not include transit signals in addition to the GP signal, therefore transits that are still undetected could be removed or their profiles altered. For this reasons, we did not adopt this light curve as our final dataset. We will adopt a more thorough model to extract the final detrended light curve, as described in Sect. \ref{sec:tessdetrending}. We searched for transit signals through the \texttt{Transit Least-Squares} (TLS) periodogram \citep{HippkeTLS2019}, which easily recovered the HIP\,29442\,$b$ signal with a high Signal Detection Efficiency (SDE). We used \texttt{starry} \citep{starry2019} to model the light curve and adopted a quadratic limb darkening law \citep{kipping2013}. For a quicker fit, we used a sub-sample of data centred around the mid-transit times spanning for 0.25 days before and after each transit, and considered a circular orbit. Using the period, time of transit centre and duration resulting from this fit, we masked the transits of HIP\,29442\,$b$ in the light curve, and ran the TLS search again. The second TLS search reveals a signal with a SDE of 12.31 at $\sim$3.54 days. The same methodology was applied once again to mask these additional transits. The third iteration with TLS resulted in the identification of a peak with SDE=7.06 at $\sim$6.43 days. Different SDE thresholds have been proposed in the literature to claim a detection. \citet{Dressing2015} consider SDE$>$6 to be a sufficient threshold, while other authors suggest fixing the threshold at SDE$>$7 \citep{Siverd12, HippkeTLS2019}. According to the TLS documentation\footnote{\url{https://transitleastsquares.readthedocs.io/en/latest/FAQ.html}}, an SDE=7 corresponds to a false alarm probability (FAP) of 1$\%$, and SDE=8.3 to a FAP of 0.1$\%$. A further iteration with TLS resulted in peaks which are not significant, and do not have a counterpart in the RVs.
In the end, we detected four transits of HIP\,29442\,$b$ (two in S6, and two in S33), twelve transits of HIP\,29442\,$c$ (six in S6, and six in S33), and seven transits of HIP\,29442\,$d$ (three in S6, and four in S33). We note that one transit of HIP\,29442\,$c$ is almost overlapped to a transit of HIP\,29442\,$b$ (around BJD 2\,459\,210.7).

\begin{figure}
    \centering
    \includegraphics[width=0.42\textwidth]{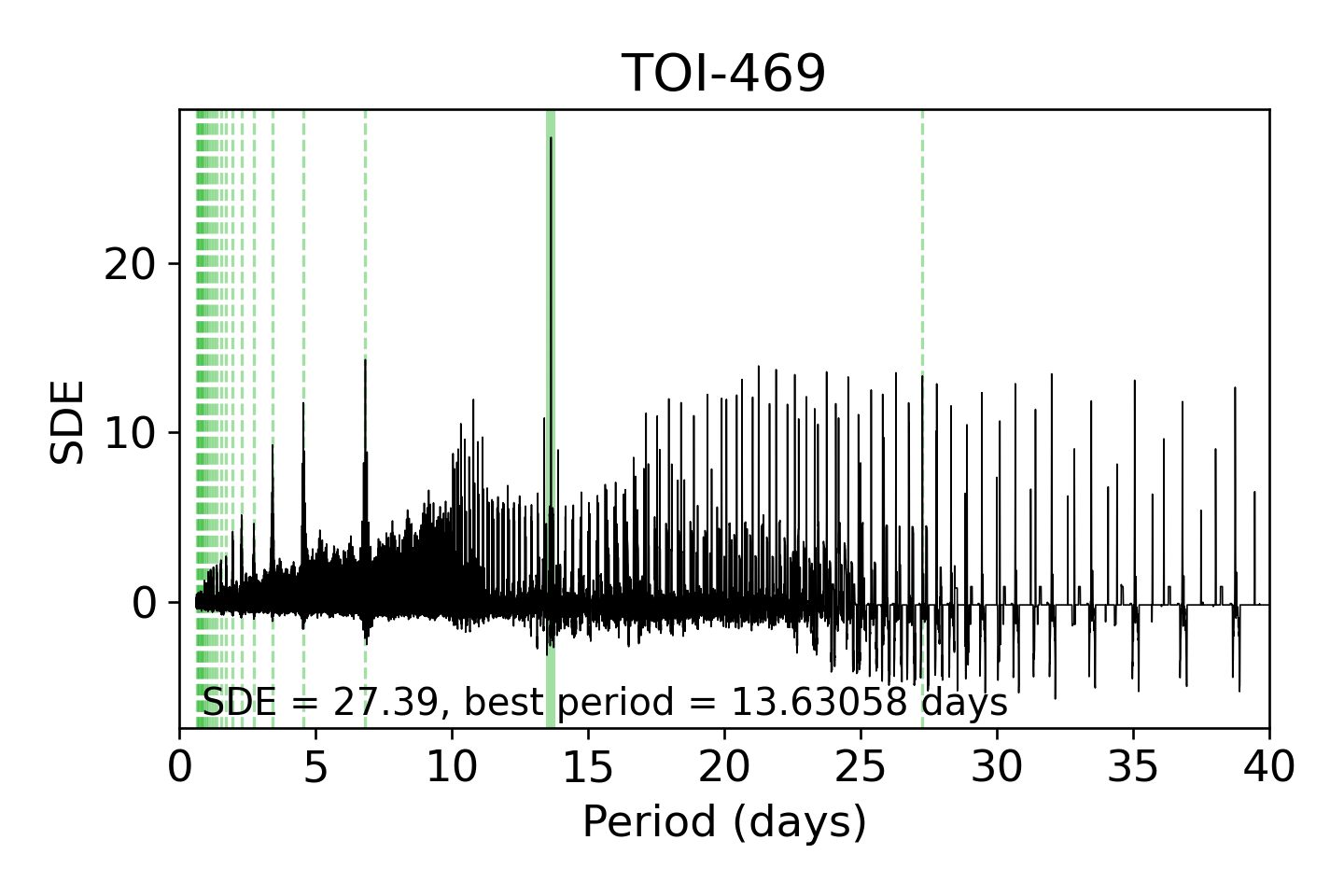}\\
    \includegraphics[width=0.42\textwidth]{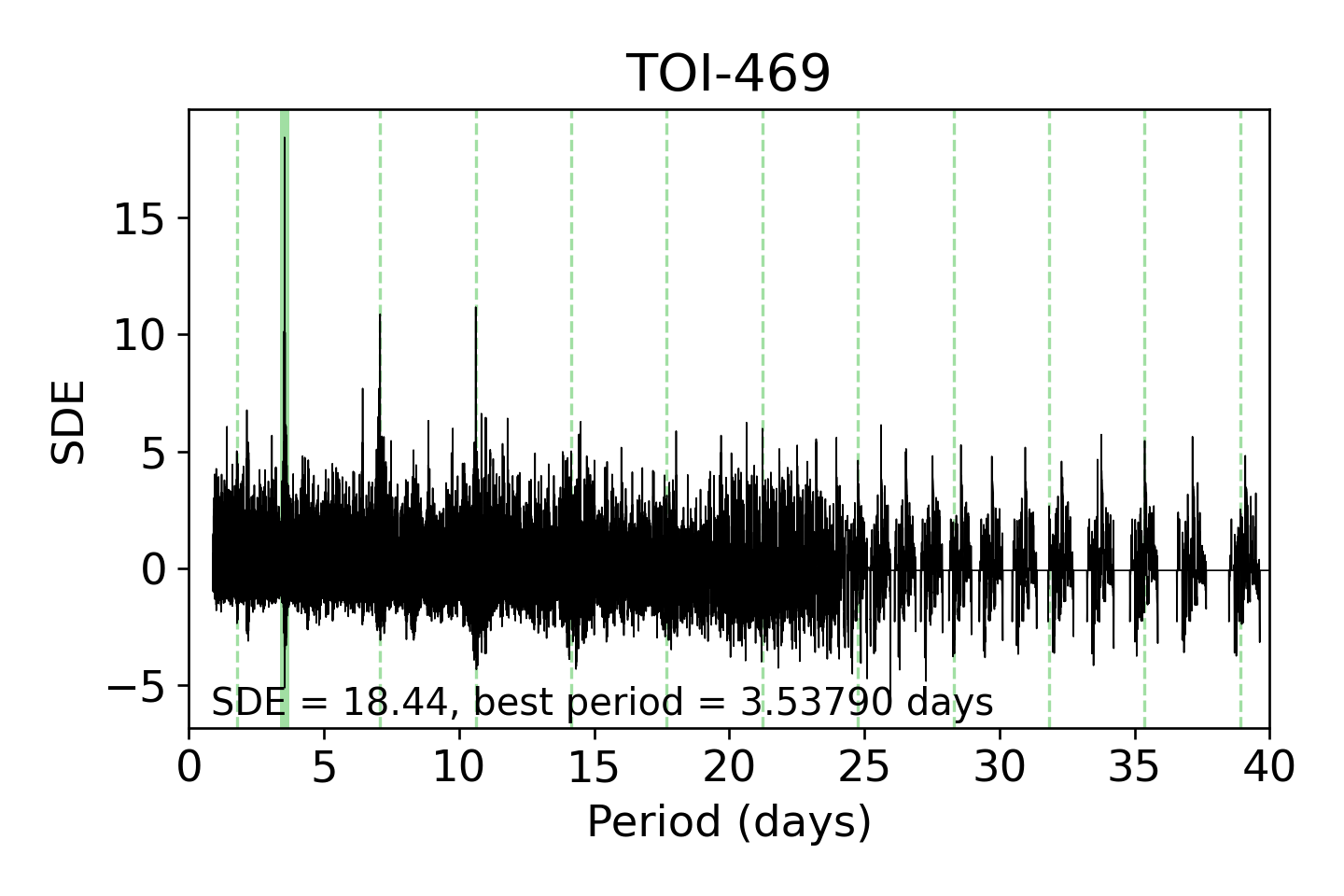}\\
    \includegraphics[width=0.43\textwidth]{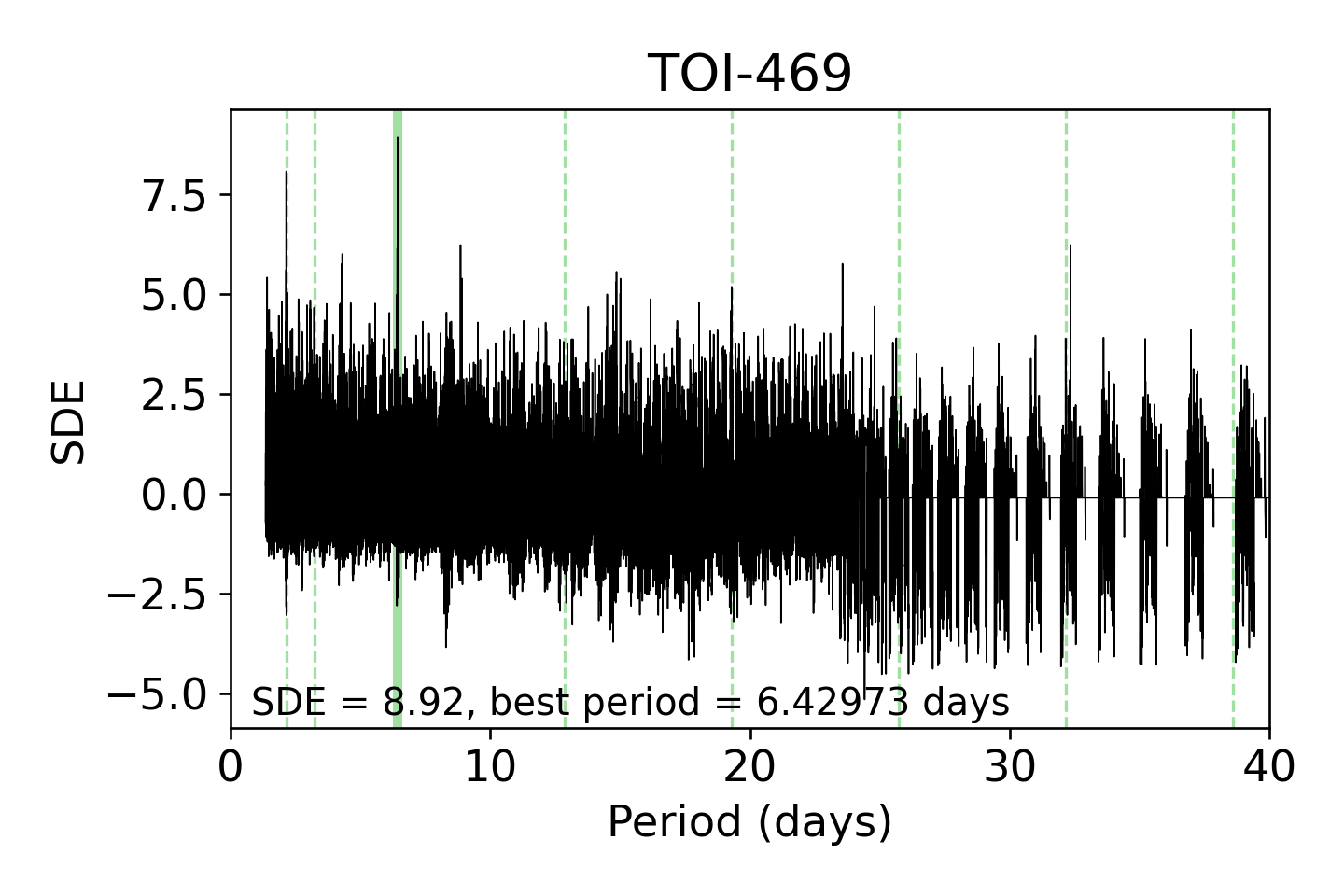}
    \caption{TLS periodograms of the detrended \textit{TESS} light curve (see Sect. \ref{sec:tess_lc}), showing peaks of the three transiting planets. The periodogram in the second panel has been obtained after masking the transits of planets $b$, and the one in the third panel has been calculated after masking transits of planets b and c. Dashed lines correspond to the harmonics of the orbital periods. Planets c and b have been identified first in the ESPRESSO RV time series, their presence confirmed by the transit signals detected in the \textit{TESS} data.}
    \label{fig:tlssde}
\end{figure}

After this first analysis with TLS, all the transits were masked in the PDCSAP light curve, and the stellar activity was modelled again using the same approach as described above, interpolating the best-fit model to correct the masked datapoints. We analysed this new version of the light curve again with TLS to re-assess the SDE of the transit signals. The final TLS periodograms are shown in Fig. \ref{fig:tlssde}. The SDE of the signals at P=3.54 and P=6.43 increased to 18.4 and 8.9, respectively. 
The transits correspond to a $R\sim$3.6 $\rearth$ (HIP\,29442\,$b$) and to two Earth-sized companions on innermost orbits. The orbital periods of these two innermost companions derived from TESS photometry are perfectly in agreement with the results of the RV frequency analysis. Therefore, we conclude that photometry and RVs both significantly reveal the presence of two additional companions that, taking into account their preliminary mass and size, can be confirmed as planets. Hereafter, we label HIP\,29442\,$c$ and HIP\,29442\,$d$ the planets with the shortest and intermediate orbital periods, respectively.     

\subsection{Modelling of the TESS data}
\label{sec:tessdetrending}

We found low S/N transits of HIP\,29442\,$c$ and HIP\,29442\,$d$ in the TESS data, which is the reason why they have low SDE and went undetected so far. This also implies that the detrending of the light curve has to be performed with special care in order to minimise the distortion of the transit profiles, which would hamper a correct retrieval of the transit parameters and the planetary radii. To this purpose, we made a step further with respect to the preliminary analysis described in Sect. \ref{sec:newtesstransits} by fitting the TESS PDCSAP curve through a model composed of two components: the transit signals, and a GP regression to model photometric correlated signals. Modelling these two components jointly has been proved to be very successful in preserving the transit shapes \citep[e.g.][]{2021A&A...645A..41L,2021A&A...649A.144S,2023A&A...669A.109L}, especially for transits with low S/N such as those of HIP\,29442\,$c$ and HIP\,29442\,$d$ \citep[e.g.][]{2022MNRAS.515.1328D, 2022A&A...668A..85C}. Besides, this approach ensures a correct propagation of the uncertainties of the model parameters \citep{2021A&A...649A..26L}.
We used the \texttt{batman} implementation \citep{batman05} of the quadratic limb darkened transit model formulated by \citet{2002ApJ...580L.171M}. The model is described by the orbital period ($P_{\rm orb}$), the time of inferior conjunction ($ T_{\rm conj}$), the orbital inclination angle ($i_p$), and the radius ($R_{\rm p}$) of the transiting planets (in units of stellar radii), as well as by the stellar quadratic limb darkening (LD) coefficients $u_{\rm 1}$ and $u_{\rm 2}$. We used the adopted stellar mass and radius to calculate the stellar density, and used it to better constrain the model parameters $a_{\rm (b,~c,~d)}/R_{*}$ \citep{sozzetti2007ApJ...664.1190S}. 

We modelled the TESS correlated noise through a GP with an approximate Matérn-3/2 kernel \citep[][]{celerite1}, which has been widely used to model TESS PDCSAP \citep[e.g.][]{2022AJ....163..298M,2023A&A...673A..32M,2023arXiv230409220M}. 
We chose this kernel because it has covariance properties that are very well matched to short-term instrumental red-noise structures \citep[e.g.][]{2017AJ....153..177P,stefansson2020AJ....160..259S}. Hence, it is very appropriate for modelling TESS data such as that of TOI-469, in which the stellar rotation is not detectable, and instead residual systematics dominate the correlated noise \citep{2023A&A...675A..52C}. 

The approximate Mat\'ern-3/2 kernel can be written in terms of the temporal separation between two data points $\tau = t_{i} - t_{j}$ as

\begin{equation}
    K_{3/2} = \eta_{\sigma}^{2} \left[ \left(1 + \frac{1}{\epsilon} \right) e^{-(1-\epsilon) \sqrt{3} \tau / \eta_{\rho}} \cdot \left(1 - \frac{1}{\epsilon} \right) e^{-(1+\epsilon) \sqrt{3} \tau / \eta_{\rho}} \right],
\end{equation}

\noindent where the hyperparameters $\eta_{\sigma}$ and $\eta_{\rho}$ are the characteristic amplitude and timescale of the correlated variations, respectively. The parameter $\epsilon$ controls the approximation to the exact Mat\'ern-3/2 kernel, and we fixed it to its default value of $10^{-2}$ \citep{celerite1}. Given that both the amplitudes and frequencies of TESS sytematics might change between sectors, we fit those parameters independently. In addition, we added a jitter term $\sigma_{\rm TESS}$ in quadrature to the flux uncertainties in order to account for the white noise not considered in our model. In order to ensure that our computed planetary parameters are independent of the kernel choice, we also tested a GP with the simple harmonic oscillator kernel described in Eq. \ref{eq:sho}, which we used for a blind transit search and it has been designed to model smooth modulations related to the stellar rotation. We found no differences in the derived model parameters. 

We sampled the posterior probability density function of the parameters in our model through a Markov chain Monte Carlo (MCMC) affine-invariant ensemble sampler \citep{2010CAMCS...5...65G} as implemented in \texttt{emcee} \citep{2013PASP..125..306F}. We performed a first run with 200$\,$000 iterations, and then reset the sampler and performed a second run with 100$\,$000 iterations starting from the maxima of the posterior probabilities as computed from the last iteration of the first run. For each run, we used four times as many walkers as the number of parameters. Once the process is finished, we estimated the autocorrelation time for each fitted parameter and verified that it is at least 30 times less than the chain length, hence ensuring the convergence. 

\begin{figure*}
    \centering
    \includegraphics[width = \textwidth]{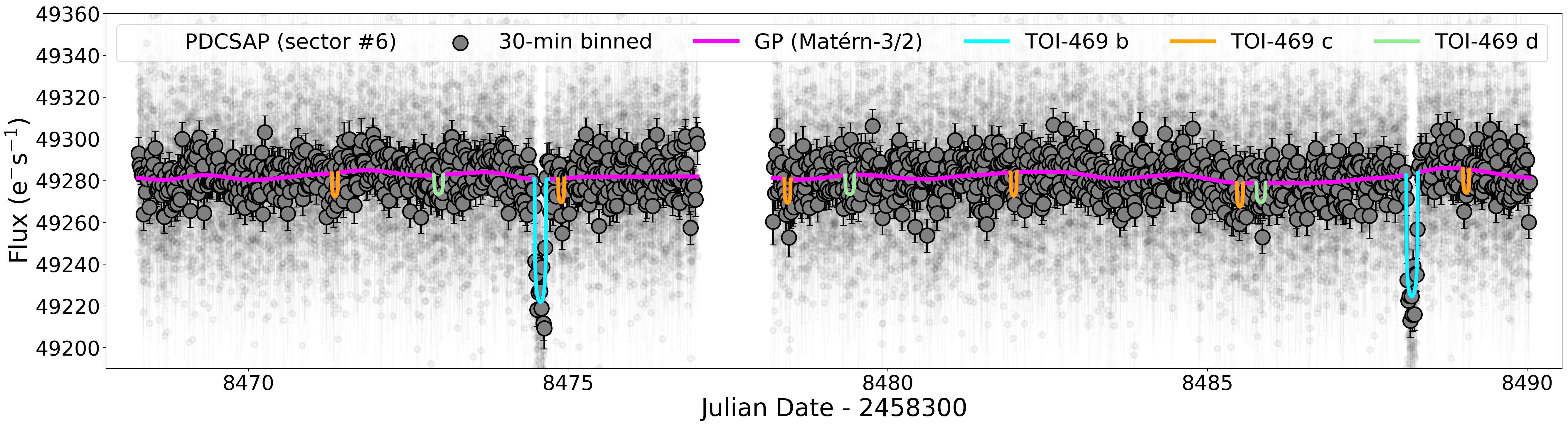}
    \includegraphics[width = \textwidth]{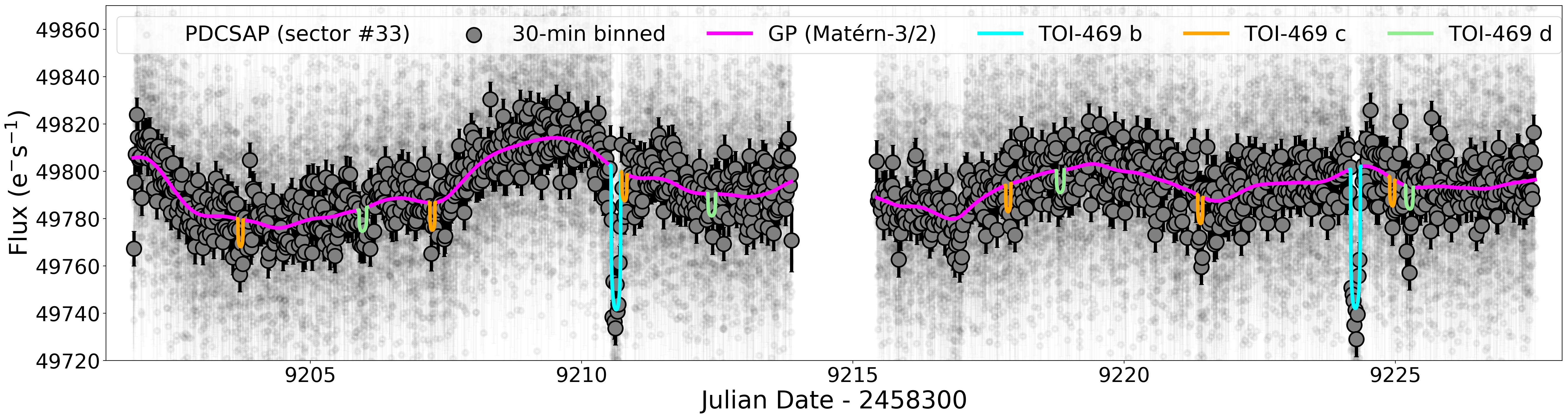}
    \caption{TESS photometry of HIP\,29442 (TOI-469) with the median posterior model (transits + GP) overplotted, showing the different locations of the transit events for the three planets over the low-frequency photometric modulation. The light gray data points correspond to the SPOC 2-minute cadence PDCSAP photometry.}
    \label{fig:complete_TESS_lc}
\end{figure*}

\begin{figure*}
    \centering
    \includegraphics[scale = 0.4]{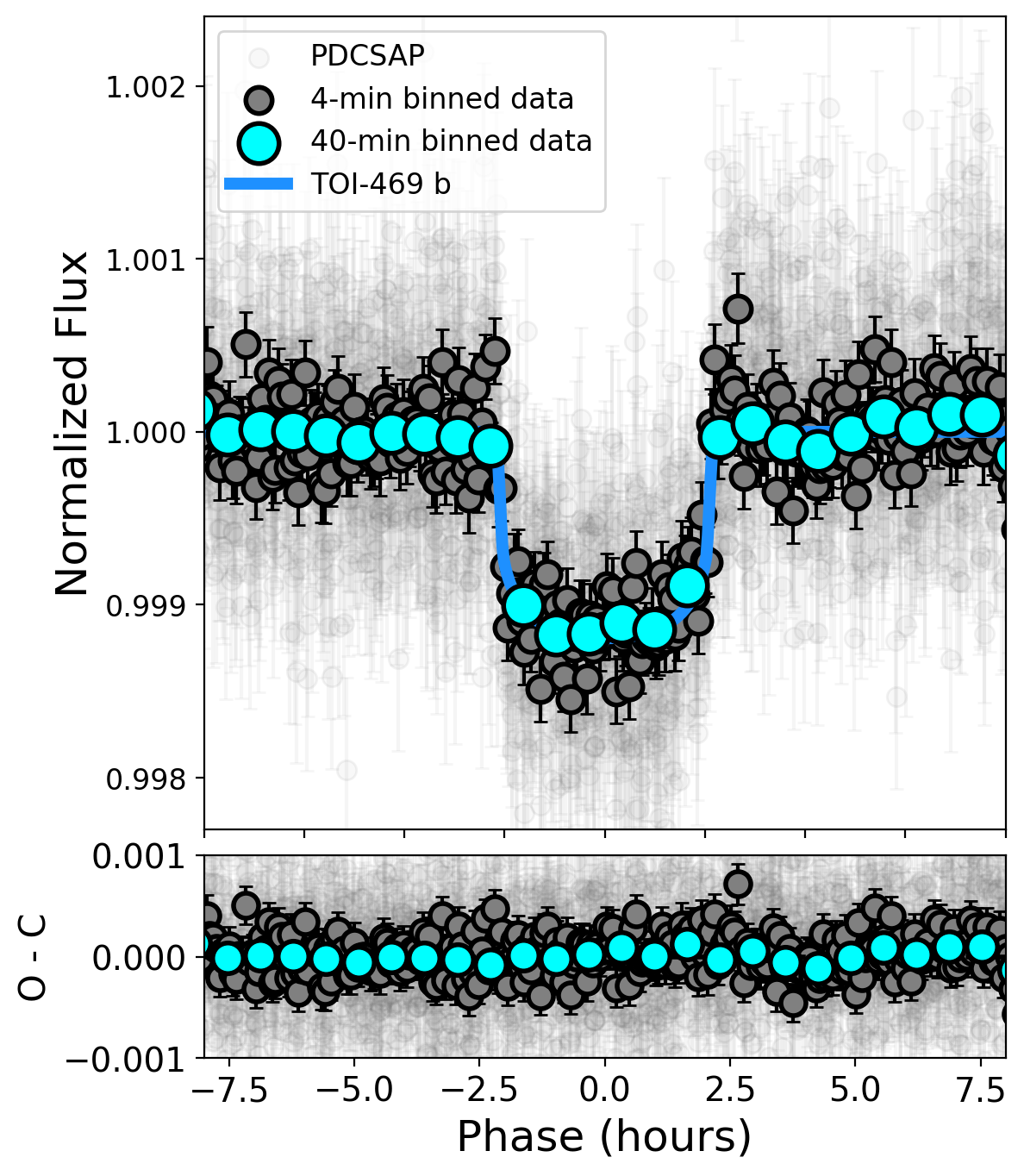}
    \includegraphics[scale = 0.4]{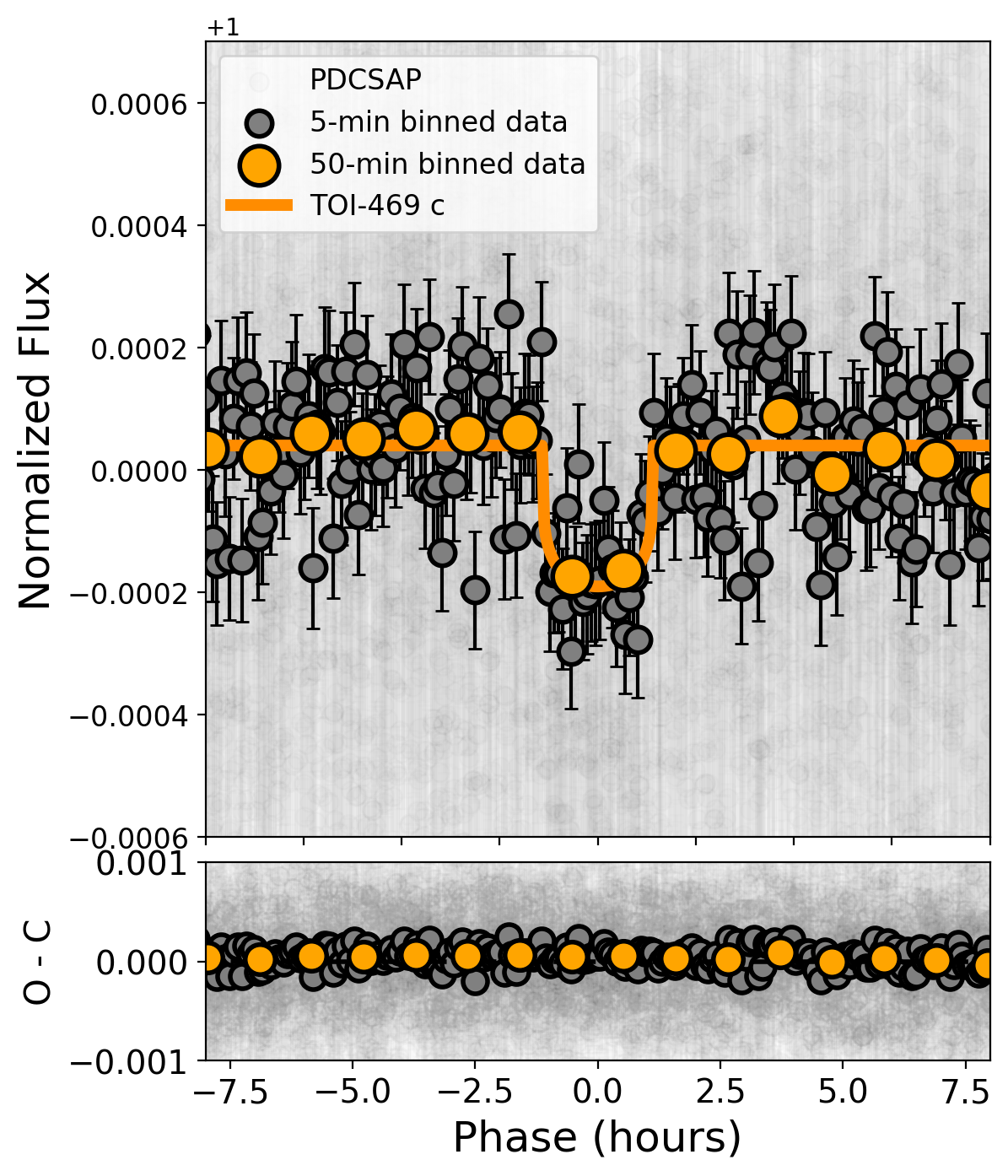}
    \includegraphics[scale = 0.4]{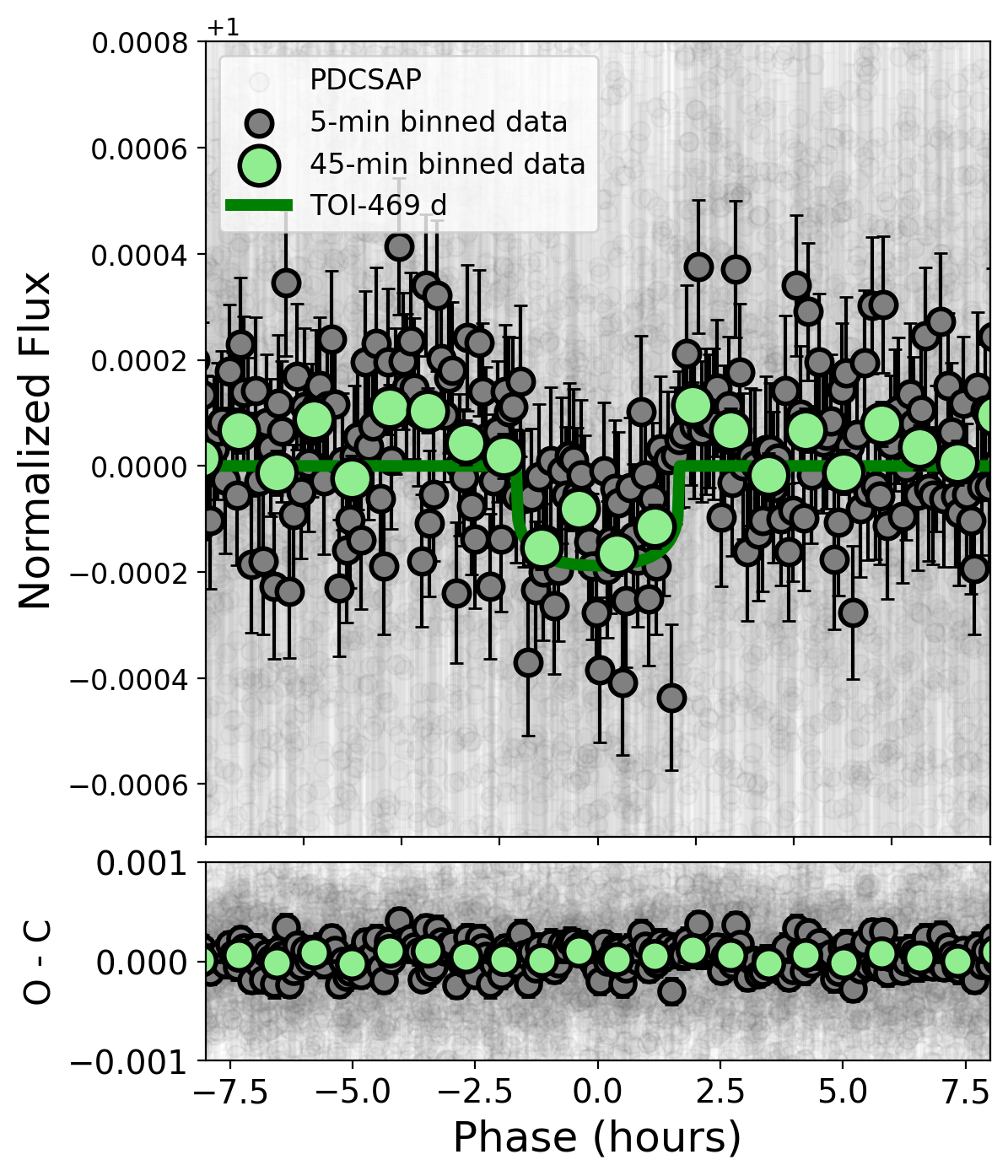}
    \caption{TESS PDCSAP light curve with the GP component substracted, and folded to the orbital periods of HIP\,29442\,$b$ (first panel), HIP\,29442\,$c$ (second panel), and HIP\,29442\,$d$ (third panel). The solid lines indicate the models calculated using the median of the posteriors. Residuals of the best-fit model are shown at the bottom of each phase-folded plot.}
    \label{fig:phase_folded_transits}
\end{figure*}

\begin{figure*}
\centering
    \includegraphics[scale=0.22]{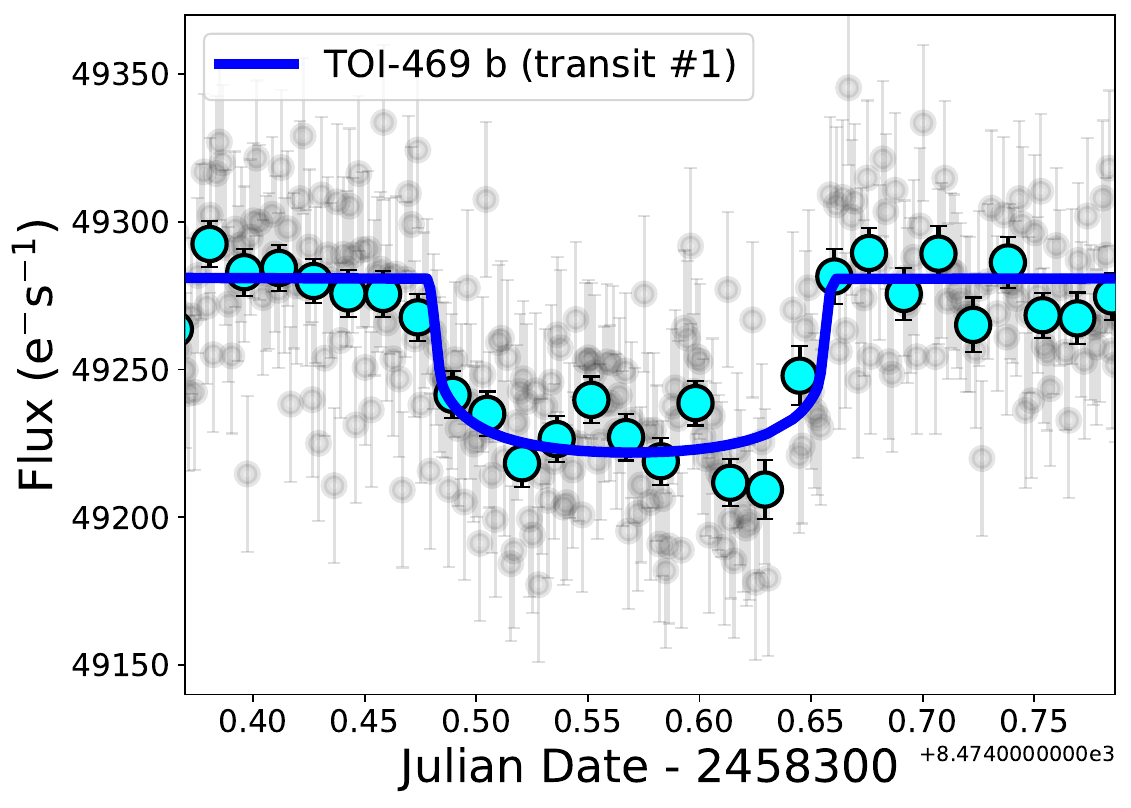}
    \includegraphics[scale=0.22]{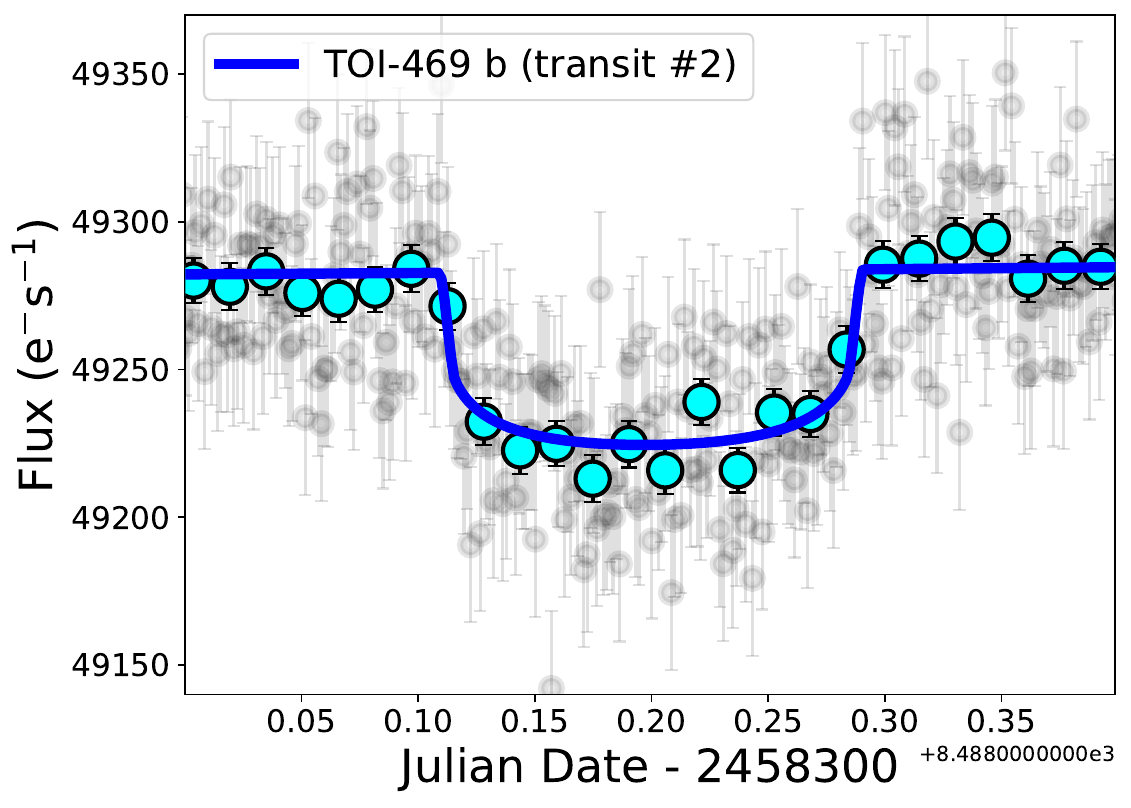}
    \includegraphics[scale=0.22]{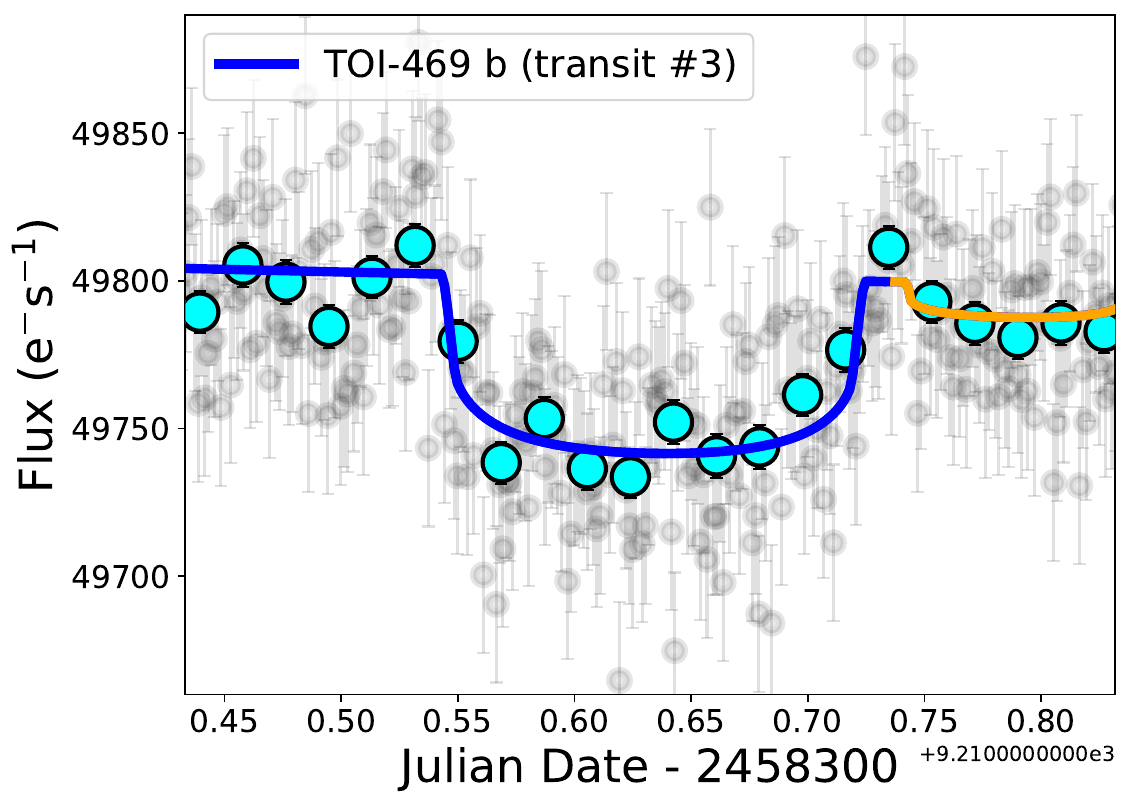}
    \includegraphics[scale=0.22]{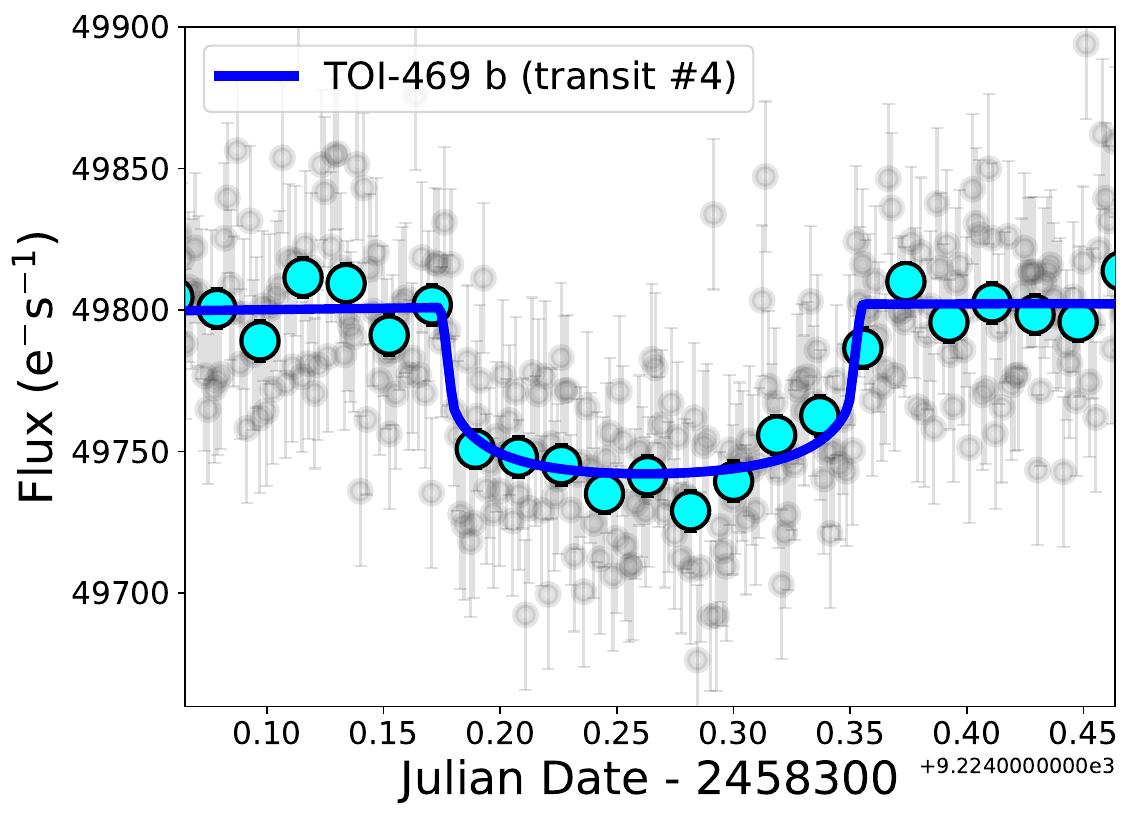}
    \includegraphics[scale=0.22]{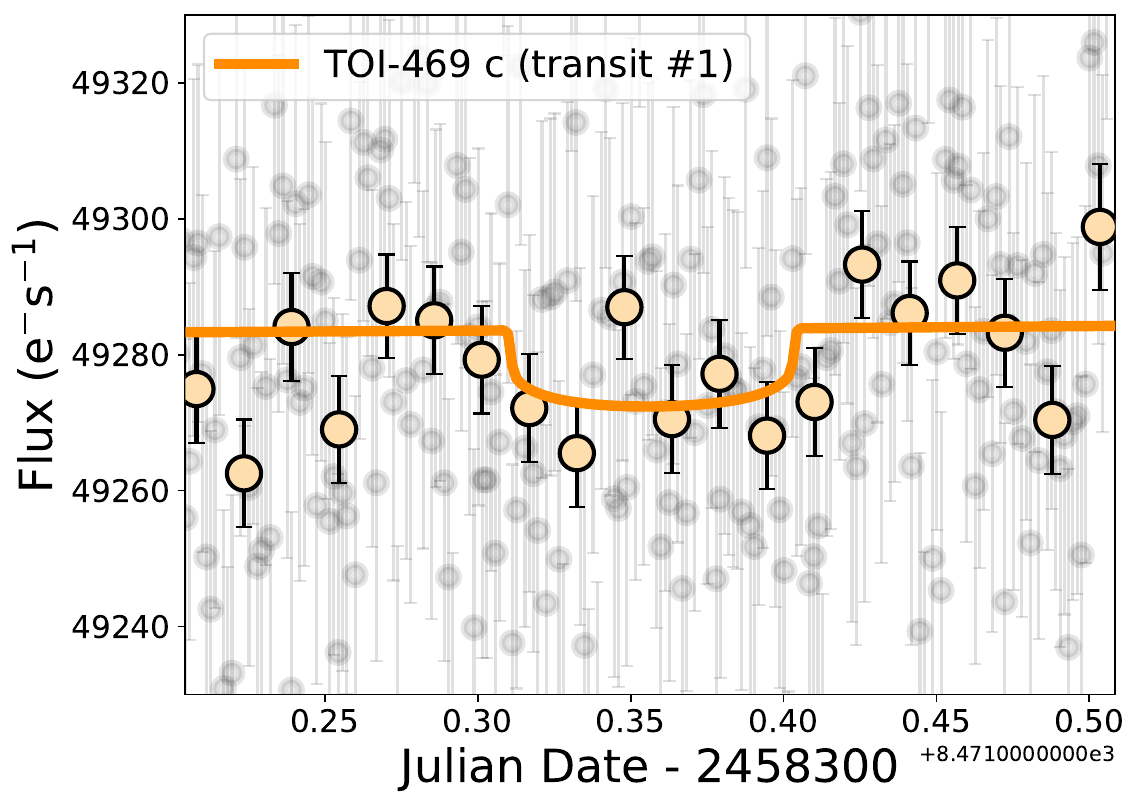}
    \includegraphics[scale=0.22]{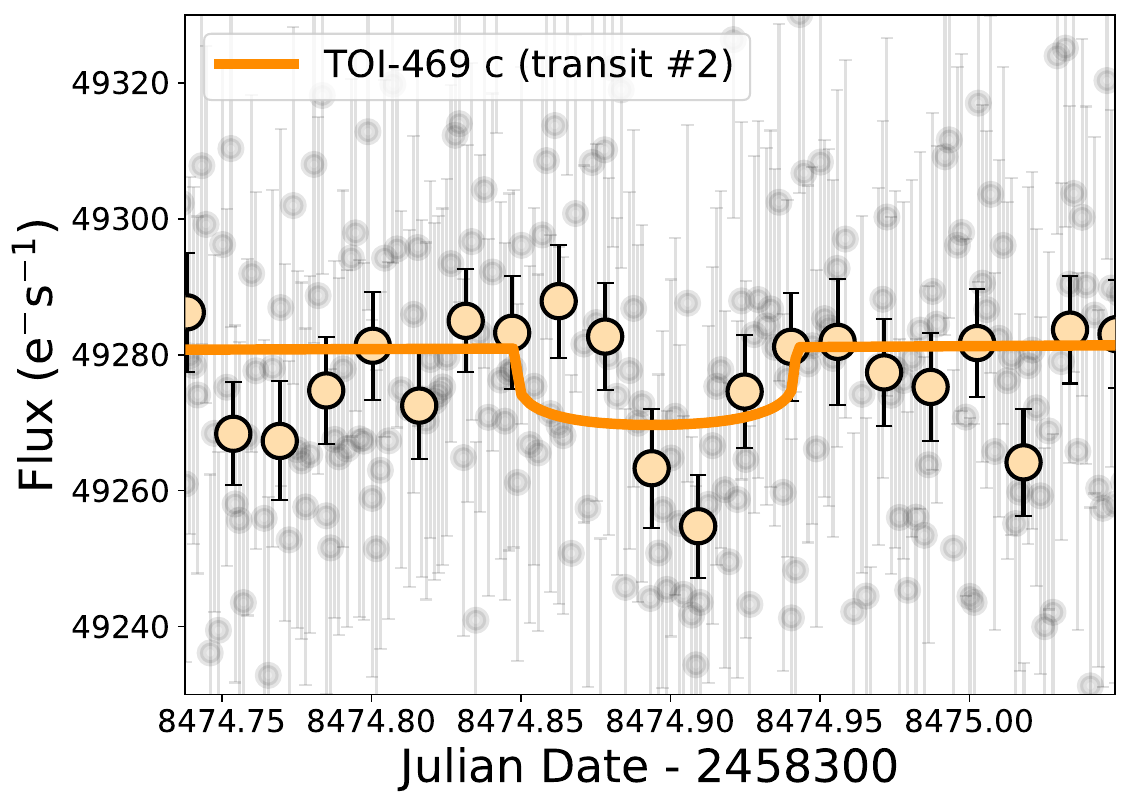}
    \includegraphics[scale=0.22]{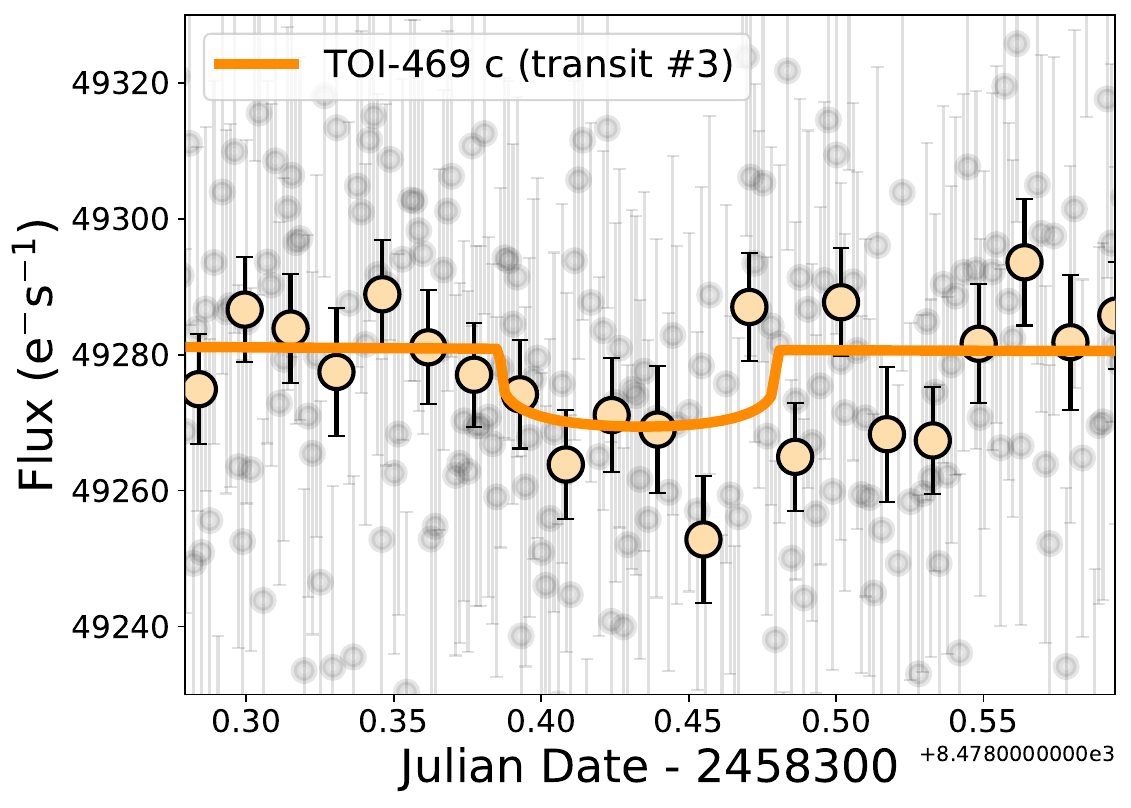}
    \includegraphics[scale=0.22]{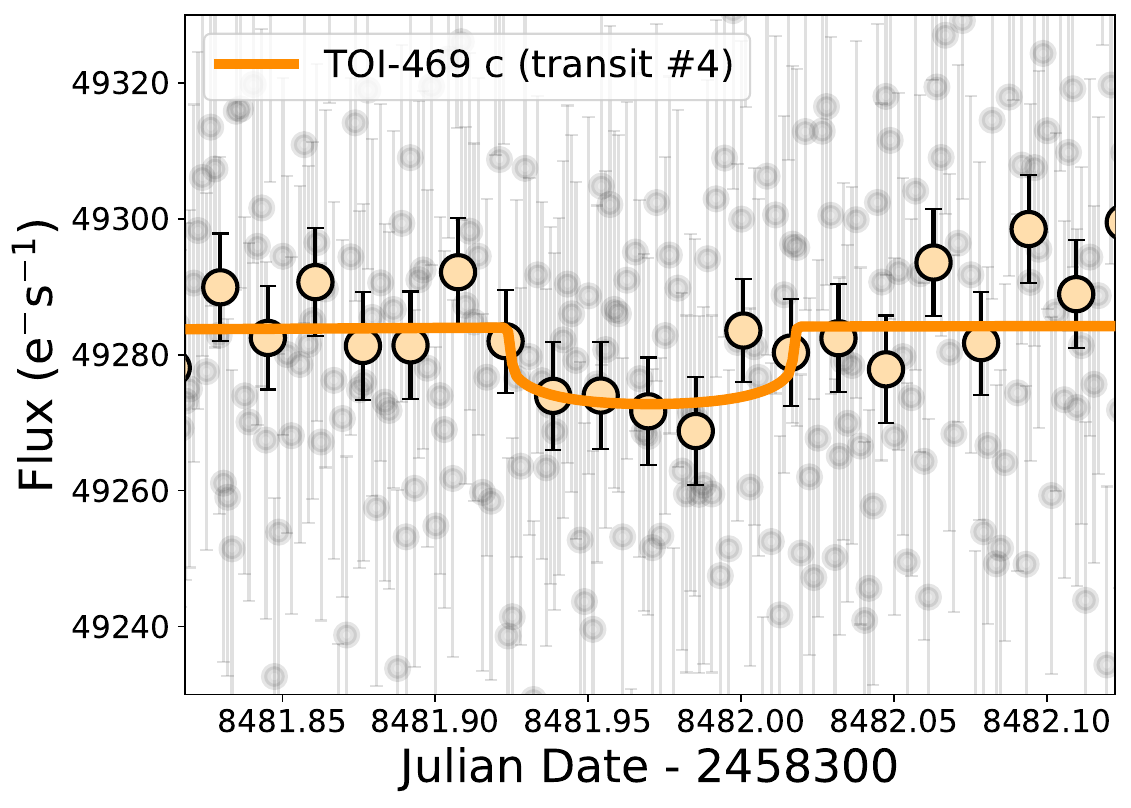}
    \includegraphics[scale=0.22]{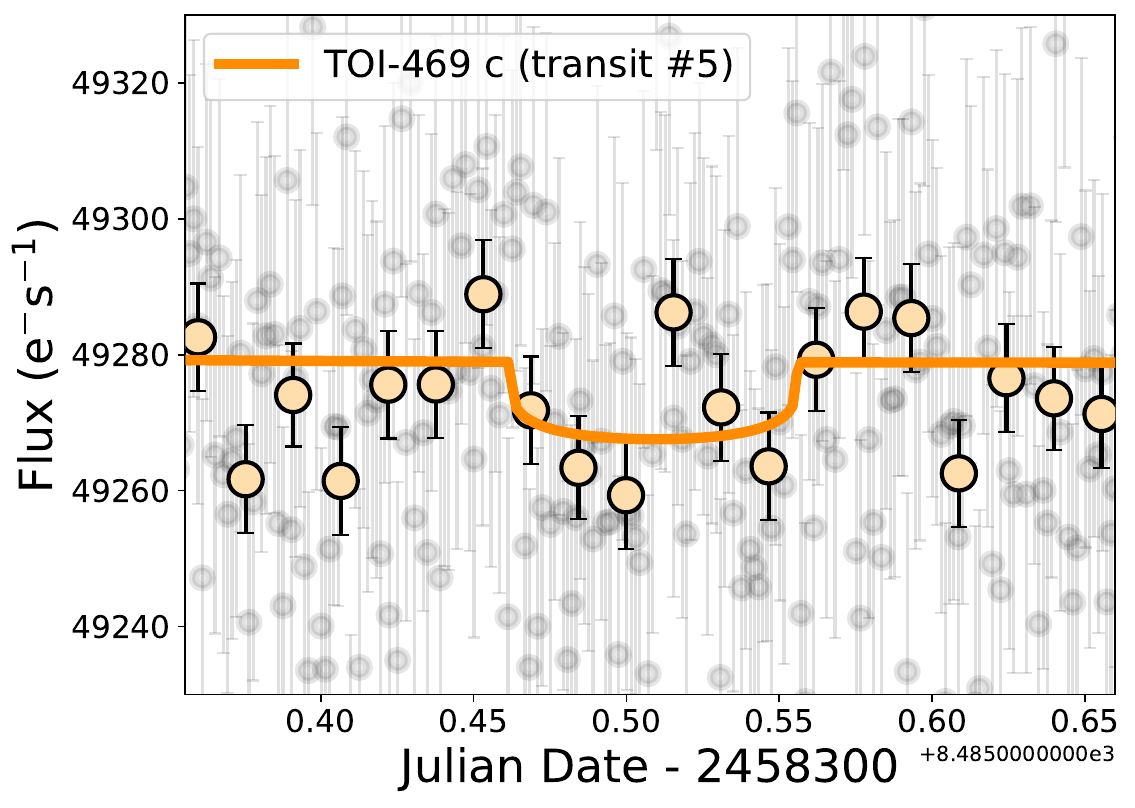}
    \includegraphics[scale=0.22]{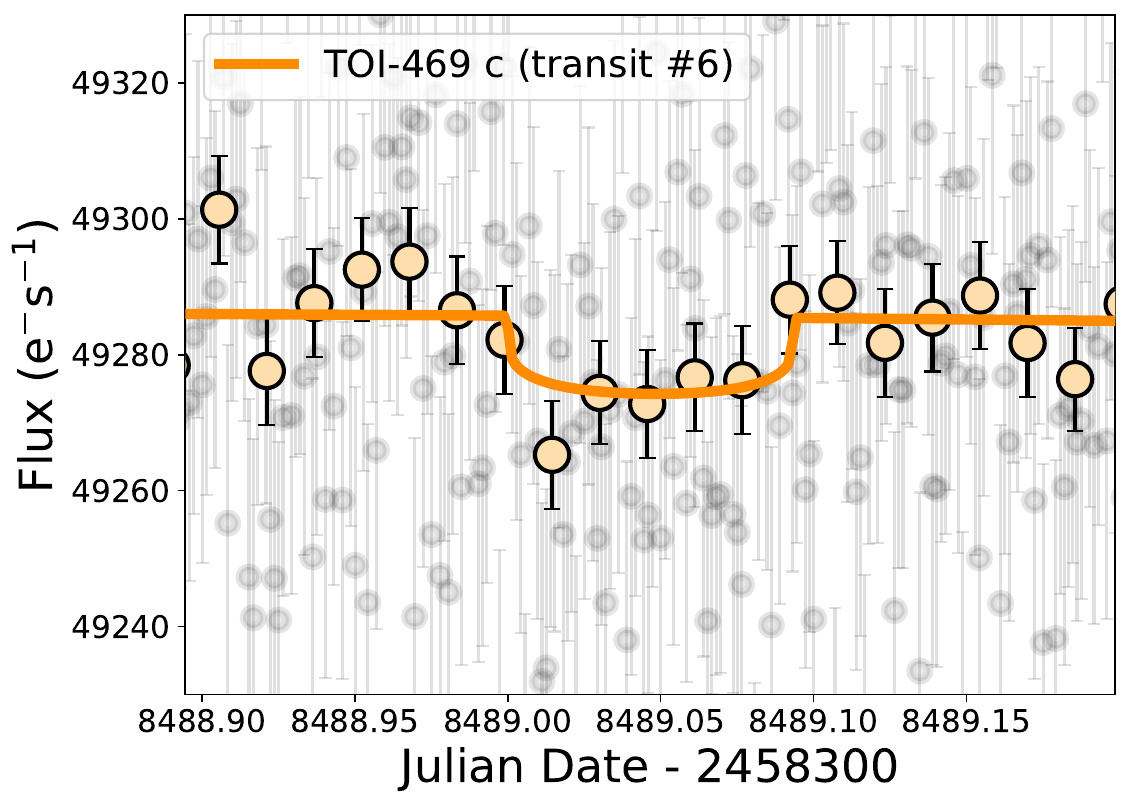}
    \includegraphics[scale=0.22]{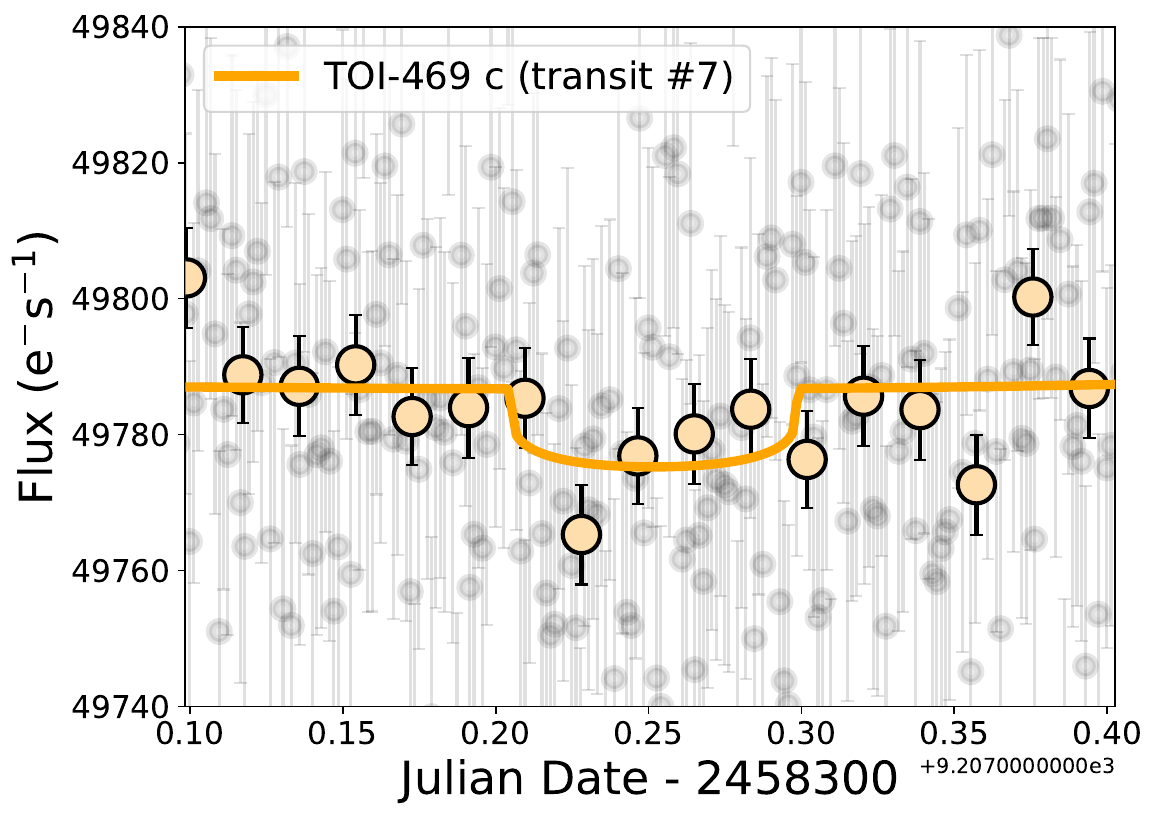}
    \includegraphics[scale=0.22]{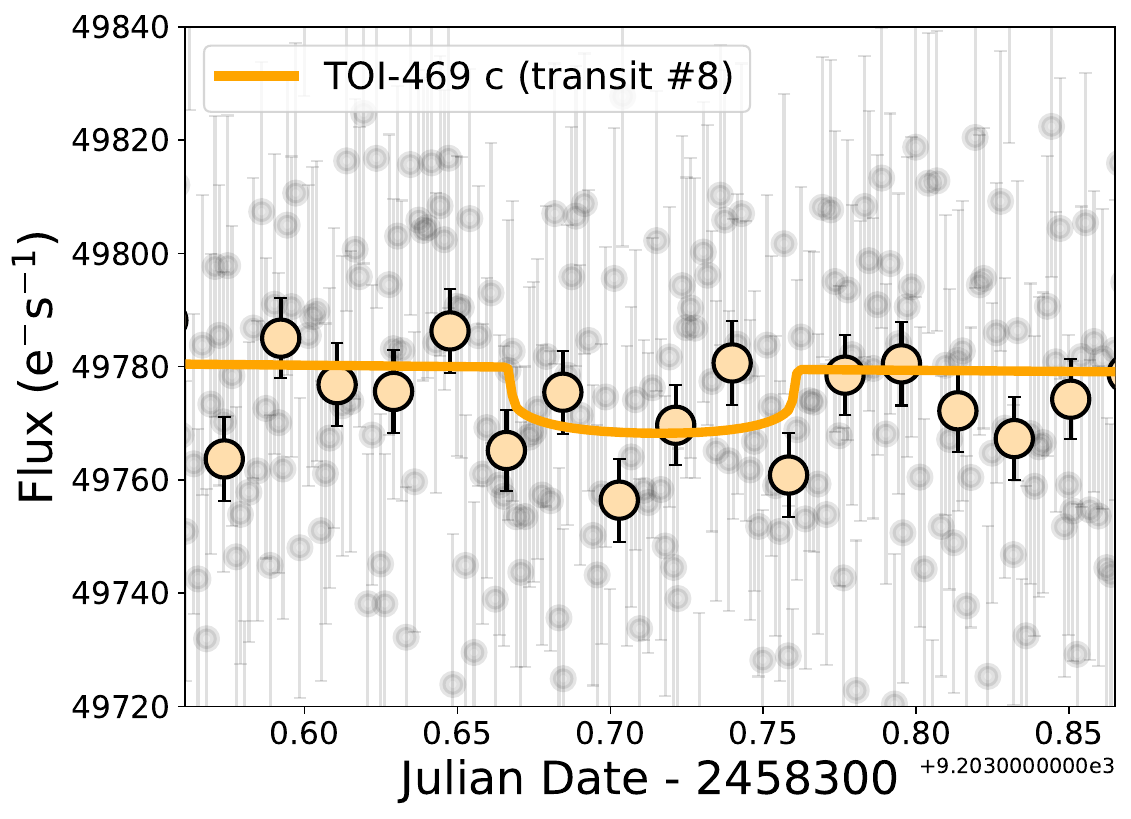}
    \includegraphics[scale=0.22]{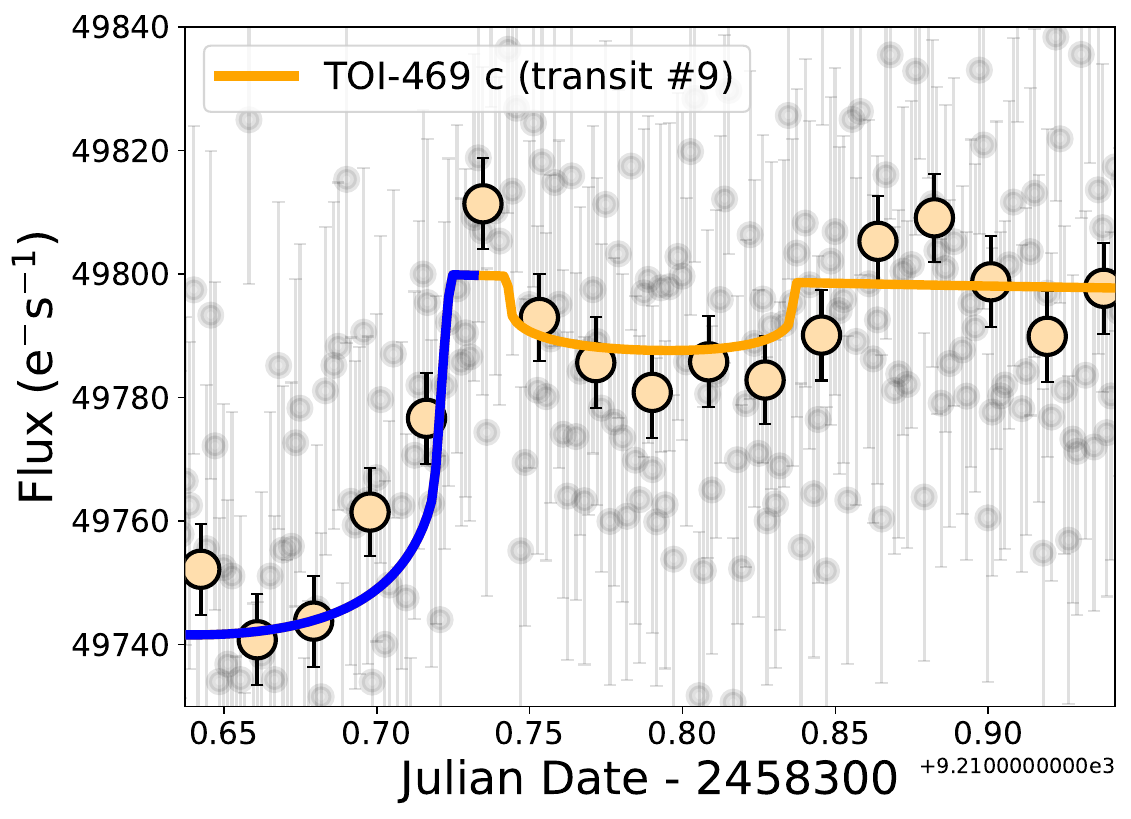}
    \includegraphics[scale=0.22]{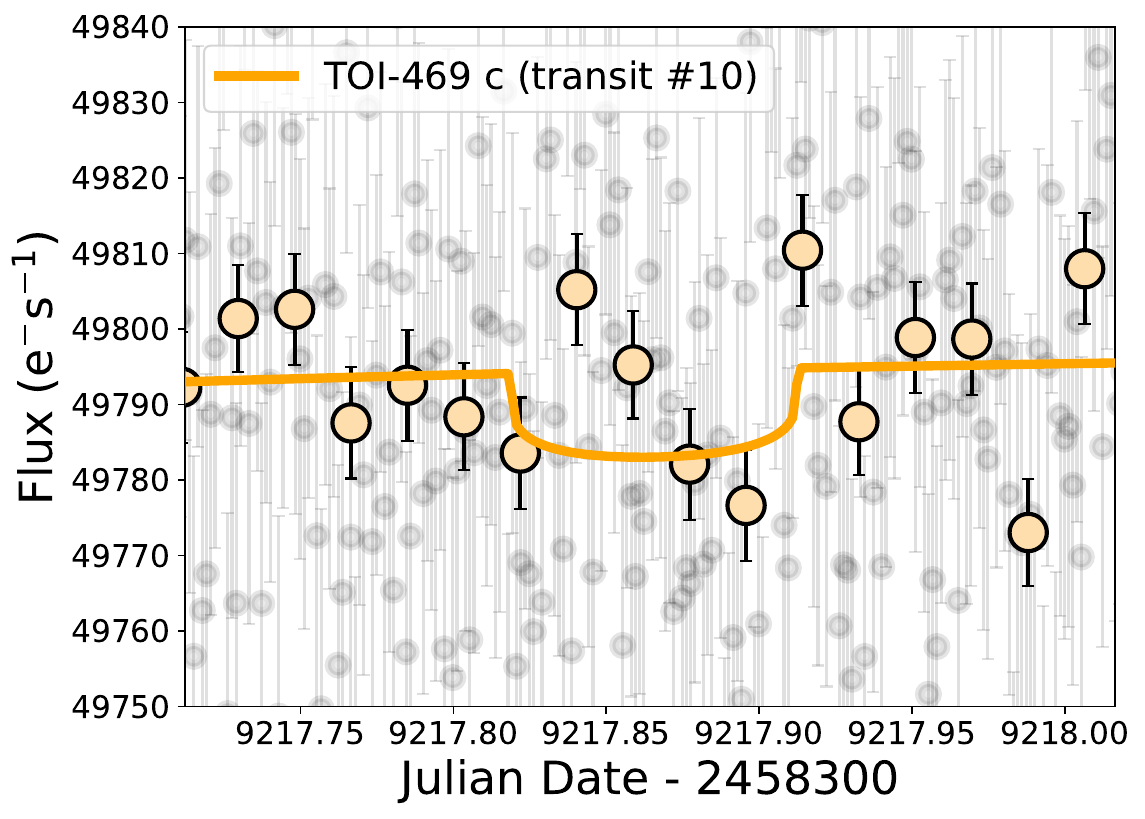}
    \includegraphics[scale=0.22]{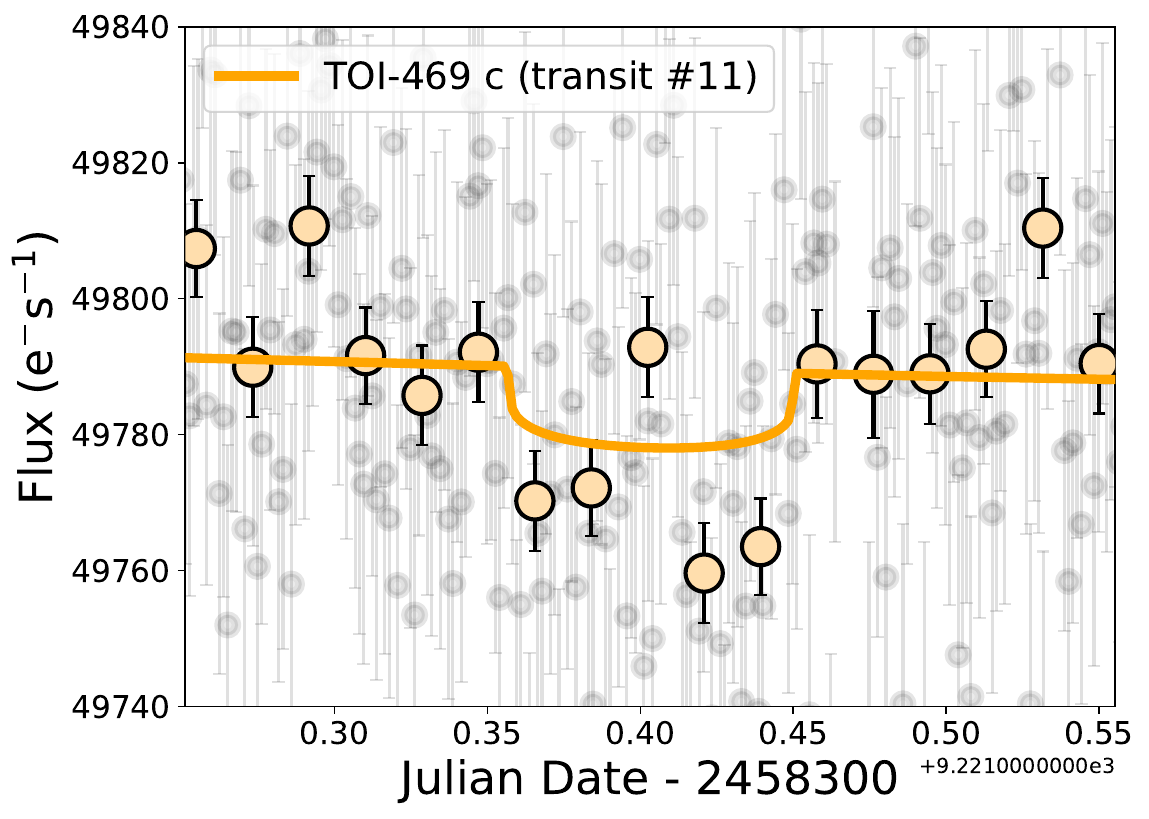}
    \includegraphics[scale=0.22]{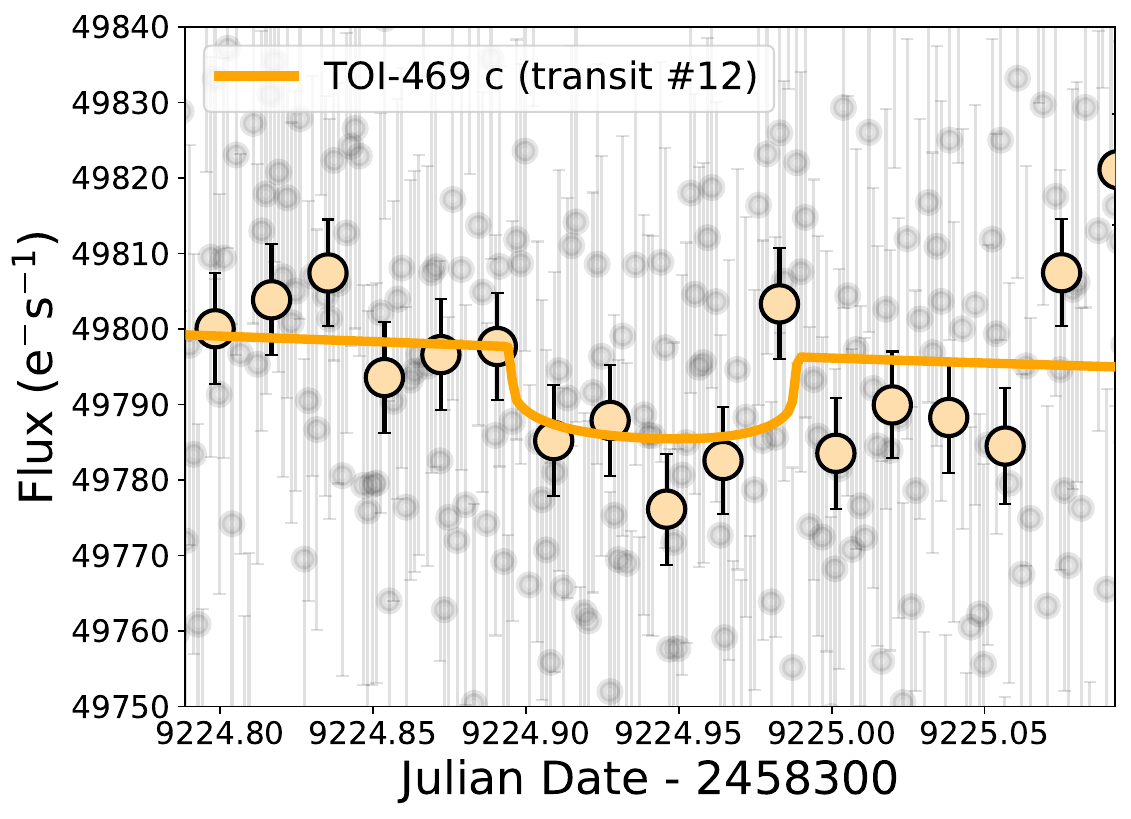}
    \includegraphics[scale=0.22]{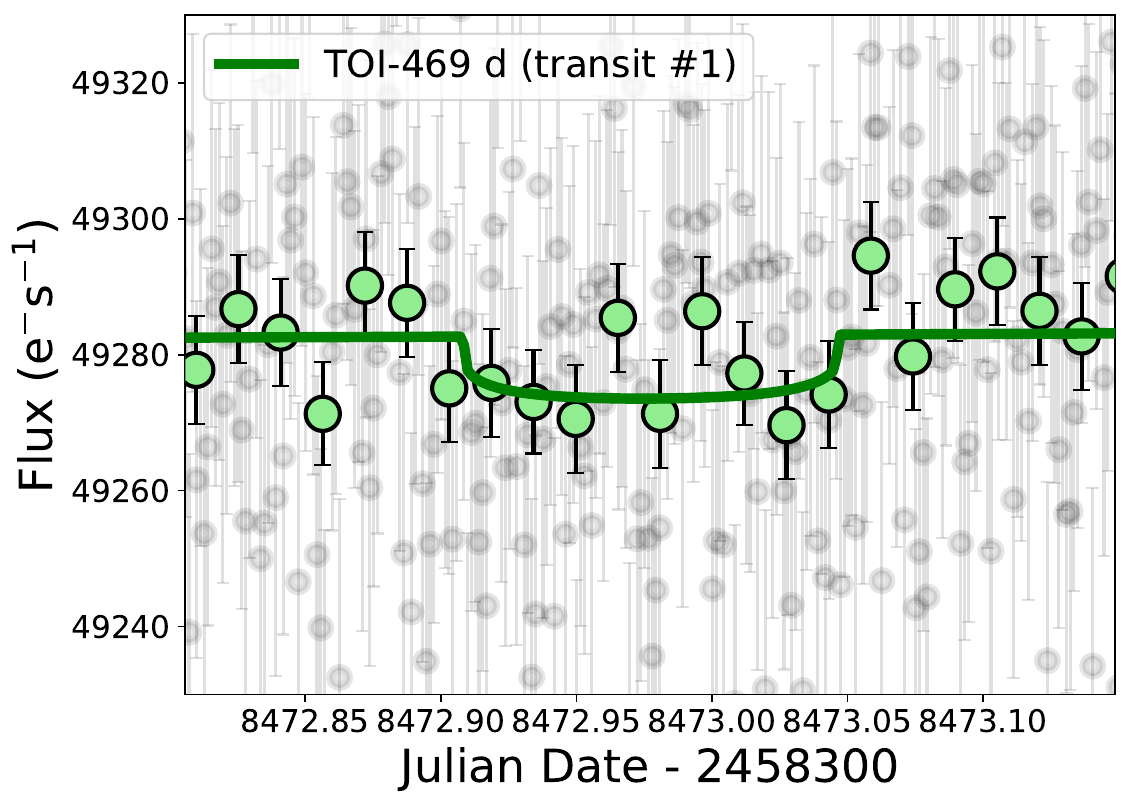}
    \includegraphics[scale=0.22]{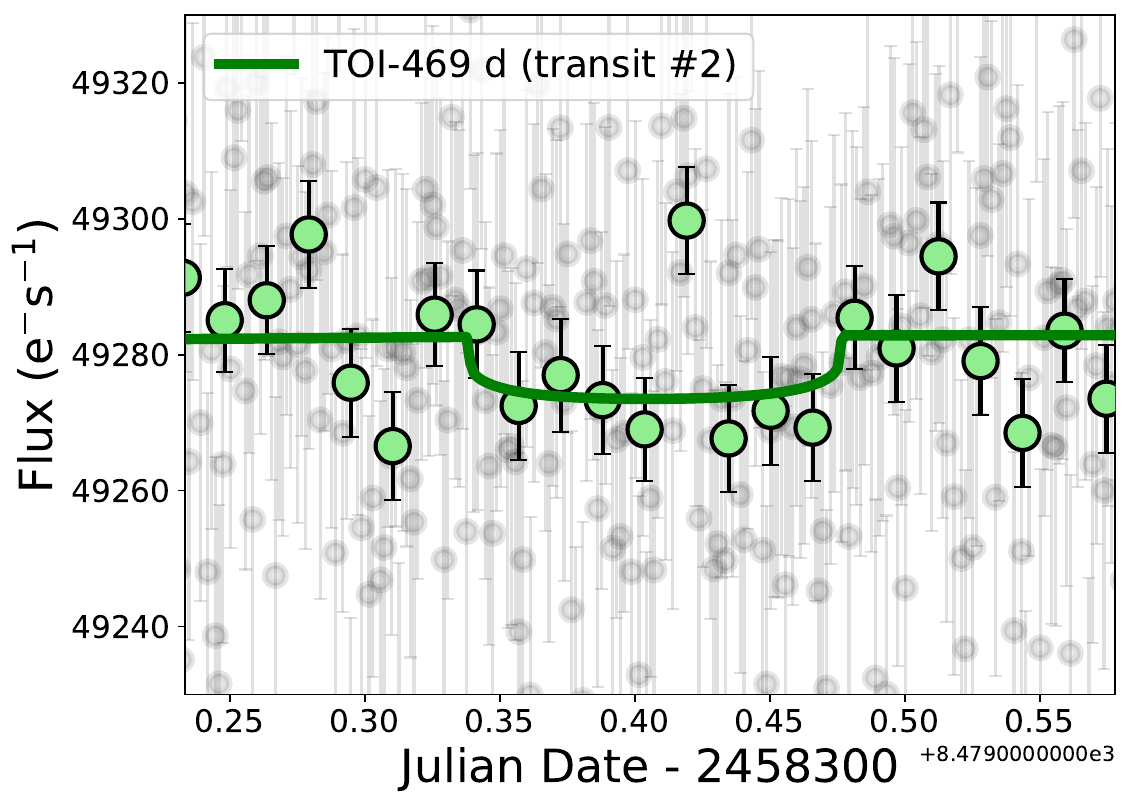}
    \includegraphics[scale=0.22]{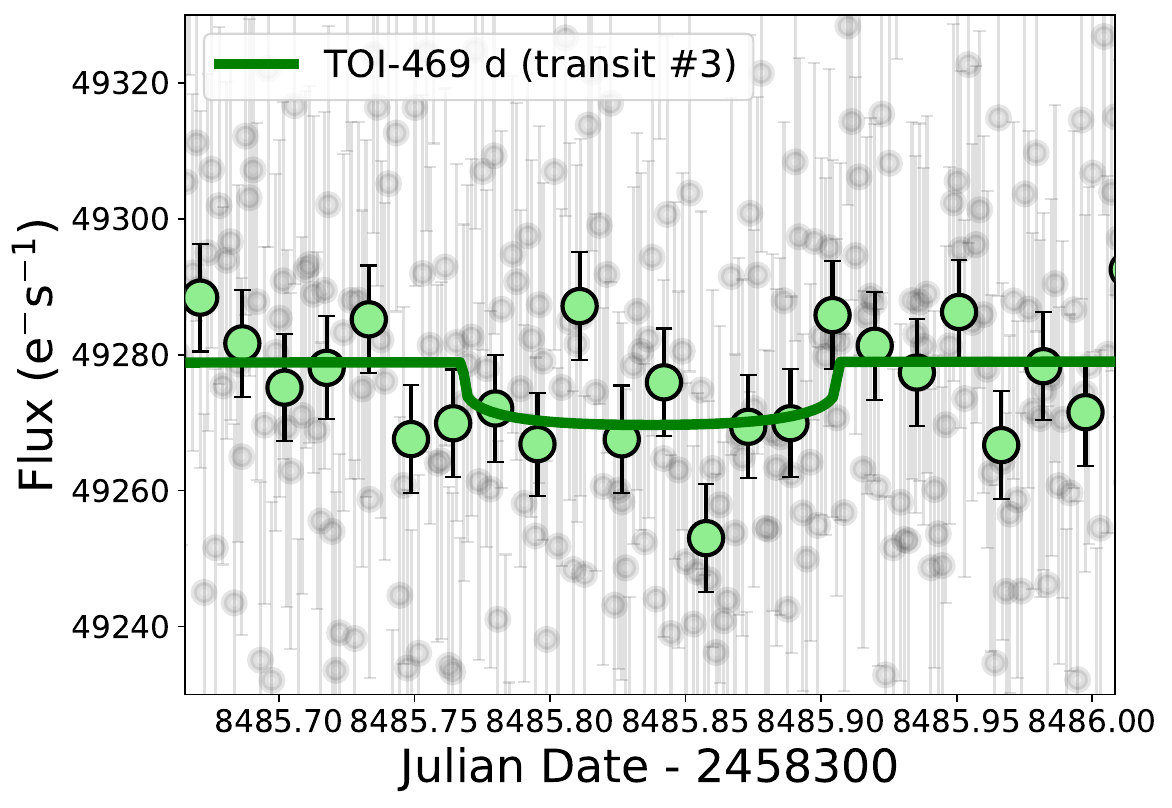}
    \includegraphics[scale=0.22]{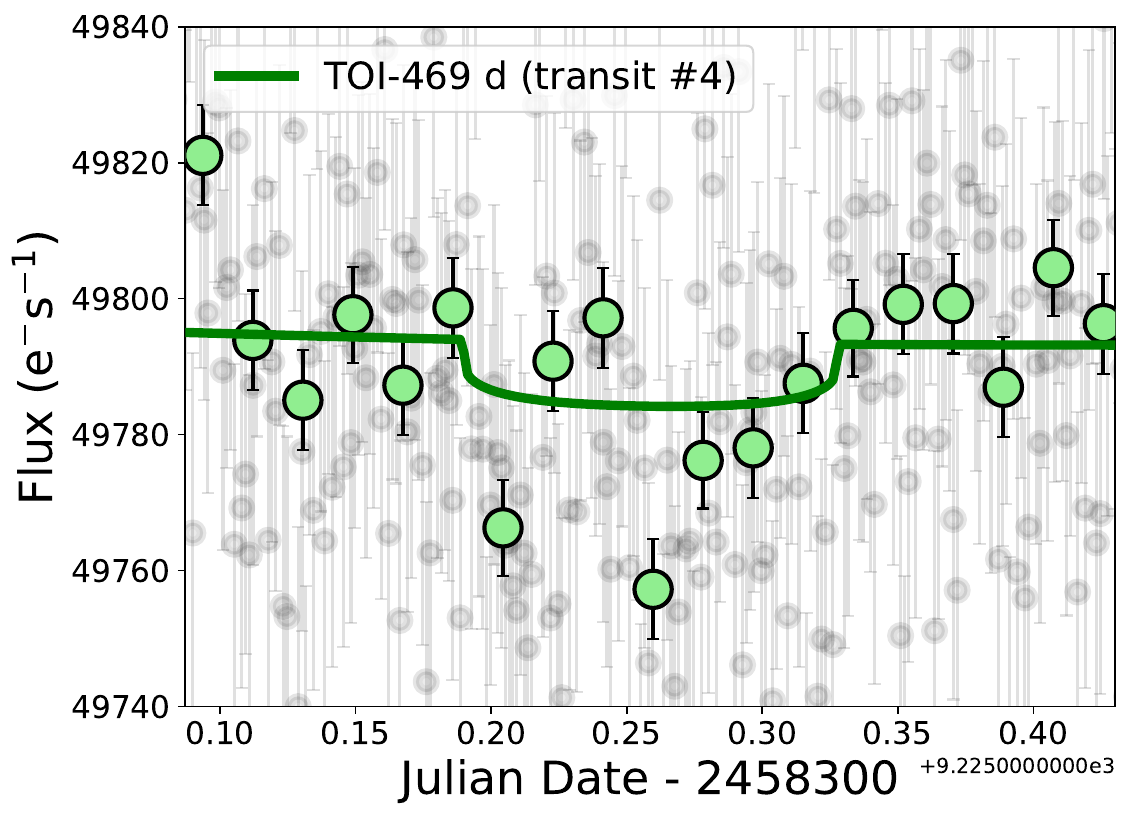}
    \includegraphics[scale=0.22]{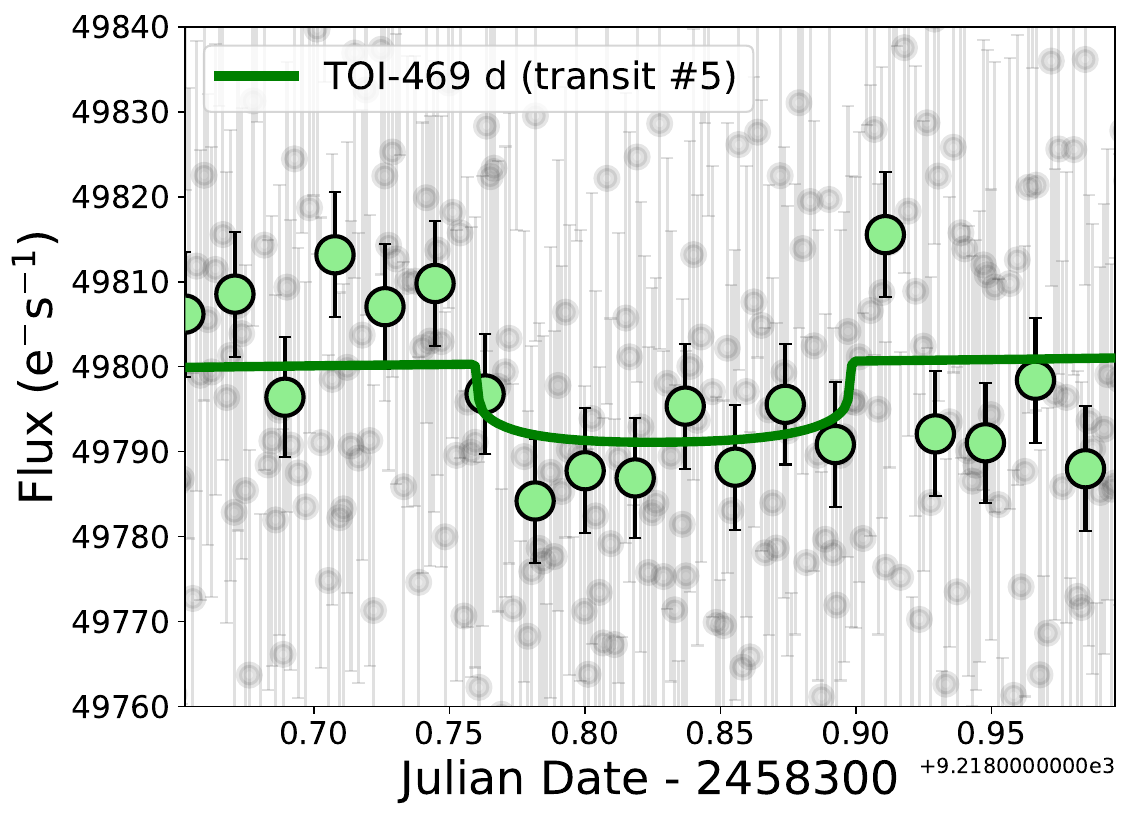}
    \includegraphics[scale=0.22]{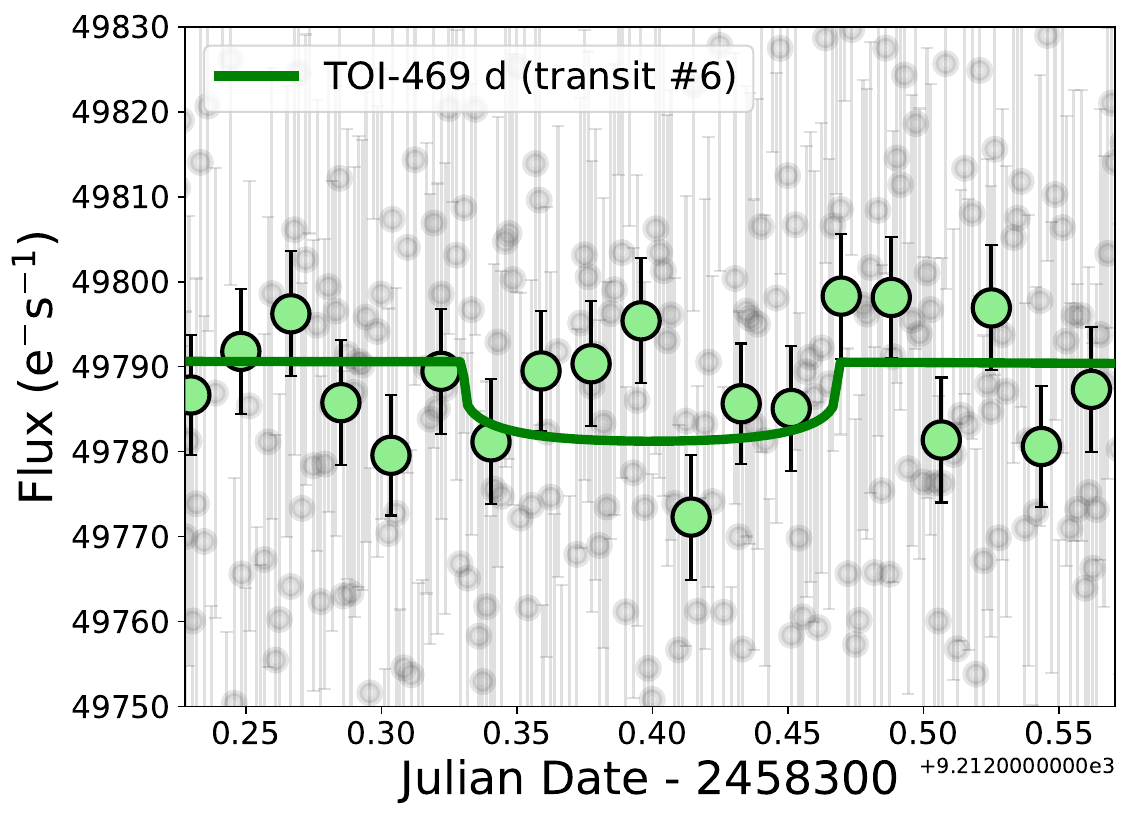}
    \includegraphics[scale=0.22]{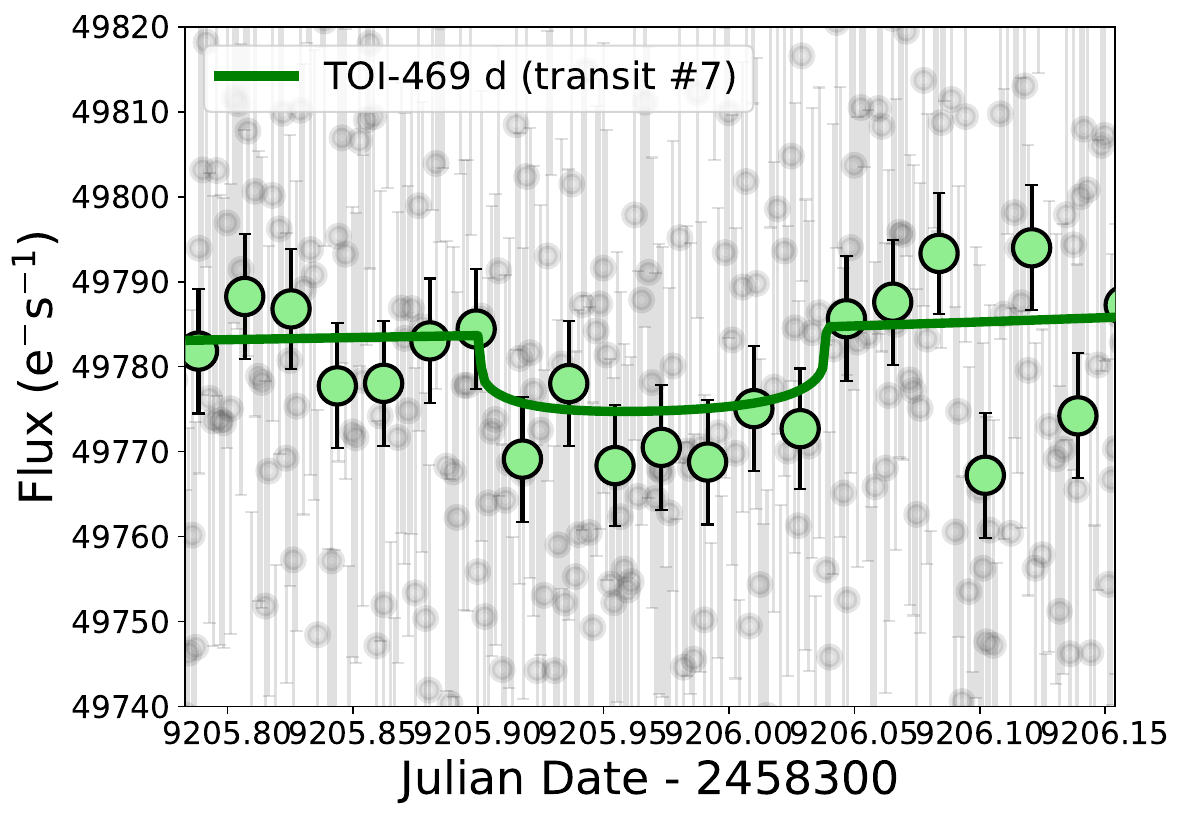}
    \caption{Individual transits of HIP\,29442\,$b$, HIP\,29442\,$c$, and HIP\,29442\,$d$ within the PDCSAP together with the best-fit transit+GP model overplotted.}
    \label{fig:individual_transits}
\end{figure*}

We ran the MCMC fit on the full S6+S33 dataset. We summarise priors and best-fit values of the free parameters in Tab. \ref{tab:params_amadeo}. We show in Fig. \ref{fig:complete_TESS_lc} the full TESS light curve with the best-fit GP and transit models, and \ref{fig:phase_folded_transits} the transit light curves of the three planets phase-folded using their best-fit transit ephemeris. The transit depths, expressed as $\frac{R_p^2}{R_\star^2}$, are $1032^{+42}_{-41}$, $213_{-28}^{+30}$, and $161_{-24}^{+26}$ ppm, respectively. As a comparison, the rms of the detrended and normalised light curve, after masking the transits, is 581 ppm and 604 ppm, for Sectors S6 and S33, respectively. The individual transits of all the planets in the system are shown in Fig. \ref{fig:individual_transits}, to emphasise that the transits of HIP\,29442\,$c$ and HIP\,29442\,$d$ are so shallow that some of them can be barely seen, or not at all, by eye (e.g. see transits nr. 1, 8, and 10 of planet $c$, and nr. 1, 3, 6 of planet $d$). The observed high variability in the S/N of the individual transits brings us to the conclusion that the 1$\sigma$ error bars for the radii of planets $c$ and $d$ derived through the MCMC simulations are likely too optimistic. Thus, in order to provide more conservative results, it seems appropriate to also adopt the 3$\sigma$ confidence interval especially, but not exclusively, for the transit-derived parameters of the two innermost planets. This, of course, has important implications for the accuracy and precision of the interior structure modelling that we describe in Sect. \ref{sec:internalstructure}. From the results with TESS data, it is clear that a high-precision photometric follow-up is needed to measure an accurate transit depth and radius of HIP\,29442\,$c$ and HIP\,29442\,$d$, and to determine their bulk structure more precisely. 

\begin{table*}
\label{tab:params_amadeo}
\centering
\small
\caption{Prior and posterior distributions of the transit+GP model parameters used to analyse the TESS PDCSAP (sectors S6+S33). }
\begin{tabular}{llc}
\hline \hline
\textbf{Parameter} & \textbf{Prior}\tablefootmark{a}  & \textbf{Best-fit values}\tablefootmark{b} \\ 
\hline
\textit{Orbital parameters}                          &                                    &   \\ 
\hline
 \noalign{\smallskip}
$P_{\rm orb,\,b}$ [days]  & $\mathcal{U} \, (13.6, 13.7)$   &  $13.63083^{+0.00003\,(+0.00009)}_{-0.00003\,(-0.00009)}$ \\
 \noalign{\smallskip}
$P_{\rm orb,\,c}$ [days]  & $\mathcal{U} \, (3.5, 3.6)$  & $3.53796^{+0.00003\,(+0.0001)}_{-0.00003\,(-0.0001)}$  \\
 \noalign{\smallskip}
$P_{\rm orb,\,d}$ [days] & $\mathcal{U} \, (6.4, 6.5)$  & $6.42975^{+0.00009\,(+0.0005)}_{-0.00010\,(+0.0005)}$  \\
 \noalign{\smallskip}
$T_{\rm conj,\,b}$ [BJD - 2\,450\,000]  & $\mathcal{U} \, (9210.4, 9210.8)$  & $9210.634\pm0.001 ^{(+0.003)}_{(-0.004)}$ \\
 \noalign{\smallskip}
$T_{\rm conj,\,c}$ [BJD - 2\,450\,000]  & $\mathcal{U} \, (9207.1, 9207.5)$  & $9207.252^{+0.004,(+0.02)}_{-0.003\,(-0.01)}$     \\
 \noalign{\smallskip}
$T_{\rm conj,\,d}$ [BJD - 2\,450\,000]  & $\mathcal{U} \, (9225.1, 9225.5)$  & $9225.259^{+0.005\,(+0.02)}_{-0.008\,(-0.04)}$     \\

 \noalign{\smallskip}
$i_{\rm p,\,b}$ [degrees] & $\mathcal{U} \, (80, 90)$ & $89.3^{+0.4}_{-0.3}(\pm0.7)$             \\
 \noalign{\smallskip}
$i_{\rm p,\,c}$ [degrees] & $\mathcal{U} \, (80, 90)$  & $86.3^{+0.8\,(+3.6)}_{-0.6\,(-1.5)}$             \\
 \noalign{\smallskip}
$i_{\rm p,\, d}$ [degrees] & $\mathcal{U} \, (80, 90)$  & $88.9^{+0.7\,(+1.1)}_{-0.8\,(-2.0)}$   \\ 
\noalign{\smallskip}
\hline
\noalign{\smallskip}
\textit{Planet parameters}                           &                                    &                                     \\ \hline
 \noalign{\smallskip}
$R_{\rm p,\,b}$/$R_{\rm \star}$  & $\mathcal{U} \, (0.0, 0.1)$  & $0.0321^{+0.0006\,(+0.0018)}_{-0.0007\,(-0.0026)}$ \\ 
 \noalign{\smallskip}
$R_{\rm p,\,c}$/$R_{\rm \star}$ & $\mathcal{U} \, (0.0, 0.1)$ &  $0.0146\pm0.0010\,(\pm0.0030)$ \\ 
 \noalign{\smallskip}
$R_{\rm p,\,d}$/$R_{\rm \star}$  & $\mathcal{U} \, (0.0, 0.1)$ & $0.0127\pm0.0010^{(+0.0030)}_{(-0.0040)}$ \\ 
 \noalign{\smallskip}
$R_{\rm p,\,b}$ [$\rearth$] & derived &  $3.48^{+0.07\,(+0.19)}_{-0.08\,(-0.28)}$ \\ 
 \noalign{\smallskip}
$R_{\rm p,\,c}$ [$\rearth$] & derived &  $1.58^{+0.10\,(+0.30)}_{-0.11\,(-0.34)}$ \\ 
 \noalign{\smallskip}
$R_{\rm p,\,d}$ [$\rearth$] & derived &  $1.37\pm0.11^{(+0.32)}_{(-0.43)}$ \\ 
\noalign{\smallskip}
\hline
\noalign{\smallskip}
\textit{Stellar parameters}                           &                                    &                                     \\ 
\noalign{\smallskip}
\hline
 \noalign{\smallskip}
$M_{\rm \star}$ $[\rm M_{\odot}]$ & $\mathcal{ZTG} \, (0.88, 0.04)$ & $0.89\pm0.04$ \\
 \noalign{\smallskip}
$R_{\rm \star}$ $[\rm R_{\odot}]$ & $\mathcal{ZTG} \, (0.993, 0.035)$ & $0.98\pm0.03$           \\
 \noalign{\smallskip}
$u_{\rm 1}$ & $\mathcal{U} \, (0, 1)$ &  $0.26\pm0.17$                     \\
 \noalign{\smallskip}
$u_{\rm 2}$ & $\mathcal{U} \, (0, 1)$ &  $0.36^{+0.30}_{-0.24}$  \\ 
\noalign{\smallskip}
\hline
\noalign{\smallskip}
\textit{Mat\'ern GP parameters}                           &                &                                     \\ \hline
 \noalign{\smallskip}
$\eta_{\sigma,\,S6}$ & $\mathcal{U} \, (0, 100)$ & $2.2^{+0.7}_{-0.4}$              \\
 \noalign{\smallskip}
$\eta_{\sigma,\,S33}$ & $\mathcal{U} \, (0, 100)$ & $10.4^{+3.7}_{-2.0}$               \\
 \noalign{\smallskip}
$\eta_{\rho,\,S6}$ & $\mathcal{U} \, (0, 100)$ & $0.8^{+0.6}_{-0.3}$              \\
 \noalign{\smallskip}
$\eta_{\rho,\,S33}$ & $\mathcal{U} \, (0, 100)$ & $1.4^{+0.6}_{-0.3}$              \\ 
\noalign{\smallskip}
\hline
\noalign{\smallskip}
\textit{Instrument-dependent parameters}     &                                    &                                     \\ \hline
 \noalign{\smallskip}
$F_{\rm 0,S6}$ [$\rm e^{-}/s$] & $\mathcal{U} \, (-100, 100)$ & $1.6\pm0.7$              \\
 \noalign{\smallskip}
$F_{\rm 0,S33}$ [$\rm e^{-}/s$] & $\mathcal{U} \, (-100, 100)$ & $2.4^{+3.9}_{-3.6}$ \\
 \noalign{\smallskip}
$\sigma_{\rm TESS,S6}$ [$\rm e^{-}/s$]  & $\mathcal{U} \, (0, 100)$  & $9.0\pm0.5$ \\
 \noalign{\smallskip}
$\sigma_{\rm TESS,S33}$ [$\rm e^{-}/s$] & $\mathcal{U} \, (0, 100)$  & $12.3\pm0.4$ \\ 
\noalign{\smallskip}
\hline \hline
\end{tabular}
\tablefoot{
\tablefoottext{a}{ The $\mathcal{U}(a,b)$ symbol indicates uniform distributions, being $a$ and $b$ the lower and upper limits. The $\mathcal{G}(\mu, \sigma)$ and $\mathcal{ZTG}(\mu, \sigma)$ symbols indicate Gaussian and zero-truncated Gaussian distributions, being $\mu$ and $\sigma$ the mean and width of the distributions.}
\tablefoottext{a}{The best-fit values are given as the median of the posterior distributions, and the uncertainties are given as the 16th and 84th percentiles (i.e. 1$\sigma$ confidence interval), and as the 0.2th and 99.8th percentiles (i.e. 3$\sigma$ confidence interval) for the planet parameters, the latter indicated in parenthesis.}
}
\end{table*}


\subsection{Monte Carlo RV fitting}
\label{sec:rvfit}
We used the results of the transit modelling to constrain the orbital periods and time of inferior conjunction by using Gaussian priors, based on the analysis of the S6+S33 dataset (see Table \ref{tab:params_amadeo}). Given the low S/N of the transits, a joint RV+light curve analysis would not help to better constraining the system's parameters, at a higher cost in terms of computing resources when different RV models are tested, as it happens in our case. We modelled the orbits of all the transiting planets with Keplerians, and adopted the parametrization $\sqrt{e_{b,c,d}}\cos\omega_{b,c,d,\,\star}$ and $\sqrt{e_{b,c,d}}\sin\omega_{b,c,d,\,\star}$, where $e_{b,c,d}$ and $\omega_{b,c,d,\,\star}$ are free parameters for the eccentricities and arguments of periastron. We also tested models with $e_{b,c,d}$ fixed to zero to assess the statistical significance of the fitted eccentricities. To investigate the existence of an additional planetary-like signal, we also tested models including a fourth Keplerian (or sinusoid), that we labelled as ``signal \textit{x}'', with period $P_x$ sampled uniformly between 15 and 1000 days. By treating pre- and post-COVID data as independent RV sub-samples (labelled as ESP19 and ESP21, respectively), the models include the pair of offsets $\gamma_{\rm ESP,\,19}$ and $\gamma_{\rm ESP,\,21}$, and two white noise jitters $\sigma_{\rm jitt. ESP,\,19}$ and $\sigma_{\rm jitt. ESP,\,21}$\footnote{We also tested a model including only one offset $\gamma_{\rm RV}$ and an acceleration term, for which we got a best-fit value $-0.0001^{+ 0.0021}_{-0.0017}$, compatible with zero. This model is also statistically not favoured, therefore we won't refer to it anymore in the paper.}. 

We tested models including a Gaussian process (GP) regression, to model a correlated signal induced by the stellar variability and modulated over the stellar rotation with a periodicity of $\sim 37$ days, which is present in the data as shown by the results of the frequency analysis (Sect. \ref{sec:rvfreqanalysis}), and similarly in the FWHM time series (Sect. \ref{sec:staractivity}). To this purpose, we adopted the frequently-adopted and well-performing quasi-periodic (QP) kernel (e.g. \citealt{haywood14}). The elements of the QP covariance matrix are defined as follows:

\begin{eqnarray} 
\label{eq:eqgpqpkernel}
k_{QP}(t, t^{\prime}) = h^2\cdot\exp\Bigg[-\frac{(t-t^{\prime})^2}{2\lambda^2} - \frac{\sin^{2}\bigg(\pi\big(t-t^{\prime})/\theta\bigg)}{2w^2}\Bigg] + \nonumber \\
+\, (\sigma^{2}_{\rm RV}(t)+\sigma^{2}_{\rm jitt. \,ESP})\cdot\delta_{t, t^{\prime}}
\end{eqnarray}

Here, $t$ and $t^{\prime}$ denote two different epochs of observations, $\sigma_{\rm RV}(t)$ represent the radial velocity uncertainty at the observing epoch $t$, and $\delta_{t, t^{\prime}}$ is the Kronecker delta. The GP hyper-parameters are $h$, which denotes the amplitude scale of the correlated signal; $\theta$, which represents the periodic time-scale of the modelled signal, and corresponds to the stellar rotation period; $w$, which describes the "weight" of the rotation period harmonic content within a complete stellar rotation (i.e. a low value of $w$ indicates that the periodic variations contain a significant contribution from the harmonics of the rotation periods); and $\lambda$, which represents the decay timescale of the correlations, and is related to the temporal evolution of the magnetically active regions responsible for the correlated signal observed in the RVs.  

We explored the full (hyper-)parameter space using the publicly available Monte Carlo (MC) nested sampler and Bayesian inference tool \textsc{MultiNest v3.10} (e.g. \citealt{Feroz2019}), through the \textsc{pyMultiNest} wrapper \citep{Buchner2014}. Our MC set-up included 300 live points and we adopted a sampling efficiency of 0.5 and a tolerance on the Bayesian evidence calculation of 0.3. To perform the GP regression, we used the publicly available \textsc{python} module \textsc{george} v0.2.1 \citep{2015ITPAM..38..252A}, integrated within the MultiNest framework. 

We summarise all the tested models in Table \ref{tab:testedRVLCmodels}, providing their Bayesian evidences $\ln \mathcal{Z}$. For the Bayesian model comparison analysis, we adopt the conventional scale used for the interpretation of model probabilities presented in Table 1 of \cite{feroz11}, according to which $\Delta \ln \mathcal{Z}\sim2.5$ and $\Delta \ln \mathcal{Z}>5$ denote, respectively, moderate and strong evidence in favour of the model with the higher value of $\ln \mathcal{Z}$. We highlight that we also tested models that do not include a GP term, and in all the cases we found that they are statistically strongly penalised. 
Table \ref{tab:testedRVLCmodels} includes two test models which incorporate the FWHM time series. In these cases, we modelled the RVs and FWHM with the same GP QP kernel, with three out of four hyper-parameters in common between the two time series (namely, $\theta$, $w$, and $\lambda$), while different scale amplitudes have been used. With this approach, we make use of the information about the stellar activity contained in the FWHM data to constrain the activity-related correlated signal in the RVs. Due to the different datasets analysed in these cases, the Bayesian evidences of these two models cannot be compared with that of the other models in Table \ref{tab:testedRVLCmodels}. 

\begin{figure*}[h]
  \centering
  \subfigure[]{\includegraphics[scale=0.43]{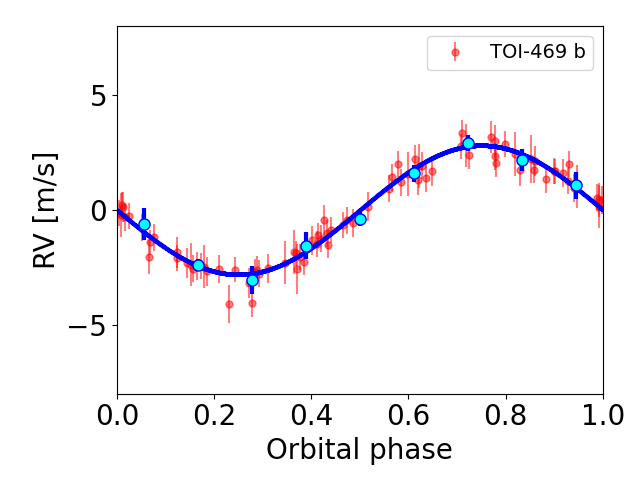}}
  \subfigure[]{\includegraphics[scale=0.42]{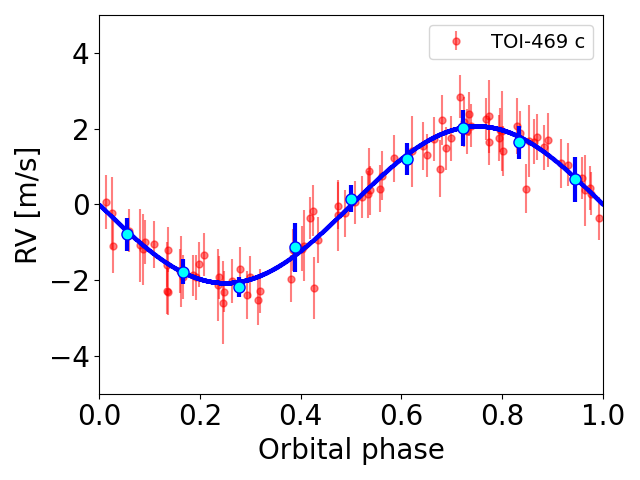}}
  \subfigure[]{\includegraphics[scale=0.42]{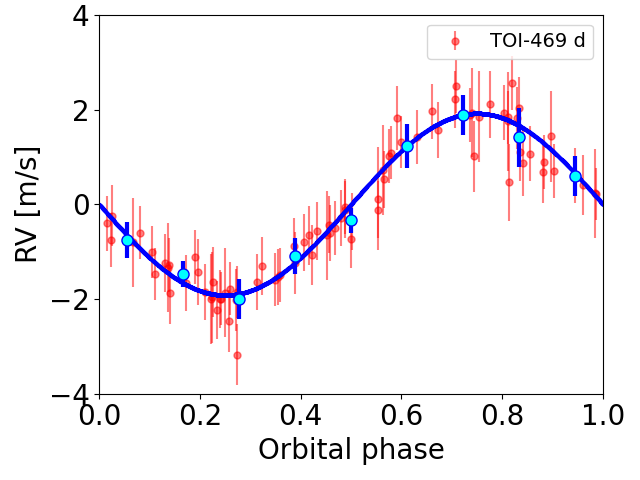}}
  \subfigure[]{\includegraphics[scale=0.43]{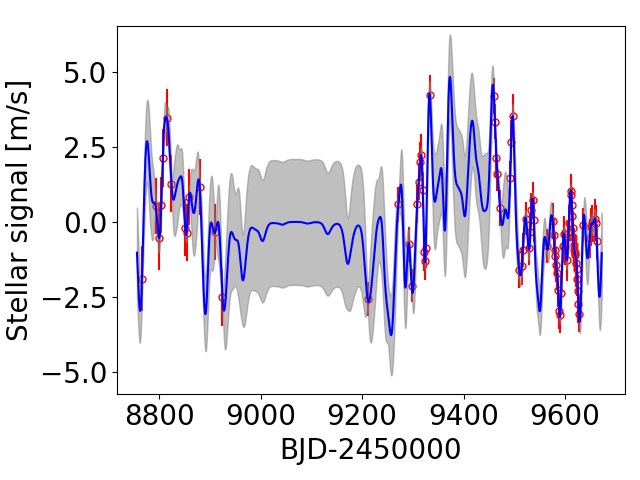}}
   \caption{RV Doppler signals due to three transiting planets orbiting HIP\,29442 (TOI-469) (panels \textit{a}, \textit{b}, and \textit{c}), and the RV activity term, as fitted through a GP QP regression (panel \textit{d}; curve in blue: best-fit model; grey area: 1$\sigma$ confidence interval). The RV error bars include a jitter term added in quadrature to the measurement uncertainties. } 
   \label{fig:planetdata}
\end{figure*}

From the results listed in Table \ref{tab:testedRVLCmodels} we can draw some conclusions. The three transiting planets are well described by assuming circular orbits, because the model with Keplerians (M1) is strongly penalised with respect to M2 (eccentricities of model M1 are all consistent with zero). Models M3 and M4, which include a fourth signal \textit{x}, are not statistically favoured over the simpler three-planet model M2. In M3, the semi-amplitude $K_x$ appears significant at a $\gtrsim2\sigma$ level, the period is not well constrained ($P_x=576^{+262}_{-48}$ d), and the signal has a large eccentricity ($e_x=0.6\pm0.2$). We note for model M2 we found no significant relative offset. For other targets followed-up by the ESPRESSO GTO collaboration we found very small RV offsets between the ESP19 and ESP21 sub-samples, and we expect the same for HIP\,29442. Moreover, we found that using the FWHM to constrain the stellar activity term in the RVs (models M$_{\rm FWHM,\,1}$ and M$_{\rm FWHM,\,2}$) does not affect the planetary parameters for HIP\,29442\,$b$, $c$, and $d$, but affects the posteriors of the parameters of signal \textit{x}: the semi-amplitude $K_x$ is lower and less significant than for models M3 and M4. From these considerations, we conclude that our data do not support the existence of a periodic signal \textit{x} that can be attributed to a fourth planet on an external orbit. We elect model M2 as our reference model, and summarise the best-fit values (the median of the marginalised posteriors, and 1$\sigma$ confidence intervals) for all the free parameters in Table \ref{tab:tessrvmodel}, together with the priors adopted in the analysis. Best-fit spectroscopic orbits, and quasi-periodic stellar activity induced signal, are shown in Fig. \ref{fig:planetdata}. We emphasise that the best-fit values found for the planetary parameters are consistent for all the test models, and they are recovered with very high significance. Thus, we consider the GP QP regression a successful and sufficient approach, and we deem it not necessary to test alternative models for the stellar activity component.

\begin{table*}[]
    \centering
    \tiny
    \caption{List of GP-based ESPRESSO RV models tested in this work, and the corresponding best-fit values of RV semi-amplitudes for the three transiting planets $b$, $c$, and $d$. }
    \begin{tabular}{cccc}
    \hline
        Model ID\tablefootmark{a} & Description & $\ln \mathcal{Z}$\tablefootmark{b} & $K_p$ [m$\,s^{-1}$] \\
    \hline    
    \noalign{\smallskip}
         M1 & Three Keplerians; two RV offsets and jitters & -167.3 & $K_{\rm b} = 2.8\pm0.2$ \\
         & & & $K_{\rm c} = 2.1\pm0.1$ \\
         & & & $K_{\rm d} = 1.9\pm0.1$ \\ 
          \noalign{\smallskip}
         M2 & As M1, but with $e_{b,c,d}$ fixed to zero (circular orbits) & -159.8 & $K_{\rm b} = 2.8\pm0.2$ \\
         & & & $K_{\rm c} = 2.1\pm0.1$ \\
         & & & $K_{\rm d} = 1.9\pm0.1$ \\ 
          \noalign{\smallskip}
         M3  & Three circular orbits (planets $b$, $c$, and $d$), one Keplerian (signal $x$); two RV offsets and jitters & -158.1 & $K_{\rm b} = 2.8\pm0.2$\\
         & & & $K_{\rm c} = 2.1\pm0.1$ \\
         & & & $K_{\rm d} = 1.9\pm0.1$ \\ 
         & & & $K_{\rm x} = 3.1\pm1.6$ \\ 
          \noalign{\smallskip}
          M4  & As M3, but with $e_{x}$ fixed to zero & -159.2 & $K_{\rm b} = 2.8\pm0.2$\\
          & & & $K_{\rm c} = 2.1\pm0.1$ \\
          & & & $K_{\rm d} = 1.9\pm0.1$ \\ 
          & & & $K_{\rm x} = 1.6\pm0.8$ \\ 
          \noalign{\smallskip}
          M$_{\rm FWHM,\,1}$   & Three planets on circular orbits; two RV offset/jitter;  & -406.9 & $K_{\rm b} = 2.8^{+0.3}_{-0.2}$\\
         & GP QP stellar activity term trained on the FWHM & & $K_{\rm c} = 2.1\pm0.1$ \\
         & & & $K_{\rm d} = 1.9\pm0.1$ \\ 
          \noalign{\smallskip}
          M$_{\rm FWHM,\,2}$  & Three planets on circular orbits, plus an additional sinusoidal signal $x$; two RV offset/jitter;  &  -406.4 & $K_{\rm b} = 2.8\pm0.2$\\
         & GP QP stellar activity term trained on the FWHM & & $K_{\rm c} = 2.1\pm0.1$ \\
         & & & $K_{\rm d} = 1.9\pm0.1$ \\ 
         & & & $K_{\rm x,\, 4^{th}\, sinusoid} = 1.7^{+1.0}_{-1.1}$ \\ 
          \noalign{\smallskip}
    \hline     
    \end{tabular}
    \tablefoot{
    \tablefoottext{a}{A few models include a fourth signal treated as planetary (labelled as signal $x$).}
    \tablefoottext{b}{ $\ln \mathcal{Z}$ denotes the natural logarithm of a model Bayesian evidence. }
    }
    \label{tab:testedRVLCmodels}
\end{table*}

\begin{table*}
       \caption{Best-fit results for model M2 (Table \ref{tab:testedRVLCmodels}).} 
       \label{tab:tessrvmodel}
       \centering
       \begin{tabular}{lcc}
            \hline
            \noalign{\smallskip}
             \textbf{Parameter} &  \textbf{Prior} & \textbf{Best-fit value}\tablefootmark{a} \\
             \noalign{\smallskip}
            \hline
            \noalign{\smallskip}
            \textit{Activity-related GP parameters }\\
            \noalign{\smallskip}
            \hline
            \noalign{\smallskip}
            $h$ [$\ms$] & $\mathcal{U}$(0,5) & $2.1^{+0.6}_{-0.4}$ \\
            \noalign{\smallskip}
            $w$ & $\mathcal{U}$(0,1) & $0.46^{+0.12}_{-0.09}$ \\
            \noalign{\smallskip}
            $\theta$ [days] & $\mathcal{U}$(33,50) & $39.9^{+2.4}_{-2.2}$ \\ 
            \noalign{\smallskip}
            $\lambda$ [days] & $\mathcal{U}$(0,1000) & $45^{+28}_{-20}$ \\ 
            \noalign{\smallskip}
            \hline
            \noalign{\smallskip}
            \textit{Estimated planetary parameters} \\ 
            \noalign{\smallskip}
            \hline
            \noalign{\smallskip}
            RV Doppler semi-amplitude, $K_{\rm p,\,b}$ [m$\,s^{-1}$] & $\mathcal{U}$(0,10) & $2.8\pm0.2$ \\
            \noalign{\smallskip}
            RV Doppler semi-amplitude, $K_{\rm p,\,c}$ [m$\,s^{-1}$] & $\mathcal{U}$(0,10) & $2.1\pm0.1$ \\ 
            \noalign{\smallskip}
            RV Doppler semi-amplitude, $K_{\rm p,\,d}$ [m$\,s^{-1}$] & $\mathcal{U}$(0,10) & $1.9\pm0.1$ \\ 
            \hline
            \noalign{\smallskip}
            \textit{Derived planetary parameters} && \\
             \noalign{\smallskip}
             \hline
             \noalign{\smallskip}
             orbital semi-major axis, $a_{\rm p,\,b}$ [au] & & $0.1070\pm0.0016$ \\
             \noalign{\smallskip}
             orbital semi-major axis, $a_{\rm p,\,c}$ [au] & & $0.0436^{+0.0006}_{-0.0007}$ \\
             \noalign{\smallskip}
             orbital semi-major axis, $a_{\rm p,\,d}$ [au] & & $0.0649^{+0.0009}_{-0.0010}$ \\
             \noalign{\smallskip} 
             mass, $m_{\rm p,\,b}$ [M$_{\rm \oplus}$] & & $9.6\pm0.8$ \\
             \noalign{\smallskip}
             mass, $m_{\rm p,\,c}$ [M$_{\rm \oplus}$] & & $4.5\pm0.3$ \\
             \noalign{\smallskip}
             mass, $m_{\rm p,\,d}$ [M$_{\rm \oplus}$] & & $5.1\pm0.4$ \\
             \noalign{\smallskip}
             average density, $\rho_{\rm p,\,b}$ [$g$ $cm^{\rm -3}$] & & $1.3\pm0.2\,(\pm0.3)$ \\
             \noalign{\smallskip}
             average density, $\rho_{\rm p,\,c}$ [$g$ $cm^{\rm -3}$] & & $6.3^{+1.7\,(+6.0)}_{-1.3\,(-2.7)}$ \\
             \noalign{\smallskip}
             average density, $\rho_{\rm p,\,d}$ [$g$ $cm^{\rm -3}$] & & $11.0^{+3.4\,(+21.0)}_{-2.4\,(-6.3)}$ \\
             \noalign{\smallskip}
             incident flux, $S_{\rm p,\,b}$ [S$_\oplus$] & & $61\pm3$ \\
             \noalign{\smallskip}
             incident flux, $S_{\rm p,\,c}$ [S$_\oplus$] & & $373\pm19$ \\
             \noalign{\smallskip}
             incident flux, $S_{\rm p,\,d}$ [S$_\oplus$] & & $167\pm9$ \\
             \noalign{\smallskip}
             equilibrium temperature\tablefootmark{b}, $T_{\rm eq,\,b}$ [K] & & $777\pm18$ \\
             \noalign{\smallskip}
             equilibrium temperature\tablefootmark{b}, $T_{\rm eq,\,c}$ [K] & & $1217\pm29$ \\
             \noalign{\smallskip}
             equilibrium temperature\tablefootmark{b}, $T_{\rm eq,\,d}$  [K] &  & $998\pm24$ \\
             \noalign{\smallskip}
             \hline
            \noalign{\smallskip} 
            \textit{Jitters and offsets } \\
            \noalign{\smallskip}
            \hline
            \noalign{\smallskip}
            $\sigma_{\rm jitt.\, ESP19}$ [m$\,s^{-1}$] & $\mathcal{U}$(0,10) & $0.9^{+0.8}_{-0.6}$ \\ 
            \noalign{\smallskip}
            $\sigma_{\rm jitt.\, ESP21}$ [m$\,s^{-1}$] & $\mathcal{U}$(0,10) & $0.5\pm0.1$ \\ 
            \noalign{\smallskip}
            $\gamma_{\rm ESP19}$ [m$\,s^{-1}$] & $\mathcal{U}$(81700,81800)  & $81728.7^{+1.0}_{-1.2}$ \\ 
            \noalign{\smallskip}
            $\gamma_{\rm ESP21}$ [m$\,s^{-1}$] & $\mathcal{U}$(81700,81800) & $81728.8\pm0.7$ \\ 
            \noalign{\smallskip}
            \noalign{\smallskip}
             \hline
\end{tabular}
\tablefoot{
\tablefoottext{a}{The uncertainties are given as the $16^{\rm th}$ and $84^{\rm th}$ percentiles of the posterior distributions (1$\sigma$ confidence interval), and as the $0.2^{\rm th}$ and $99.8^{\rm th}$ percentiles (3$\sigma$ confidence interval) for the bulk densities (indicated in parenthesis), because the poorly determined radius measurements.}
\tablefoottext{c}{Derived from the relation $T_{\rm eq}$=$T_{\rm eff}\cdot\sqrt{\frac{R_\star}{2a_b}}\cdot(1-A_B)^{0.25}$, assuming Bond albedo $A_B$=0.}
}
\end{table*}

\section{Modelling of the planetary internal structure and composition} \label{sec:internalstructure}

\begin{figure*}
    \centering
    \includegraphics[scale=0.35]{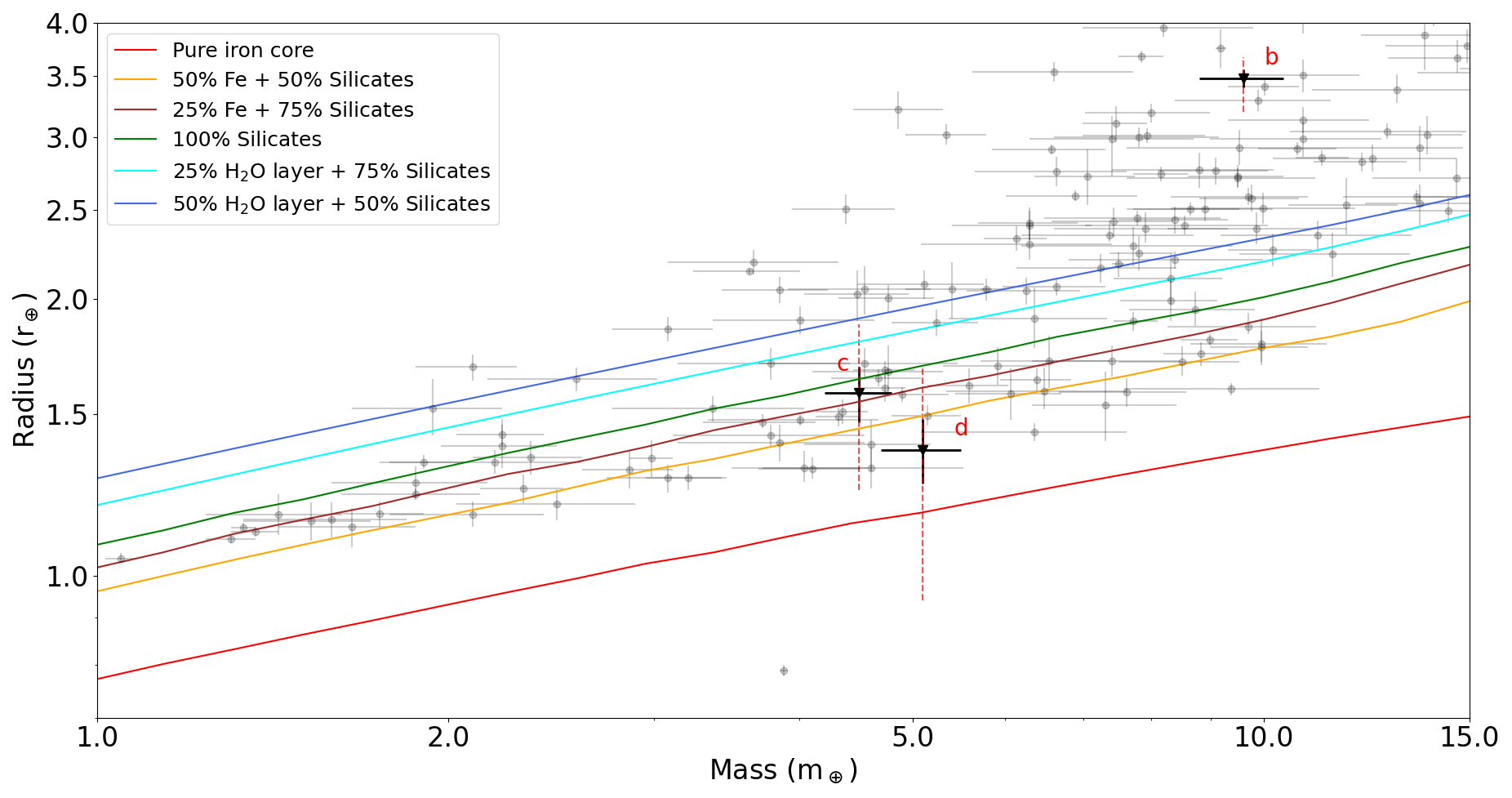}
    \caption{Mass-radius diagram for planets with mass $<$20 $\mearth$ and radius $<$4 $\rearth$, selected from the TEPCAT sample, available at \url{https://www.astro.keele.ac.uk/jkt/tepcat/} (updated to 22 March 2023; \citealt{2011MNRAS.417.2166S}). Gray dots represent planets with mass and radius measured with a relative precision lower than 20$\%$ and 10$\%$, respectively. The planets of the HIP\,29442 system are indicated by black triangles. 1$\sigma$ confidence intervals for the radius measurements are shown as black lines, while red dashed lines indicate the 3$\sigma$ confidence intervals. The overplotted theoretical curves are derived by assuming equations of state detailed in \cite{Leleu2021} }
    \label{fig:mass_radius_diag}
\end{figure*}

Following the discussion in Sect. \ref{sec:tessdetrending}, we modelled the planets' internal structure and composition adopting the $3\sigma$ confidence intervals for the  planet-to-star ratio (Tab. \ref{tab:params_amadeo}). Before entering the details of the analysis, we show in Fig. \ref{fig:mass_radius_diag} the location of the HIP\,29442 planets on a mass-radius diagram for a selected sample of planets with precise mass and radius measurements. HIP\,29442\,$b$ appears as a typical sub-Neptune, with a structure and composition consistent with an Earth-like rocky core covered by a massive H$_2$O layer, plus a small amount of gaseous envelope. We note that our mass measurement significantly differs from that obtained by \cite{akana2023arXiv230616587A} ($5.8\pm2.4$ $\mearth$), likely due mainly to undetected planets $c$ and $d$ in their work. The composition of planets $c$ and $d$ is very uncertain, as expected if one adopts the 3$\sigma$ error bars for the radii. The two planets could share a similar Earth-like rocky composition, and it cannot be ruled out that HIP\,29442\,$d$ is a high-density iron-dominated core. 

We performed a quantitative Bayesian analysis of the internal structure of the three planets, with the purpose of improving the results that can be extrapolated from Fig. \ref{fig:mass_radius_diag} by putting tighter constraints. The method is described in detail in \cite{Dorn2015,Dorn2017} and has already been used to study systems such as L98-59 \citep{Demangeon2021}, TOI-178 \citep{Leleu2021} and $\nu^2$ Lupi \citep{Delrez2021}. The model of planetary interiors assumes four layers: an inner core made of iron and sulfur, a mantle of silicates (Si, Mg, and Fe), a water layer, and a gaseous envelope of pure H-He. For the iron core, we used the equation of state from \cite{Hakim2018}, the equation of state for the silicate mantle comes from \cite{Sotin2007}, and for the water layer we adopted the equation of state from \citep{Haldemann2020}. These three layers constitute the ``solid'' part of a planet, and the thickness of the gaseous envelope depends on its mass and radius as well as the stellar age and irradiation \citep{Lopez2014}.

For the Bayesian analysis, we followed two steps. First, we generated 8000 synthetic stars, taking randomly their masses, radii, effective temperatures, ages and Mg/Si bulk molar ratios within the range of the stellar parameters derived in Sect. \ref{sec:stellarparameters}. Then, for each simulated star, we generated 6000 planetary systems, varying the internal structure parameters of all planets, and assuming that the bulk Fe/Si/Mg molar ratios are equal to the stellar ones. The transit depth and RV semi-amplitude are then computed for each of the planets, and we retained the models that fitted at 2-$\sigma$ the observed data within the error bars. By generating planetary systems around each simulated star, we include the fact that all synthetic planets orbit a star with exactly the same parameters since transit depth and RV semi-amplitude depend on the stellar radius and mass. The modelled planetary parameters are the mass fraction of each layer, the iron molar fraction in the core, the silicon and magnesium molar fraction in the mantle, the equilibrium temperature, and the age of the planet (equal to the age of the star). Uniform priors are used for these parameters, except for the mass of the gas layer which is assumed to follow a uniform-in-log prior, with the water mass fraction having an upper boundary of 0.5 \citep{Thiabaud2014,Marboeuf2014}. For more details related to the link between observed data and derived parameters, we refer to \cite{Leleu2021}. 

The results of the internal structure modelling of the planets $b$, $c$, and $d$ are shown in Table \ref{tab:structureresults}. The corner plots with the marginalised posteriors of the model free parameters are shown in Fig. \ref{fig:corner_internal_structure}. As for HIP\,29442\,$b$, they are consistent with a planet having a core with a mass fraction of $\sim$ 0.14, and a mantle with a mass fraction of $\sim$0.61 surrounded by a water layer representing 24$\%$ of the mass of the solid part of the planet. The planet could be surrounded by a gas layer of pure H-He with a mass of 0.27$^{+0.24}_{-0.17}$ M$_{\oplus}$, and a thickness of 1.37$_{0.54}^{0.49}$ R$_{\oplus}$. For HIP\,29442\,$c$ and $d$ the model suggests an internal structure consistent to that of planet $b$ and likely a Earth-like composition: $\sim$15--16$\%$ of their mass could be concentrated in an iron core, and $\sim$66--68$\%$ of the mass in the mantle. The amount of mass in form of a water layer surrounding the mantle is found to be compatible with zero for both planets (Fig. \ref{fig:corner_internal_structure}), and in Table \ref{tab:structureresults} we report the upper limits at a 95\% level of confidence. According to our results, neither HIP~29442\,$c$ nor $d$ are expected to harbour a H-He dominated atmosphere.
The model uses a scientific approach to analyse all possible configurations of planets based on their given characteristics, such as mass and radii, while considering the associated error bars. The resulting modelled planets generally exhibit lower densities compared to those estimated using the median values provided in Table \ref{tab:params_amadeo}. This tendency towards lower density allows for compositions that are more similar to Earth-like planets, especially for planet $d$ in this case. A more accurate and precise radius determination will allow to verify the model prediction.

\begin{table}
     \begin{threeparttable}
    \caption{Results of the internal structure modelling of the planets orbiting HIP\,29442. }
     \centering
	\tiny
		\begin{tabular}{l c c c}
		    \hline
		    \noalign{\smallskip}
			\textbf{Parameter} & \textbf{HIP\,29442\,$b$} & \textbf{HIP\,29442\,$c$} & \textbf{HIP\,29442\,$d$} \\
			\hline
		    \noalign{\smallskip}
			Core mass fract., $mf_{\rm core}$ &  $0.14^{+0.14}_{-0.12}$ & $0.15^{+0.14}_{-0.13}$ & $0.16^{+0.14}_{-0.13}$ \\
			 \noalign{\smallskip} 
			Mantle mass fract. $mf_{\rm mantle}$ & 0.61$^{+0.24}_{-0.20}$ & 0.66$^{+0.22}_{-0.22}$ & 0.68$^{+0.21}_{-0.23}$ \\
			 \noalign{\smallskip} 
                Water mass fract. $mf_{\rm water}$ & 0.24$^{+0.24}_{-0.21}$ & < 0.45 & < 0.42 \\
                \noalign{\smallskip} 
                Gas mass (M$_{\oplus}$) & 0.27$^{+0.24}_{-0.17}$ & 0 & 0 \\
			\noalign{\smallskip}
			\hline
		\end{tabular}  
 \begin{tablenotes}
 \tiny
\item \textbf{Notes.} For the analysis, we assumed the 3$\sigma$ confidence intervals for the planets' radii, as given in Tab \ref{tab:params_amadeo}. The posteriors are shown in Fig. \ref{fig:corner_internal_structure}. The best-fit values are given as the median of the posteriors, with the error bars defined from the 0.05-th and 0.95-th percentiles, except for the water fraction of planets $c$ and $d$, whose upper limit corresponds to the 0.95-th percentile.     
 \end{tablenotes}
\end{threeparttable}
\label{tab:structureresults}
\end{table}

\section{Conclusions}
\label{sec:conclusion}

The main achievements of this work are the discovery and the characterisation of a compact multi-planet system orbiting the evolved K0V star HIP\,29442 (TOI-469), based on data collected with the TESS space telescope and the ESPRESSO spectrograph. Thanks to the RVs collected with ESPRESSO, we first detected the Doppler signals due to the companions HIP\,29442\,$c$ and $d$. The results from the RV analysis allowed us to uncover the shallow transit signals in the TESS light curve. Using the ESPRESSO RVs we measured the dynamical masses of the planets with the very high precision of 6.7$\%$, 7.8$\%$, and 8.3$\%$ for planets $c$, $d$, and $b$ respectively. The low S/N of the transit signals of HIP\,29442\,$c$ and $d$, which made their blind detection in TESS data so far elusive, compelled us to adopt the more conservative 3$\sigma$ error bars for their radii (included planet $b$) when performing the internal structure analysis ($R_{\rm p,\,b}=3.48^{+0.19}_{-0.28} ~\rearth$, $R_{\rm p,\,c}=1.58^{+0.30}_{-0.34}~\rearth$, and  $R_{\rm p,\,d}=1.37^{+0.32}_{-0.43}~\rearth$). Photometric follow-up is indeed necessary to determine more accurate and precise radii, and consequently more accurate and precise bulk densities, which will enable to better constrain planets' internal structure and composition. Such a follow-up has been conducted with CHEOPS at the time of writing (paper in preparation).  


Albeit the uncertainty on the radii, nonetheless we can draw interesting conclusions about the system. It is composed of planets that span the bi-modal radius distribution, with the outermost planet $b$ being a sub-Neptune, possibly surrounded by a water layer for more than 20\% of the total mass, and by a gaseous envelope with $\sim30\%$ of the Earth's mass and a thickness of nearly 1.5 $\rearth$. The innermost companions $c$ and $d$ could reside at the rocky planet/super-Earth side of the distribution, but they could be located within the radius gap. Their composition is likely Earth-like, and they are not expected to have a H-He gaseous envelope. Our results with the TESS light curve do not exclude the possibility that they are high-density cores, especially planet $d$, which could be a pure iron core. This would make HIP\,29442\,$d$ a potential member of the now emerging family of super-Mercuries (i.e. higher-mass analogs of Mercury; e.g. \citealt{adibe2021}). 

\begin{figure}
    \centering
    \includegraphics[width=\columnwidth]{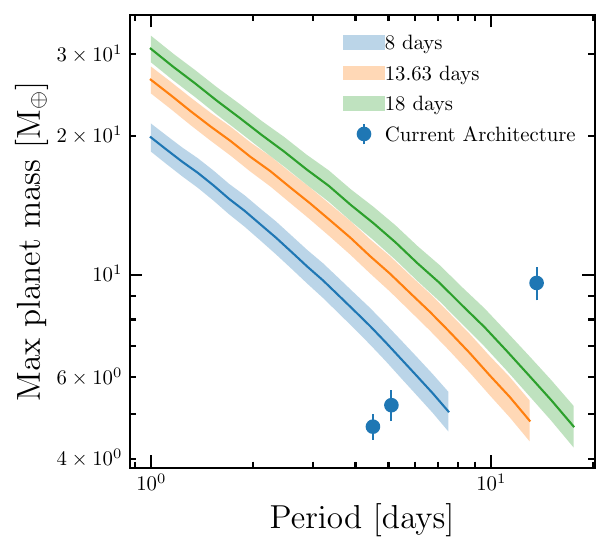}
    \caption{The maximum possible initial planet mass an interior planet (to planet $b$) could have had and be consistent with photo-evaporation, assuming planet $b$ retained its atmosphere against photo-evaporation. Given the possibility that dynamical re-arrangement within the system might have occurred, we performed this calculation assuming that planet $b$ had an initial period identical to its current period (orange), a shorter initial orbital period of 8~days (blue) or a longer initial orbital period of 18~days (green). The shaded region encompasses the 1-$\sigma$ error bars, mainly arising from uncertainties in the properties of planet $b$. The current architecture of the system is shown by the blue points. The fact the two interior planets to $b$ lie below the curves indicate the system is consistent with photo-evaporation sculpting the system before any dynamical re-arrangement. }
    \label{fig:evap_plot}
\end{figure}

HIP\,29442 represents an interesting system to explore the role of photo-evaporation during its evolution. It provides an excellent test of the photo-evaporation scenario as uncertainty in the history of the star's high-energy output can be overcome by scaling the planet's relative to each other \citep{owen2020}. To check consistency with the photo-evaporation scenario, we apply the test described in \citet{owen2020}, and we estimate the mass that planet $b$ must have to retain a hydrogen-dominated atmosphere given planet $c$ and $d$ have lost their atmosphere. We use the updated version of {\sc evapmass}\footnote{\url{https://github.com/jo276/EvapMass}}, where mass-loss efficiencies from hydrodynamic simulations are used instead of the original power-law scaling (see \citealt{Rogers2021b}). In doing this calculation we find that planet $b$ must have a mass of $\gtrsim 2.6$~M$_\oplus$ at the 95\% confidence level, consistent with its measured mass. 
We also made a reverse test to that performed above, asking: assuming that planet $b$ has retained a hydrogen-dominated atmosphere against photo-evaporation, how massive both planets $c$ and $d$ have originally been and still have lost their atmospheres? Due to the possibility of dynamical changes in the planet's orbit triggered by mass-loss \citep[e.g.][]{Boue2012,Teyssandier2015,Fujita2022}, we did this calculation assuming planet $b$ originally had an orbital period of 8~days, 13.63~days or 18~days before any dynamical re-arrangement took place. We calculated the maximum planet mass any interior planet (in this case either $c$ or $d$) could have had as a function of different original periods. Namely, if we assume planet $b$ at an original period of 8 days and the interior planet had an original period of 2 days, we ask how massive could the interior planet has been such that it could have lost its atmosphere while $b$ retained it. This calculation is then repeated for a range of possible initial periods for the interior planets. The results of this analysis are shown in Figure~\ref{fig:evap_plot}. They show that the inner planets ($c$ and $d$) could have been considerably more massive than currently, and are still consistent with a system that underwent photo-evaporation while leaving planet $b$ with a H-He-dominated atmosphere. 

Our analysis indicates that the architecture of the HIP\,29442 system is consistent with a picture in which all planets accreted a primordial hydrogen dominated atmosphere, which the inner two planets then lost due to their high irradiation levels. We leave to a future work a more detailed investigation of the possible formation and evolutionary histories of the system HIP~29442, once the planets' size and bulk physical structure will be pinned down more precisely.

\begin{acknowledgements}
The INAF authors acknowledge financial support of the Italian Ministry of Education, University, and Research through PRIN 201278X4FL and the ``Progetti Premiali'' funding scheme.
JD and YA acknowledge support from the Swiss National Science Foundation (SNSF) under grant 200020$\_$192038. Part of this work has been carried out within the framework of the NCCR PlanetS supported by the Swiss National Science Foundation under grants 51NF40$\_$182901 and 51NF40$\_$205606. J.L.-B. and A.C.-G. are partly funded by grants LCF/BQ/PI20/11760023, Ram\'on y Cajal fellowship with code RYC2021-031640-I, and the Spanish MCIN/AEI/10.13039/501100011033 grant PID2019-107061GB-C61.
JDou would like to acknowledge the funding support from the Chinese Scholarship Council (nr. 202008610218,). The giant impact simulations were carried out using the computational facilities of the Advanced Computing Research Centre, University of Bristol (\url{https://www.bristol.ac.uk/acrc/}). FPE and CLO would like to acknowledge the Swiss National Science Foundation (SNSF) for supporting research with ESPRESSO through the SNSF grants nr. 140649, 152721, 166227 and 184618. The ESPRESSO Instrument Project was partially funded through SNSF’s FLARE Programme for large infrastructures. This work has received funding from the European Research Council (ERC) under the European Union’s Horizon 2020 research and innovation programme (grant agreement SCORE No 851555). This study was funded/co-funded by the European Union (ERC, FIERCE, 101052347). Views and opinions expressed are however those of the author(s) only and do not necessarily reflect those of the European Union or the European Research Council. Neither the European Union nor the granting authority can be held responsible for them. The team of the Universidade do Porto acknowledge the support from FCT-Fundaç\~{a}o para a Ci\^{e}ncia e a Tecnologia through national funds and by FEDER through COMPETE2020-Programa Operacional Competitividade e Internacionalizaç\~{a}o by these grants: UIDB/04434/2020; UIDP/04434/2020; 2022.04048.PTDC. EDM acknowledges the support from FCT through Stimulus FCT contract 2021.01294.CEECIND. CJM also acknowledges FCT and POCH/FSE (EC) support through Investigador. FCT Contract 2021.01214.CEECIND/CP1658/CT0001. 
JIGH and ASM acknowledge financial support from the Spanish Ministry of Science and Innovation (MICINN) project PID2020-117493GB-I00, and also from the Government of the Canary Islands project ProID2020010129.
This work is based in part on observations obtained at the Southern Astrophysical Research (SOAR) telescope, which is a joint project of the Minist\'{e}rio da Ci\^{e}ncia, Tecnologia e Inova\c{c}\~{o}es (MCTI/LNA) do Brasil, the US National Science Foundation’s NOIRLab, the University of North Carolina at Chapel Hill (UNC), and Michigan State University (MSU).
This research has made use of the Exoplanet Follow-up Observation Program (ExoFOP; DOI: 10.26134/ExoFOP5) website, which is operated by the California Institute of Technology, under contract with the National Aeronautics and Space Administration under the Exoplanet Exploration Program. This work has made use of data from the European Space Agency (ESA) mission {\it Gaia} (\url{https://www.cosmos.esa.int/gaia}), processed by the {\it Gaia} Data Processing and Analysis Consortium (DPAC, \url{https://www.cosmos.esa.int/web/gaia/dpac/consortium}). Funding
for the DPAC has been provided by national institutions, in particular the institutions participating in the {\it Gaia} Multilateral Agreement.
\end{acknowledgements}

%
%

\bibliographystyle{aa} 
\bibliography{toi469_biblio} 

\begin{thebibliography}{127}
\expandafter\ifx\csname natexlab\endcsname\relax\def\natexlab#1{#1}\fi

\bibitem[{{Adibekyan} {et~al.}(2021){Adibekyan}, {Dorn}, {Sousa}, {Santos},
  {Bitsch}, {Israelian}, {Mordasini}, {Barros}, {Delgado Mena}, {Demangeon},
  {Faria}, {Figueira}, {Hakobyan}, {Oshagh}, {Soares}, {Kunitomo}, {Takeda},
  {Jofr{\'e}}, {Petrucci}, \& {Martioli}}]{adibe2021}
{Adibekyan}, V., {Dorn}, C., {Sousa}, S.~G., {et~al.} 2021, Science, 374, 330

\bibitem[{{Adibekyan} {et~al.}(2015){Adibekyan}, {Figueira}, {Santos}, {Sousa},
  {Faria}, {Delgado-Mena}, {Oshagh}, {Tsantaki}, {Hakobyan}, {Gonz{\'a}lez
  Hern{\'a}ndez}, {Su{\'a}rez-Andr{\'e}s}, \& {Israelian}}]{Adibekyan-15}
{Adibekyan}, V., {Figueira}, P., {Santos}, N.~C., {et~al.} 2015, \aap, 583, A94

\bibitem[{{Adibekyan} {et~al.}(2012){Adibekyan}, {Sousa}, {Santos}, {Delgado
  Mena}, {Gonz{\'a}lez Hern{\'a}ndez}, {Israelian}, {Mayor}, \&
  {Khachatryan}}]{Adibekyan-12}
{Adibekyan}, V.~Z., {Sousa}, S.~G., {Santos}, N.~C., {et~al.} 2012, \aap, 545,
  A32

\bibitem[{{Akana Murphy} {et~al.}(2023){Akana Murphy}, {Batalha}, {Scarsdale},
  {Isaacson}, {Ciardi}, {Gonzales}, {Giacalone}, {Twicken}, {Dattilo},
  {Fetherolf}, {Rubenzahl}, {Crossfield}, {Dressing}, {Fulton}, {Howard},
  {Huber}, {Kane}, {Petigura}, {Robertson}, {Roy}, {Weiss}, {Beard}, {Chontos},
  {Dai}, {Rice}, {Van Zandt}, {Lubin}, {Blunt}, {Polanski}, {Behmard}, {Dalba},
  {Hill}, {Rosenthal}, {Brinkman}, {Mayo}, {Turtelboom}, {Angelo},
  {Mo{\v{c}}nik}, {MacDougall}, {Pidhorodetska}, {Tyler}, {Kosiarek},
  {Holcomb}, {Louden}, {Hirsch}, {Anderson}, \&
  {Valenti}}]{akana2023arXiv230616587A}
{Akana Murphy}, J.~M., {Batalha}, N.~M., {Scarsdale}, N., {et~al.} 2023, arXiv
  e-prints, arXiv:2306.16587

\bibitem[{{Aller} {et~al.}(2020){Aller}, {Lillo-Box}, {Jones}, {Miranda}, \&
  {Barcel{\'o} Forteza}}]{tpfplotter2020}
{Aller}, A., {Lillo-Box}, J., {Jones}, D., {Miranda}, L.~F., \& {Barcel{\'o}
  Forteza}, S. 2020, \aap, 635, A128

\bibitem[{{Ambikasaran} {et~al.}(2015){Ambikasaran}, {Foreman-Mackey},
  {Greengard}, {Hogg}, \& {O'Neil}}]{2015ITPAM..38..252A}
{Ambikasaran}, S., {Foreman-Mackey}, D., {Greengard}, L., {Hogg}, D.~W., \&
  {O'Neil}, M. 2015, IEEE Transactions on Pattern Analysis and Machine
  Intelligence, 38 [\eprint[arXiv]{1403.6015}]

\bibitem[{{Bailer-Jones} {et~al.}(2021){Bailer-Jones}, {Rybizki}, {Fouesneau},
  {Demleitner}, \& {Andrae}}]{bailer2021AJ....161..147B}
{Bailer-Jones}, C.~A.~L., {Rybizki}, J., {Fouesneau}, M., {Demleitner}, M., \&
  {Andrae}, R. 2021, \aj, 161, 147

\bibitem[{{Barros} {et~al.}(2022){Barros}, {Demangeon}, {Alibert}, {Leleu},
  {Adibekyan}, {Lovis}, {Bossini}, {Sousa}, {Hara}, {Bouchy}, {Lavie},
  {Rodrigues}, {Gomes da Silva}, {Lillo-Box}, {Pepe}, {Tabernero}, {Zapatero
  Osorio}, {Sozzetti}, {Su{\'a}rez Mascare{\~n}o}, {Micela}, {Allende Prieto},
  {Cristiani}, {Damasso}, {Di Marcantonio}, {Ehrenreich}, {Faria}, {Figueira},
  {Gonz{\'a}lez Hern{\'a}ndez}, {Jenkins}, {Lo Curto}, {Martins}, {Micela},
  {Nunes}, {Pall{\'e}}, {Santos}, {Rebolo}, {Seager}, {Twicken}, {Udry},
  {Vanderspek}, \& {Winn}}]{barros2022}
{Barros}, S.~C.~C., {Demangeon}, O.~D.~S., {Alibert}, Y., {et~al.} 2022, \aap,
  665, A154

\bibitem[{{Bensby} {et~al.}(2003){Bensby}, {Feltzing}, \&
  {Lundstr{\"o}m}}]{2003A&A...410..527B}
{Bensby}, T., {Feltzing}, S., \& {Lundstr{\"o}m}, I. 2003, \aap, 410, 527

\bibitem[{{Bensby} {et~al.}(2014){Bensby}, {Feltzing}, \&
  {Oey}}]{bensby2014A&A...562A..71B}
{Bensby}, T., {Feltzing}, S., \& {Oey}, M.~S. 2014, \aap, 562, A71

\bibitem[{{Bertran de Lis} {et~al.}(2015){Bertran de Lis}, {Delgado Mena},
  {Adibekyan}, {Santos}, \& {Sousa}}]{Bertrandelis-15}
{Bertran de Lis}, S., {Delgado Mena}, E., {Adibekyan}, V.~Z., {Santos}, N.~C.,
  \& {Sousa}, S.~G. 2015, \aap, 576, A89

\bibitem[{{Bou{\'e}} {et~al.}(2012){Bou{\'e}}, {Figueira}, {Correia}, \&
  {Santos}}]{Boue2012}
{Bou{\'e}}, G., {Figueira}, P., {Correia}, A.~C.~M., \& {Santos}, N.~C. 2012,
  \aap, 537, L3

\bibitem[{{Bressan} {et~al.}(2012){Bressan}, {Marigo}, {Girardi}, {Salasnich},
  {Dal Cero}, {Rubele}, \& {Nanni}}]{Bressan2012}
{Bressan}, A., {Marigo}, P., {Girardi}, L., {et~al.} 2012, MNRAS, 427, 127

\bibitem[{{Buchner} {et~al.}(2014){Buchner}, {Georgakakis}, {Nandra}, {Hsu},
  {Rangel}, {Brightman}, {Merloni}, {Salvato}, {Donley}, \&
  {Kocevski}}]{Buchner2014}
{Buchner}, J., {Georgakakis}, A., {Nandra}, K., {et~al.} 2014, \aap, 564, A125

\bibitem[{{Cacciapuoti} {et~al.}(2022){Cacciapuoti}, {Inno}, {Covone},
  {Kostov}, {Barclay}, {Quintana}, {Colon}, {Stassun}, {Hord}, {Giacalone},
  {Kane}, {Hoffman}, {Rowe}, {Wang}, {Collins}, {Collins}, {Tan}, {Gallo},
  {Magliano}, {Ienco}, {Rabus}, {Ciardi}, {Furlan}, {Howell}, {Gnilka},
  {Scott}, {Lester}, {Ziegler}, {Brice{\~n}o}, {Law}, {Mann}, {Burke}, {Quinn},
  {Ciaramella}, {De Luca}, {Fiscale}, {Rotundi}, {Marcellino}, {Galletti},
  {Bifulco}, {Oliva}, {Spencer}, {Kaltenegger}, {McDermott}, {Essack},
  {Jenkins}, {Wohler}, {Winn}, {Seager}, {Vanderspek}, {Zhou}, {Shporer},
  {Dragomir}, \& {Fong}}]{2022A&A...668A..85C}
{Cacciapuoti}, L., {Inno}, L., {Covone}, G., {et~al.} 2022, \aap, 668, A85

\bibitem[{{Carter} {et~al.}(2012){Carter}, {Agol}, {Chaplin}, {Basu},
  {Bedding}, {Buchhave}, {Christensen-Dalsgaard}, {Deck}, {Elsworth},
  {Fabrycky}, {Ford}, {Fortney}, {Hale}, {Handberg}, {Hekker}, {Holman},
  {Huber}, {Karoff}, {Kawaler}, {Kjeldsen}, {Lissauer}, {Lopez}, {Lund},
  {Lundkvist}, {Metcalfe}, {Miglio}, {Rogers}, {Stello}, {Borucki}, {Bryson},
  {Christiansen}, {Cochran}, {Geary}, {Gilliland}, {Haas}, {Hall}, {Howard},
  {Jenkins}, {Klaus}, {Koch}, {Latham}, {MacQueen}, {Sasselov}, {Steffen},
  {Twicken}, \& {Winn}}]{Carter2012}
{Carter}, J.~A., {Agol}, E., {Chaplin}, W.~J., {et~al.} 2012, Science, 337, 556

\bibitem[{{Casali} {et~al.}(2020){Casali}, {Magrini}, {Frasca}, {Bragaglia},
  {Catanzaro}, {D'Orazi}, {Sordo}, {Carretta}, {Origlia}, {Andreuzzi}, {Fu}, \&
  {Vallenari}}]{casali2020A&A...643A..12C}
{Casali}, G., {Magrini}, L., {Frasca}, A., {et~al.} 2020, \aap, 643, A12

\bibitem[{{Castro-Gonz{\'a}lez} {et~al.}(2023){Castro-Gonz{\'a}lez},
  {Demangeon}, {Lillo-Box}, {Lovis}, {Lavie}, {Adibekyan}, {Acu{\~n}a},
  {Deleuil}, {Aguichine}, {Zapatero Osorio}, {Tabernero}, {Davoult}, {Alibert},
  {Santos}, {Sousa}, {Antoniadis-Karnavas}, {Borsa}, {Winn}, {Allende Prieto},
  {Figueira}, {Jenkins}, {Sozzetti}, {Damasso}, {Silva}, {Astudillo-Defru},
  {Barros}, {Bonfils}, {Cristiani}, {Di Marcantonio}, {Gonz{\'a}lez
  Hern{\'a}ndez}, {Curto}, {Martins}, {Nunes}, {Palle}, {Pepe}, {Seager}, \&
  {Su{\'a}rez Mascare{\~n}o}}]{2023A&A...675A..52C}
{Castro-Gonz{\'a}lez}, A., {Demangeon}, O.~D.~S., {Lillo-Box}, J., {et~al.}
  2023, \aap, 675, A52

\bibitem[{{Chaturvedi} {et~al.}(2022){Chaturvedi}, {Bluhm}, {Nagel}, {Hatzes},
  {Morello}, {Brady}, {Korth}, {Molaverdikhani}, {Kossakowski}, {Caballero},
  {Guenther}, {Pall{\'e}}, {Espinoza}, {Seifahrt}, {Lodieu}, {Cifuentes},
  {Furlan}, {Amado}, {Barclay}, {Bean}, {B{\'e}jar}, {Bergond}, {Boyle},
  {Ciardi}, {Collins}, {Collins}, {Esparza-Borges}, {Fukui}, {Gnilka}, {Goeke},
  {Guerra}, {Henning}, {Herrero}, {Howell}, {Jeffers}, {Jenkins}, {Jensen},
  {Kasper}, {Kodama}, {Latham}, {L{\'o}pez-Gonz{\'a}lez}, {Luque}, {Montes},
  {Morales}, {Mori}, {Murgas}, {Narita}, {Nowak}, {Parviainen}, {Passegger},
  {Quirrenbach}, {Reffert}, {Reiners}, {Ribas}, {Ricker}, {Rodriguez},
  {Rodr{\'\i}guez-L{\'o}pez}, {Schlecker}, {Schwarz}, {Schweitzer}, {Seager},
  {Stef{\'a}nsson}, {Stockdale}, {Tal-Or}, {Twicken}, {Vanaverbeke}, {Wang},
  {Watanabe}, {Winn}, \& {Zechmeister}}]{chaturvedi2022A&A...666A.155C}
{Chaturvedi}, P., {Bluhm}, P., {Nagel}, E., {et~al.} 2022, \aap, 666, A155

\bibitem[{{Chen} {et~al.}(2019){Chen}, {Girardi}, {Fu}, {Bressan}, {Aringer},
  {Dal Tio}, {Pastorelli}, {Marigo}, {Costa}, \& {Zhang}}]{Chen2019}
{Chen}, Y., {Girardi}, L., {Fu}, X., {et~al.} 2019, \aap, 632, A105

\bibitem[{{Ciardi} {et~al.}(2015){Ciardi}, {Beichman}, {Horch}, \&
  {Howell}}]{ciardi2015}
{Ciardi}, D.~R., {Beichman}, C.~A., {Horch}, E.~P., \& {Howell}, S.~B. 2015,
  \apj, 805, 16

\bibitem[{{Cloutier} {et~al.}(2020){Cloutier}, {Eastman}, {Rodriguez},
  {Astudillo-Defru}, {Bonfils}, {Mortier}, {Watson}, {Stalport}, {Pinamonti},
  {Lienhard}, {Harutyunyan}, {Damasso}, {Latham}, {Collins}, {Massey}, {Irwin},
  {Winters}, {Charbonneau}, {Ziegler}, {Matthews}, {Crossfield}, {Kreidberg},
  {Quinn}, {Ricker}, {Vanderspek}, {Seager}, {Winn}, {Jenkins}, {Vezie},
  {Udry}, {Twicken}, {Tenenbaum}, {Sozzetti}, {S{\'e}gransan}, {Schlieder},
  {Sasselov}, {Santos}, {Rice}, {Rackham}, {Poretti}, {Piotto}, {Phillips},
  {Pepe}, {Molinari}, {Mignon}, {Micela}, {Melo}, {de Medeiros}, {Mayor},
  {Matson}, {Martinez Fiorenzano}, {Mann}, {Magazz{\'u}}, {Lovis},
  {L{\'o}pez-Morales}, {Lopez}, {Lissauer}, {L{\'e}pine}, {Law}, {Kielkopf},
  {Johnson}, {Jensen}, {Howell}, {Gonzales}, {Ghedina}, {Forveille},
  {Figueira}, {Dumusque}, {Dressing}, {Doyon}, {D{\'\i}az}, {Fabrizio},
  {Delfosse}, {Cosentino}, {Conti}, {Collins}, {Cameron}, {Ciardi}, {Caldwell},
  {Burke}, {Buchhave}, {Brice{\~n}o}, {Boyd}, {Bouchy}, {Beichman}, {Artigau},
  \& {Almenara}}]{cloutier2020AJ....160....3C}
{Cloutier}, R., {Eastman}, J.~D., {Rodriguez}, J.~E., {et~al.} 2020, \aj, 160,
  3

\bibitem[{{Cutri} {et~al.}(2003){Cutri}, {Skrutskie}, {van Dyk}, {Beichman},
  {Carpenter}, {Chester}, {Cambresy}, {Evans}, {Fowler}, {Gizis}, {Howard},
  {Huchra}, {Jarrett}, {Kopan}, {Kirkpatrick}, {Light}, {Marsh}, {McCallon},
  {Schneider}, {Stiening}, {Sykes}, {Weinberg}, {Wheaton}, {Wheelock}, \&
  {Zacarias}}]{2003yCat.2246....0C}
{Cutri}, R.~M., {Skrutskie}, M.~F., {van Dyk}, S., {et~al.} 2003, VizieR Online
  Data Catalog, II/246

\bibitem[{{da Silva} {et~al.}(2006){da Silva}, {Girardi}, {Pasquini},
  {Setiawan}, {von der L{\"u}he}, {de Medeiros}, {Hatzes}, {D{\"o}llinger}, \&
  {Weiss}}]{PARAM}
{da Silva}, L., {Girardi}, L., {Pasquini}, L., {et~al.} 2006, A\&A, 458, 609

\bibitem[{{da Silva} {et~al.}(2012){da Silva}, {Porto de Mello}, {Milone}, {da
  Silva}, {Ribeiro}, \& {Rocha-Pinto}}]{dasilva2012A&A...542A..84D}
{da Silva}, R., {Porto de Mello}, G.~F., {Milone}, A.~C., {et~al.} 2012, \aap,
  542, A84

\bibitem[{{Damasso} {et~al.}(2018){Damasso}, {Bonomo}, {Astudillo-Defru},
  {Bonfils}, {Malavolta}, {Sozzetti}, {Lopez}, {Zeng}, {Haywood}, {Irwin},
  {Mortier}, {Vanderburg}, {Maldonado}, {Lanza}, {Affer}, {Almenara},
  {Benatti}, {Biazzo}, {Bignamini}, {Borsa}, {Bouchy}, {Buchhave}, {Cameron},
  {Carleo}, {Charbonneau}, {Claudi}, {Cosentino}, {Covino}, {Delfosse},
  {Desidera}, {Di Fabrizio}, {Dressing}, {Esposito}, {Fares}, {Figueira},
  {Fiorenzano}, {Forveille}, {Giacobbe}, {Gonz{\'a}lez-{\'A}lvarez}, {Gratton},
  {Harutyunyan}, {Johnson}, {Latham}, {Leto}, {Lopez-Morales}, {Lovis},
  {Maggio}, {Mancini}, {Masiero}, {Mayor}, {Micela}, {Molinari}, {Motalebi},
  {Murgas}, {Nascimbeni}, {Pagano}, {Pepe}, {Phillips}, {Piotto}, {Poretti},
  {Rainer}, {Rice}, {Santos}, {Sasselov}, {Scandariato}, {S{\'e}gransan},
  {Smareglia}, {Udry}, {Watson}, \& {W{\"u}nsche}}]{damasso2018A&A...615A..69D}
{Damasso}, M., {Bonomo}, A.~S., {Astudillo-Defru}, N., {et~al.} 2018, \aap,
  615, A69

\bibitem[{{Damasso} {et~al.}(2019){Damasso}, {Zeng}, {Malavolta}, {Mayo},
  {Sozzetti}, {Mortier}, {Buchhave}, {Vanderburg}, {Lopez-Morales}, {Bonomo},
  {Cameron}, {Coffinet}, {Figueira}, {Latham}, {Mayor}, {Molinari}, {Pepe},
  {Phillips}, {Poretti}, {Rice}, {Udry}, \&
  {Watson}}]{damasso2019A&A...624A..38D}
{Damasso}, M., {Zeng}, L., {Malavolta}, L., {et~al.} 2019, \aap, 624, A38

\bibitem[{{Delgado Mena} {et~al.}(2021){Delgado Mena}, {Adibekyan}, {Santos},
  {Tsantaki}, {Gonz{\'a}lez Hern{\'a}ndez}, {Sousa}, \& {Bertr{\'a}n de
  Lis}}]{delgado2021}
{Delgado Mena}, E., {Adibekyan}, V., {Santos}, N.~C., {et~al.} 2021, \aap, 655,
  A99

\bibitem[{{Delrez} {et~al.}(2021){Delrez}, {Ehrenreich}, {Alibert}, {Bonfanti},
  {Borsato}, {Fossati}, {Hooton}, {Hoyer}, {Pozuelos}, {Salmon}, {Sulis},
  {Wilson}, {Adibekyan}, {Bourrier}, {Brandeker}, {Charnoz}, {Deline},
  {Guterman}, {Haldemann}, {Hara}, {Oshagh}, {Sousa}, {Van Grootel}, {Alonso},
  {Anglada-Escud{\'e}}, {B{\'a}rczy}, {Barrado}, {Barros}, {Baumjohann},
  {Beck}, {Bekkelien}, {Benz}, {Billot}, {Bonfils}, {Broeg}, {Cabrera},
  {Collier Cameron}, {Davies}, {Deleuil}, {Delisle}, {Demangeon}, {Demory},
  {Erikson}, {Fortier}, {Fridlund}, {Futyan}, {Gandolfi}, {Garcia Mu{\~n}oz},
  {Gillon}, {Guedel}, {Heng}, {Kiss}, {Laskar}, {Lecavelier des Etangs},
  {Lendl}, {Lovis}, {Maxted}, {Nascimbeni}, {Olofsson}, {Osborn}, {Pagano},
  {Pall{\'e}}, {Piotto}, {Pollacco}, {Queloz}, {Rauer}, {Ragazzoni}, {Ribas},
  {Santos}, {Scandariato}, {S{\'e}gransan}, {Simon}, {Smith}, {Steller},
  {Szab{\'o}}, {Thomas}, {Udry}, \& {Walton}}]{Delrez2021}
{Delrez}, L., {Ehrenreich}, D., {Alibert}, Y., {et~al.} 2021, Nature Astronomy,
  5, 775

\bibitem[{{Demangeon} {et~al.}(2021){Demangeon}, {Zapatero Osorio}, {Alibert},
  {Barros}, {Adibekyan}, {Tabernero}, {Antoniadis-Karnavas}, {Camacho},
  {Su{\'a}rez Mascare{\~n}o}, {Oshagh}, {Micela}, {Sousa}, {Lovis}, {Pepe},
  {Rebolo}, {Cristiani}, {Santos}, {Allart}, {Allende Prieto}, {Bossini},
  {Bouchy}, {Cabral}, {Damasso}, {Di Marcantonio}, {D'Odorico}, {Ehrenreich},
  {Faria}, {Figueira}, {G{\'e}nova Santos}, {Haldemann}, {Hara}, {Gonz{\'a}lez
  Hern{\'a}ndez}, {Lavie}, {Lillo-Box}, {Lo Curto}, {Martins}, {M{\'e}gevand},
  {Mehner}, {Molaro}, {Nunes}, {Pall{\'e}}, {Pasquini}, {Poretti}, {Sozzetti},
  \& {Udry}}]{Demangeon2021}
{Demangeon}, O.~D.~S., {Zapatero Osorio}, M.~R., {Alibert}, Y., {et~al.} 2021,
  Astronomy and Astrophysics, 653, A41

\bibitem[{{Diamond-Lowe} {et~al.}(2022){Diamond-Lowe}, {Kreidberg}, {Harman},
  {Kempton}, {Rogers}, {Joyce}, {Eastman}, {King}, {Kopparapu}, {Youngblood},
  {Kosiarek}, {Livingston}, {Hardegree-Ullman}, \&
  {Crossfield}}]{diamond2022AJ....164..172D}
{Diamond-Lowe}, H., {Kreidberg}, L., {Harman}, C.~E., {et~al.} 2022, \aj, 164,
  172

\bibitem[{{Dorn} {et~al.}(2015){Dorn}, {Khan}, {Heng}, {Connolly}, {Alibert},
  {Benz}, \& {Tackley}}]{Dorn2015}
{Dorn}, C., {Khan}, A., {Heng}, K., {et~al.} 2015, Astronomy and Astrophysics,
  577, A83

\bibitem[{{Dorn} {et~al.}(2017){Dorn}, {Venturini}, {Khan}, {Heng}, {Alibert},
  {Helled}, {Rivoldini}, \& {Benz}}]{Dorn2017}
{Dorn}, C., {Venturini}, J., {Khan}, A., {et~al.} 2017, Astronomy and
  Astrophysics, 597, A37

\bibitem[{{Dransfield} {et~al.}(2022){Dransfield}, {Triaud}, {Guillot},
  {Mekarnia}, {Nesvorn{\'y}}, {Crouzet}, {Abe}, {Agabi}, {Buttu}, {Cabrera},
  {Gandolfi}, {G{\"u}nther}, {Rodler}, {Schmider}, {Stee}, {Suarez}, {Collins},
  {D{\'e}vora-Pajares}, {Howell}, {Matthews}, {Standing}, {Stassun},
  {Stockdale}, {Quinn}, {Ziegler}, {Crossfield}, {Lissauer}, {Mann}, {Matson},
  {Schlieder}, \& {Zhou}}]{2022MNRAS.515.1328D}
{Dransfield}, G., {Triaud}, A. H.~M.~J., {Guillot}, T., {et~al.} 2022, \mnras,
  515, 1328

\bibitem[{{Dressing} \& {Charbonneau}(2015)}]{Dressing2015}
{Dressing}, C.~D. \& {Charbonneau}, D. 2015, \apj, 807, 45

\bibitem[{Feroz {et~al.}(2011)Feroz, Balan, \& Hobson}]{feroz11}
Feroz, F., Balan, S.~T., \& Hobson, M.~P. 2011, Monthly Notices of the Royal
  Astronomical Society, 415, 3462

\bibitem[{{Feroz} {et~al.}(2019){Feroz}, {Hobson}, {Cameron}, \&
  {Pettitt}}]{Feroz2019}
{Feroz}, F., {Hobson}, M.~P., {Cameron}, E., \& {Pettitt}, A.~N. 2019, The Open
  Journal of Astrophysics, 2, 10

\bibitem[{{Foreman-Mackey}(2018)}]{celerite2}
{Foreman-Mackey}, D. 2018, Research Notes of the American Astronomical Society,
  2, 31

\bibitem[{{Foreman-Mackey} {et~al.}(2017){Foreman-Mackey}, {Agol},
  {Ambikasaran}, \& {Angus}}]{celerite1}
{Foreman-Mackey}, D., {Agol}, E., {Ambikasaran}, S., \& {Angus}, R. 2017, \aj,
  154, 220

\bibitem[{{Foreman-Mackey} {et~al.}(2013){Foreman-Mackey}, {Hogg}, {Lang}, \&
  {Goodman}}]{2013PASP..125..306F}
{Foreman-Mackey}, D., {Hogg}, D.~W., {Lang}, D., \& {Goodman}, J. 2013, \pasp,
  125, 306

\bibitem[{{Foreman-Mackey} {et~al.}(2021){Foreman-Mackey}, {Luger}, {Agol},
  {Barclay}, {Bouma}, {Brandt}, {Czekala}, {David}, {Dong}, {Gilbert},
  {Gordon}, {Hedges}, {Hey}, {Morris}, {Price-Whelan}, \&
  {Savel}}]{exoplanet:joss}
{Foreman-Mackey}, D., {Luger}, R., {Agol}, E., {et~al.} 2021, The Journal of
  Open Source Software, 6, 3285

\bibitem[{{Fujita} {et~al.}(2022){Fujita}, {Hori}, \& {Sasaki}}]{Fujita2022}
{Fujita}, N., {Hori}, Y., \& {Sasaki}, T. 2022, \apj, 928, 105

\bibitem[{Fulton {et~al.}(2017)Fulton, Petigura, Howard, Isaacson, Marcy,
  Cargile, Hebb, Weiss, Johnson, Morton, Sinukoff, Crossfield, \&
  Hirsch}]{Fulton_2017}
Fulton, B.~J., Petigura, E.~A., Howard, A.~W., {et~al.} 2017, The Astronomical
  Journal, 154, 109

\bibitem[{{Furlan} {et~al.}(2017){Furlan}, {Ciardi}, {Everett}, {Saylors},
  {Teske}, {Horch}, {Howell}, {van Belle}, {Hirsch}, {Gautier}, {Adams},
  {Barrado}, {Cartier}, {Dressing}, {Dupree}, {Gilliland}, {Lillo-Box},
  {Lucas}, \& {Wang}}]{furlan2017}
{Furlan}, E., {Ciardi}, D.~R., {Everett}, M.~E., {et~al.} 2017, \aj, 153, 71

\bibitem[{{Gaia Collaboration} {et~al.}(2021){Gaia Collaboration}, {Brown},
  {Vallenari}, {Prusti}, {de Bruijne}, {Babusiaux}, {Biermann}, {Creevey},
  {Evans}, {Eyer}, {Hutton}, {Jansen}, {Jordi}, {Klioner}, {Lammers},
  {Lindegren}, {Luri}, {Mignard}, {Panem}, {Pourbaix}, {Randich}, {Sartoretti},
  {Soubiran}, {Walton}, {Arenou}, {Bailer-Jones}, {Bastian}, {Cropper},
  {Drimmel}, {Katz}, {Lattanzi}, {van Leeuwen}, {Bakker}, {Cacciari},
  {Casta{\~n}eda}, {De Angeli}, {Ducourant}, {Fabricius}, {Fouesneau},
  {Fr{\'e}mat}, {Guerra}, {Guerrier}, {Guiraud}, {Jean-Antoine Piccolo},
  {Masana}, {Messineo}, {Mowlavi}, {Nicolas}, {Nienartowicz}, {Pailler},
  {Panuzzo}, {Riclet}, {Roux}, {Seabroke}, {Sordo}, {Tanga}, {Th{\'e}venin},
  {Gracia-Abril}, {Portell}, {Teyssier}, {Altmann}, {Andrae}, {Bellas-Velidis},
  {Benson}, {Berthier}, {Blomme}, {Brugaletta}, {Burgess}, {Busso}, {Carry},
  {Cellino}, {Cheek}, {Clementini}, {Damerdji}, {Davidson}, {Delchambre},
  {Dell'Oro}, {Fern{\'a}ndez-Hern{\'a}ndez}, {Galluccio}, {Garc{\'\i}a-Lario},
  {Garcia-Reinaldos}, {Gonz{\'a}lez-N{\'u}{\~n}ez}, {Gosset}, {Haigron},
  {Halbwachs}, {Hambly}, {Harrison}, {Hatzidimitriou}, {Heiter},
  {Hern{\'a}ndez}, {Hestroffer}, {Hodgkin}, {Holl}, {Jan{\ss}en}, {Jevardat de
  Fombelle}, {Jordan}, {Krone-Martins}, {Lanzafame}, {L{\"o}ffler}, {Lorca},
  {Manteiga}, {Marchal}, {Marrese}, {Moitinho}, {Mora}, {Muinonen}, {Osborne},
  {Pancino}, {Pauwels}, {Petit}, {Recio-Blanco}, {Richards}, {Riello},
  {Rimoldini}, {Robin}, {Roegiers}, {Rybizki}, {Sarro}, {Siopis}, {Smith},
  {Sozzetti}, {Ulla}, {Utrilla}, {van Leeuwen}, {van Reeven}, {Abbas}, {Abreu
  Aramburu}, {Accart}, {Aerts}, {Aguado}, {Ajaj}, {Altavilla}, {{\'A}lvarez},
  {{\'A}lvarez Cid-Fuentes}, {Alves}, {Anderson}, {Anglada Varela}, {Antoja},
  {Audard}, {Baines}, {Baker}, {Balaguer-N{\'u}{\~n}ez}, {Balbinot}, {Balog},
  {Barache}, {Barbato}, {Barros}, {Barstow}, {Bartolom{\'e}}, {Bassilana},
  {Bauchet}, {Baudesson-Stella}, {Becciani}, {Bellazzini}, {Bernet}, {Bertone},
  {Bianchi}, {Blanco-Cuaresma}, {Boch}, {Bombrun}, {Bossini}, {Bouquillon},
  {Bragaglia}, {Bramante}, {Breedt}, {Bressan}, {Brouillet}, {Bucciarelli},
  {Burlacu}, {Busonero}, {Butkevich}, {Buzzi}, {Caffau}, {Cancelliere},
  {C{\'a}novas}, {Cantat-Gaudin}, {Carballo}, {Carlucci}, {Carnerero},
  {Carrasco}, {Casamiquela}, {Castellani}, {Castro-Ginard}, {Castro Sampol},
  {Chaoul}, {Charlot}, {Chemin}, {Chiavassa}, {Cioni}, {Comoretto}, {Cooper},
  {Cornez}, {Cowell}, {Crifo}, {Crosta}, {Crowley}, {Dafonte}, {Dapergolas},
  {David}, {David}, {de Laverny}, {De Luise}, {De March}, {De Ridder}, {de
  Souza}, {de Teodoro}, {de Torres}, {del Peloso}, {del Pozo}, {Delbo},
  {Delgado}, {Delgado}, {Delisle}, {Di Matteo}, {Diakite}, {Diener},
  {Distefano}, {Dolding}, {Eappachen}, {Edvardsson}, {Enke}, {Esquej}, {Fabre},
  {Fabrizio}, {Faigler}, {Fedorets}, {Fernique}, {Fienga}, {Figueras},
  {Fouron}, {Fragkoudi}, {Fraile}, {Franke}, {Gai}, {Garabato},
  {Garcia-Gutierrez}, {Garc{\'\i}a-Torres}, {Garofalo}, {Gavras}, {Gerlach},
  {Geyer}, {Giacobbe}, {Gilmore}, {Girona}, {Giuffrida}, {Gomel}, {Gomez},
  {Gonzalez-Santamaria}, {Gonz{\'a}lez-Vidal}, {Granvik},
  {Guti{\'e}rrez-S{\'a}nchez}, {Guy}, {Hauser}, {Haywood}, {Helmi}, {Hidalgo},
  {Hilger}, {H{\l}adczuk}, {Hobbs}, {Holland}, {Huckle}, {Jasniewicz},
  {Jonker}, {Juaristi Campillo}, {Julbe}, {Karbevska}, {Kervella}, {Khanna},
  {Kochoska}, {Kontizas}, {Kordopatis}, {Korn}, {Kostrzewa-Rutkowska},
  {Kruszy{\'n}ska}, {Lambert}, {Lanza}, {Lasne}, {Le Campion}, {Le Fustec},
  {Lebreton}, {Lebzelter}, {Leccia}, {Leclerc}, {Lecoeur-Taibi}, {Liao},
  {Licata}, {Lindstr{\o}m}, {Lister}, {Livanou}, {Lobel}, {Madrero Pardo},
  {Managau}, {Mann}, {Marchant}, {Marconi}, {Marcos Santos}, {Marinoni},
  {Marocco}, {Marshall}, {Martin Polo}, {Mart{\'\i}n-Fleitas}, {Masip},
  {Massari}, {Mastrobuono-Battisti}, {Mazeh}, {McMillan}, {Messina},
  {Michalik}, {Millar}, {Mints}, {Molina}, {Molinaro}, {Moln{\'a}r},
  {Montegriffo}, {Mor}, {Morbidelli}, {Morel}, {Morris}, {Mulone}, {Munoz},
  {Muraveva}, {Murphy}, {Musella}, {Noval}, {Ord{\'e}novic}, {Orr{\`u}},
  {Osinde}, {Pagani}, {Pagano}, {Palaversa}, {Palicio}, {Panahi}, {Pawlak},
  {Pe{\~n}alosa Esteller}, {Penttil{\"a}}, {Piersimoni}, {Pineau}, {Plachy},
  {Plum}, {Poggio}, {Poretti}, {Poujoulet}, {Pr{\v{s}}a}, {Pulone}, {Racero},
  {Ragaini}, {Rainer}, {Raiteri}, {Rambaux}, {Ramos}, {Ramos-Lerate}, {Re
  Fiorentin}, {Regibo}, {Reyl{\'e}}, {Ripepi}, {Riva}, {Rixon}, {Robichon},
  {Robin}, {Roelens}, {Rohrbasser}, {Romero-G{\'o}mez}, {Rowell}, {Royer},
  {Rybicki}, {Sadowski}, {Sagrist{\`a} Sell{\'e}s}, {Sahlmann}, {Salgado},
  {Salguero}, {Samaras}, {Sanchez Gimenez}, {Sanna}, {Santove{\~n}a},
  {Sarasso}, {Schultheis}, {Sciacca}, {Segol}, {Segovia}, {S{\'e}gransan},
  {Semeux}, {Shahaf}, {Siddiqui}, {Siebert}, {Siltala}, {Slezak}, {Smart},
  {Solano}, {Solitro}, {Souami}, {Souchay}, {Spagna}, {Spoto}, {Steele},
  {Steidelm{\"u}ller}, {Stephenson}, {S{\"u}veges}, {Szabados}, {Szegedi-Elek},
  {Taris}, {Tauran}, {Taylor}, {Teixeira}, {Thuillot}, {Tonello}, {Torra},
  {Torra}, {Turon}, {Unger}, {Vaillant}, {van Dillen}, {Vanel}, {Vecchiato},
  {Viala}, {Vicente}, {Voutsinas}, {Weiler}, {Wevers}, {Wyrzykowski}, {Yoldas},
  {Yvard}, {Zhao}, {Zorec}, {Zucker}, {Zurbach}, \& {Zwitter}}]{GAIAEDR3}
{Gaia Collaboration}, {Brown}, A.~G.~A., {Vallenari}, A., {et~al.} 2021, \aap,
  649, A1

\bibitem[{{Gaia Collaboration} {et~al.}(2016){Gaia Collaboration}, {Prusti},
  {de Bruijne}, {Brown}, {Vallenari}, {Babusiaux}, {Bailer-Jones}, {Bastian},
  {Biermann}, {Evans}, {Eyer}, {Jansen}, {Jordi}, {Klioner}, {Lammers},
  {Lindegren}, {Luri}, {Mignard}, {Milligan}, {Panem}, {Poinsignon},
  {Pourbaix}, {Randich}, {Sarri}, {Sartoretti}, {Siddiqui}, {Soubiran},
  {Valette}, {van Leeuwen}, {Walton}, {Aerts}, {Arenou}, {Cropper}, {Drimmel},
  {H{\o}g}, {Katz}, {Lattanzi}, {O'Mullane}, {Grebel}, {Holland}, {Huc},
  {Passot}, {Bramante}, {Cacciari}, {Casta{\~n}eda}, {Chaoul}, {Cheek}, {De
  Angeli}, {Fabricius}, {Guerra}, {Hern{\'a}ndez}, {Jean-Antoine-Piccolo},
  {Masana}, {Messineo}, {Mowlavi}, {Nienartowicz}, {Ord{\'o}{\~n}ez-Blanco},
  {Panuzzo}, {Portell}, {Richards}, {Riello}, {Seabroke}, {Tanga},
  {Th{\'e}venin}, {Torra}, {Els}, {Gracia-Abril}, {Comoretto},
  {Garcia-Reinaldos}, {Lock}, {Mercier}, {Altmann}, {Andrae}, {Astraatmadja},
  {Bellas-Velidis}, {Benson}, {Berthier}, {Blomme}, {Busso}, {Carry},
  {Cellino}, {Clementini}, {Cowell}, {Creevey}, {Cuypers}, {Davidson}, {De
  Ridder}, {de Torres}, {Delchambre}, {Dell'Oro}, {Ducourant}, {Fr{\'e}mat},
  {Garc{\'\i}a-Torres}, {Gosset}, {Halbwachs}, {Hambly}, {Harrison}, {Hauser},
  {Hestroffer}, {Hodgkin}, {Huckle}, {Hutton}, {Jasniewicz}, {Jordan},
  {Kontizas}, {Korn}, {Lanzafame}, {Manteiga}, {Moitinho}, {Muinonen},
  {Osinde}, {Pancino}, {Pauwels}, {Petit}, {Recio-Blanco}, {Robin}, {Sarro},
  {Siopis}, {Smith}, {Smith}, {Sozzetti}, {Thuillot}, {van Reeven}, {Viala},
  {Abbas}, {Abreu Aramburu}, {Accart}, {Aguado}, {Allan}, {Allasia},
  {Altavilla}, {{\'A}lvarez}, {Alves}, {Anderson}, {Andrei}, {Anglada Varela},
  {Antiche}, {Antoja}, {Ant{\'o}n}, {Arcay}, {Atzei}, {Ayache}, {Bach},
  {Baker}, {Balaguer-N{\'u}{\~n}ez}, {Barache}, {Barata}, {Barbier}, {Barblan},
  {Baroni}, {Barrado y Navascu{\'e}s}, {Barros}, {Barstow}, {Becciani},
  {Bellazzini}, {Bellei}, {Bello Garc{\'\i}a}, {Belokurov}, {Bendjoya},
  {Berihuete}, {Bianchi}, {Bienaym{\'e}}, {Billebaud}, {Blagorodnova},
  {Blanco-Cuaresma}, {Boch}, {Bombrun}, {Borrachero}, {Bouquillon}, {Bourda},
  {Bouy}, {Bragaglia}, {Breddels}, {Brouillet}, {Br{\"u}semeister},
  {Bucciarelli}, {Budnik}, {Burgess}, {Burgon}, {Burlacu}, {Busonero}, {Buzzi},
  {Caffau}, {Cambras}, {Campbell}, {Cancelliere}, {Cantat-Gaudin}, {Carlucci},
  {Carrasco}, {Castellani}, {Charlot}, {Charnas}, {Charvet}, {Chassat},
  {Chiavassa}, {Clotet}, {Cocozza}, {Collins}, {Collins}, {Costigan}, {Crifo},
  {Cross}, {Crosta}, {Crowley}, {Dafonte}, {Damerdji}, {Dapergolas}, {David},
  {David}, {De Cat}, {de Felice}, {de Laverny}, {De Luise}, {De March}, {de
  Martino}, {de Souza}, {Debosscher}, {del Pozo}, {Delbo}, {Delgado},
  {Delgado}, {di Marco}, {Di Matteo}, {Diakite}, {Distefano}, {Dolding}, {Dos
  Anjos}, {Drazinos}, {Dur{\'a}n}, {Dzigan}, {Ecale}, {Edvardsson}, {Enke},
  {Erdmann}, {Escolar}, {Espina}, {Evans}, {Eynard Bontemps}, {Fabre},
  {Fabrizio}, {Faigler}, {Falc{\~a}o}, {Farr{\`a}s Casas}, {Faye}, {Federici},
  {Fedorets}, {Fern{\'a}ndez-Hern{\'a}ndez}, {Fernique}, {Fienga}, {Figueras},
  {Filippi}, {Findeisen}, {Fonti}, {Fouesneau}, {Fraile}, {Fraser}, {Fuchs},
  {Furnell}, {Gai}, {Galleti}, {Galluccio}, {Garabato}, {Garc{\'\i}a-Sedano},
  {Gar{\'e}}, {Garofalo}, {Garralda}, {Gavras}, {Gerssen}, {Geyer}, {Gilmore},
  {Girona}, {Giuffrida}, {Gomes}, {Gonz{\'a}lez-Marcos},
  {Gonz{\'a}lez-N{\'u}{\~n}ez}, {Gonz{\'a}lez-Vidal}, {Granvik}, {Guerrier},
  {Guillout}, {Guiraud}, {G{\'u}rpide}, {Guti{\'e}rrez-S{\'a}nchez}, {Guy},
  {Haigron}, {Hatzidimitriou}, {Haywood}, {Heiter}, {Helmi}, {Hobbs},
  {Hofmann}, {Holl}, {Holland}, {Hunt}, {Hypki}, {Icardi}, {Irwin}, {Jevardat
  de Fombelle}, {Jofr{\'e}}, {Jonker}, {Jorissen}, {Julbe}, {Karampelas},
  {Kochoska}, {Kohley}, {Kolenberg}, {Kontizas}, {Koposov}, {Kordopatis},
  {Koubsky}, {Kowalczyk}, {Krone-Martins}, {Kudryashova}, {Kull}, {Bachchan},
  {Lacoste-Seris}, {Lanza}, {Lavigne}, {Le Poncin-Lafitte}, {Lebreton},
  {Lebzelter}, {Leccia}, {Leclerc}, {Lecoeur-Taibi}, {Lemaitre}, {Lenhardt},
  {Leroux}, {Liao}, {Licata}, {Lindstr{\o}m}, {Lister}, {Livanou}, {Lobel},
  {L{\"o}ffler}, {L{\'o}pez}, {Lopez-Lozano}, {Lorenz}, {Loureiro},
  {MacDonald}, {Magalh{\~a}es Fernandes}, {Managau}, {Mann}, {Mantelet},
  {Marchal}, {Marchant}, {Marconi}, {Marie}, {Marinoni}, {Marrese},
  {Marschalk{\'o}}, {Marshall}, {Mart{\'\i}n-Fleitas}, {Martino}, {Mary},
  {Matijevi{\v{c}}}, {Mazeh}, {McMillan}, {Messina}, {Mestre}, {Michalik},
  {Millar}, {Miranda}, {Molina}, {Molinaro}, {Molinaro}, {Moln{\'a}r},
  {Moniez}, {Montegriffo}, {Monteiro}, {Mor}, {Mora}, {Morbidelli}, {Morel},
  {Morgenthaler}, {Morley}, {Morris}, {Mulone}, {Muraveva}, {Musella},
  {Narbonne}, {Nelemans}, {Nicastro}, {Noval}, {Ord{\'e}novic},
  {Ordieres-Mer{\'e}}, {Osborne}, {Pagani}, {Pagano}, {Pailler}, {Palacin},
  {Palaversa}, {Parsons}, {Paulsen}, {Pecoraro}, {Pedrosa}, {Pentik{\"a}inen},
  {Pereira}, {Pichon}, {Piersimoni}, {Pineau}, {Plachy}, {Plum}, {Poujoulet},
  {Pr{\v{s}}a}, {Pulone}, {Ragaini}, {Rago}, {Rambaux}, {Ramos-Lerate},
  {Ranalli}, {Rauw}, {Read}, {Regibo}, {Renk}, {Reyl{\'e}}, {Ribeiro},
  {Rimoldini}, {Ripepi}, {Riva}, {Rixon}, {Roelens}, {Romero-G{\'o}mez},
  {Rowell}, {Royer}, {Rudolph}, {Ruiz-Dern}, {Sadowski}, {Sagrist{\`a}
  Sell{\'e}s}, {Sahlmann}, {Salgado}, {Salguero}, {Sarasso}, {Savietto},
  {Schnorhk}, {Schultheis}, {Sciacca}, {Segol}, {Segovia}, {Segransan},
  {Serpell}, {Shih}, {Smareglia}, {Smart}, {Smith}, {Solano}, {Solitro},
  {Sordo}, {Soria Nieto}, {Souchay}, {Spagna}, {Spoto}, {Stampa}, {Steele},
  {Steidelm{\"u}ller}, {Stephenson}, {Stoev}, {Suess}, {S{\"u}veges}, {Surdej},
  {Szabados}, {Szegedi-Elek}, {Tapiador}, {Taris}, {Tauran}, {Taylor},
  {Teixeira}, {Terrett}, {Tingley}, {Trager}, {Turon}, {Ulla}, {Utrilla},
  {Valentini}, {van Elteren}, {Van Hemelryck}, {van Leeuwen}, {Varadi},
  {Vecchiato}, {Veljanoski}, {Via}, {Vicente}, {Vogt}, {Voss}, {Votruba},
  {Voutsinas}, {Walmsley}, {Weiler}, {Weingrill}, {Werner}, {Wevers},
  {Whitehead}, {Wyrzykowski}, {Yoldas}, {{\v{Z}}erjal}, {Zucker}, {Zurbach},
  {Zwitter}, {Alecu}, {Allen}, {Allende Prieto}, {Amorim},
  {Anglada-Escud{\'e}}, {Arsenijevic}, {Azaz}, {Balm}, {Beck}, {Bernstein},
  {Bigot}, {Bijaoui}, {Blasco}, {Bonfigli}, {Bono}, {Boudreault}, {Bressan},
  {Brown}, {Brunet}, {Bunclark}, {Buonanno}, {Butkevich}, {Carret}, {Carrion},
  {Chemin}, {Ch{\'e}reau}, {Corcione}, {Darmigny}, {de Boer}, {de Teodoro}, {de
  Zeeuw}, {Delle Luche}, {Domingues}, {Dubath}, {Fodor}, {Fr{\'e}zouls},
  {Fries}, {Fustes}, {Fyfe}, {Gallardo}, {Gallegos}, {Gardiol}, {Gebran},
  {Gomboc}, {G{\'o}mez}, {Grux}, {Gueguen}, {Heyrovsky}, {Hoar}, {Iannicola},
  {Isasi Parache}, {Janotto}, {Joliet}, {Jonckheere}, {Keil}, {Kim},
  {Klagyivik}, {Klar}, {Knude}, {Kochukhov}, {Kolka}, {Kos}, {Kutka}, {Lainey},
  {LeBouquin}, {Liu}, {Loreggia}, {Makarov}, {Marseille}, {Martayan},
  {Martinez-Rubi}, {Massart}, {Meynadier}, {Mignot}, {Munari}, {Nguyen},
  {Nordlander}, {Ocvirk}, {O'Flaherty}, {Olias Sanz}, {Ortiz}, {Osorio},
  {Oszkiewicz}, {Ouzounis}, {Palmer}, {Park}, {Pasquato}, {Peltzer}, {Peralta},
  {P{\'e}turaud}, {Pieniluoma}, {Pigozzi}, {Poels}, {Prat}, {Prod'homme},
  {Raison}, {Rebordao}, {Risquez}, {Rocca-Volmerange}, {Rosen}, {Ruiz-Fuertes},
  {Russo}, {Sembay}, {Serraller Vizcaino}, {Short}, {Siebert}, {Silva},
  {Sinachopoulos}, {Slezak}, {Soffel}, {Sosnowska}, {Strai{\v{z}}ys}, {ter
  Linden}, {Terrell}, {Theil}, {Tiede}, {Troisi}, {Tsalmantza}, {Tur},
  {Vaccari}, {Vachier}, {Valles}, {Van Hamme}, {Veltz}, {Virtanen}, {Wallut},
  {Wichmann}, {Wilkinson}, {Ziaeepour}, \&
  {Zschocke}}]{gaia2016A&A...595A...1G}
{Gaia Collaboration}, {Prusti}, T., {de Bruijne}, J.~H.~J., {et~al.} 2016,
  \aap, 595, A1

\bibitem[{{Gaia Collaboration} {et~al.}(2023){Gaia Collaboration}, {Vallenari},
  {Brown}, {Prusti}, {de Bruijne}, {Arenou}, {Babusiaux}, {Biermann},
  {Creevey}, {Ducourant}, {Evans}, {Eyer}, {Guerra}, {Hutton}, {Jordi},
  {Klioner}, {Lammers}, {Lindegren}, {Luri}, {Mignard}, {Panem}, {Pourbaix},
  {Randich}, {Sartoretti}, {Soubiran}, {Tanga}, {Walton}, {Bailer-Jones},
  {Bastian}, {Drimmel}, {Jansen}, {Katz}, {Lattanzi}, {van Leeuwen}, {Bakker},
  {Cacciari}, {Casta{\~n}eda}, {De Angeli}, {Fabricius}, {Fouesneau},
  {Fr{\'e}mat}, {Galluccio}, {Guerrier}, {Heiter}, {Masana}, {Messineo},
  {Mowlavi}, {Nicolas}, {Nienartowicz}, {Pailler}, {Panuzzo}, {Riclet}, {Roux},
  {Seabroke}, {Sordo}, {Th{\'e}venin}, {Gracia-Abril}, {Portell}, {Teyssier},
  {Altmann}, {Andrae}, {Audard}, {Bellas-Velidis}, {Benson}, {Berthier},
  {Blomme}, {Burgess}, {Busonero}, {Busso}, {C{\'a}novas}, {Carry}, {Cellino},
  {Cheek}, {Clementini}, {Damerdji}, {Davidson}, {de Teodoro}, {Nu{\~n}ez
  Campos}, {Delchambre}, {Dell'Oro}, {Esquej}, {Fern{\'a}ndez-Hern{\'a}ndez},
  {Fraile}, {Garabato}, {Garc{\'\i}a-Lario}, {Gosset}, {Haigron}, {Halbwachs},
  {Hambly}, {Harrison}, {Hern{\'a}ndez}, {Hestroffer}, {Hodgkin}, {Holl},
  {Jan{\ss}en}, {Jevardat de Fombelle}, {Jordan}, {Krone-Martins}, {Lanzafame},
  {L{\"o}ffler}, {Marchal}, {Marrese}, {Moitinho}, {Muinonen}, {Osborne},
  {Pancino}, {Pauwels}, {Recio-Blanco}, {Reyl{\'e}}, {Riello}, {Rimoldini},
  {Roegiers}, {Rybizki}, {Sarro}, {Siopis}, {Smith}, {Sozzetti}, {Utrilla},
  {van Leeuwen}, {Abbas}, {{\'A}brah{\'a}m}, {Abreu Aramburu}, {Aerts},
  {Aguado}, {Ajaj}, {Aldea-Montero}, {Altavilla}, {{\'A}lvarez}, {Alves},
  {Anders}, {Anderson}, {Anglada Varela}, {Antoja}, {Baines}, {Baker},
  {Balaguer-N{\'u}{\~n}ez}, {Balbinot}, {Balog}, {Barache}, {Barbato},
  {Barros}, {Barstow}, {Bartolom{\'e}}, {Bassilana}, {Bauchet}, {Becciani},
  {Bellazzini}, {Berihuete}, {Bernet}, {Bertone}, {Bianchi}, {Binnenfeld},
  {Blanco-Cuaresma}, {Blazere}, {Boch}, {Bombrun}, {Bossini}, {Bouquillon},
  {Bragaglia}, {Bramante}, {Breedt}, {Bressan}, {Brouillet}, {Brugaletta},
  {Bucciarelli}, {Burlacu}, {Butkevich}, {Buzzi}, {Caffau}, {Cancelliere},
  {Cantat-Gaudin}, {Carballo}, {Carlucci}, {Carnerero}, {Carrasco},
  {Casamiquela}, {Castellani}, {Castro-Ginard}, {Chaoul}, {Charlot}, {Chemin},
  {Chiaramida}, {Chiavassa}, {Chornay}, {Comoretto}, {Contursi}, {Cooper},
  {Cornez}, {Cowell}, {Crifo}, {Cropper}, {Crosta}, {Crowley}, {Dafonte},
  {Dapergolas}, {David}, {David}, {de Laverny}, {De Luise}, {De March}, {De
  Ridder}, {de Souza}, {de Torres}, {del Peloso}, {del Pozo}, {Delbo},
  {Delgado}, {Delisle}, {Demouchy}, {Dharmawardena}, {Di Matteo}, {Diakite},
  {Diener}, {Distefano}, {Dolding}, {Edvardsson}, {Enke}, {Fabre}, {Fabrizio},
  {Faigler}, {Fedorets}, {Fernique}, {Fienga}, {Figueras}, {Fournier},
  {Fouron}, {Fragkoudi}, {Gai}, {Garcia-Gutierrez}, {Garcia-Reinaldos},
  {Garc{\'\i}a-Torres}, {Garofalo}, {Gavel}, {Gavras}, {Gerlach}, {Geyer},
  {Giacobbe}, {Gilmore}, {Girona}, {Giuffrida}, {Gomel}, {Gomez},
  {Gonz{\'a}lez-N{\'u}{\~n}ez}, {Gonz{\'a}lez-Santamar{\'\i}a},
  {Gonz{\'a}lez-Vidal}, {Granvik}, {Guillout}, {Guiraud},
  {Guti{\'e}rrez-S{\'a}nchez}, {Guy}, {Hatzidimitriou}, {Hauser}, {Haywood},
  {Helmer}, {Helmi}, {Sarmiento}, {Hidalgo}, {Hilger}, {H{\l}adczuk}, {Hobbs},
  {Holland}, {Huckle}, {Jardine}, {Jasniewicz}, {Jean-Antoine Piccolo},
  {Jim{\'e}nez-Arranz}, {Jorissen}, {Juaristi Campillo}, {Julbe}, {Karbevska},
  {Kervella}, {Khanna}, {Kontizas}, {Kordopatis}, {Korn}, {K{\'o}sp{\'a}l},
  {Kostrzewa-Rutkowska}, {Kruszy{\'n}ska}, {Kun}, {Laizeau}, {Lambert},
  {Lanza}, {Lasne}, {Le Campion}, {Lebreton}, {Lebzelter}, {Leccia}, {Leclerc},
  {Lecoeur-Taibi}, {Liao}, {Licata}, {Lindstr{\o}m}, {Lister}, {Livanou},
  {Lobel}, {Lorca}, {Loup}, {Madrero Pardo}, {Magdaleno Romeo}, {Managau},
  {Mann}, {Manteiga}, {Marchant}, {Marconi}, {Marcos}, {Marcos Santos},
  {Mar{\'\i}n Pina}, {Marinoni}, {Marocco}, {Marshall}, {Martin Polo},
  {Mart{\'\i}n-Fleitas}, {Marton}, {Mary}, {Masip}, {Massari},
  {Mastrobuono-Battisti}, {Mazeh}, {McMillan}, {Messina}, {Michalik}, {Millar},
  {Mints}, {Molina}, {Molinaro}, {Moln{\'a}r}, {Monari}, {Mongui{\'o}},
  {Montegriffo}, {Montero}, {Mor}, {Mora}, {Morbidelli}, {Morel}, {Morris},
  {Muraveva}, {Murphy}, {Musella}, {Nagy}, {Noval}, {Oca{\~n}a}, {Ogden},
  {Ordenovic}, {Osinde}, {Pagani}, {Pagano}, {Palaversa}, {Palicio},
  {Pallas-Quintela}, {Panahi}, {Payne-Wardenaar}, {Pe{\~n}alosa Esteller},
  {Penttil{\"a}}, {Pichon}, {Piersimoni}, {Pineau}, {Plachy}, {Plum}, {Poggio},
  {Pr{\v{s}}a}, {Pulone}, {Racero}, {Ragaini}, {Rainer}, {Raiteri}, {Rambaux},
  {Ramos}, {Ramos-Lerate}, {Re Fiorentin}, {Regibo}, {Richards}, {Rios Diaz},
  {Ripepi}, {Riva}, {Rix}, {Rixon}, {Robichon}, {Robin}, {Robin}, {Roelens},
  {Rogues}, {Rohrbasser}, {Romero-G{\'o}mez}, {Rowell}, {Royer}, {Ruz Mieres},
  {Rybicki}, {Sadowski}, {S{\'a}ez N{\'u}{\~n}ez}, {Sagrist{\`a} Sell{\'e}s},
  {Sahlmann}, {Salguero}, {Samaras}, {Sanchez Gimenez}, {Sanna},
  {Santove{\~n}a}, {Sarasso}, {Schultheis}, {Sciacca}, {Segol}, {Segovia},
  {S{\'e}gransan}, {Semeux}, {Shahaf}, {Siddiqui}, {Siebert}, {Siltala},
  {Silvelo}, {Slezak}, {Slezak}, {Smart}, {Snaith}, {Solano}, {Solitro},
  {Souami}, {Souchay}, {Spagna}, {Spina}, {Spoto}, {Steele},
  {Steidelm{\"u}ller}, {Stephenson}, {S{\"u}veges}, {Surdej}, {Szabados},
  {Szegedi-Elek}, {Taris}, {Taylor}, {Teixeira}, {Tolomei}, {Tonello}, {Torra},
  {Torra}, {Torralba Elipe}, {Trabucchi}, {Tsounis}, {Turon}, {Ulla}, {Unger},
  {Vaillant}, {van Dillen}, {van Reeven}, {Vanel}, {Vecchiato}, {Viala},
  {Vicente}, {Voutsinas}, {Weiler}, {Wevers}, {Wyrzykowski}, {Yoldas}, {Yvard},
  {Zhao}, {Zorec}, {Zucker}, \& {Zwitter}}]{gaiadr32023A&A...674A...1G}
{Gaia Collaboration}, {Vallenari}, A., {Brown}, A.~G.~A., {et~al.} 2023, \aap,
  674, A1

\bibitem[{Giacalone {et~al.}(2020)Giacalone, Dressing, Jensen, Collins, Ricker,
  Vanderspek, Seager, Winn, Jenkins, Barclay, Barkaoui, Cadieux, Charbonneau,
  Collins, Conti, Doyon, Evans, Ghachoui, Gillon, Guerrero, Hart, Jehin,
  Kielkopf, McLean, Murgas, Palle, Parviainen, Pozuelos, Relles, Shporer,
  Socia, Stockdale, Tan, Torres, Twicken, Waalkes, \& Waite}]{Giacalone2021}
Giacalone, S., Dressing, C.~D., Jensen, E. L.~N., {et~al.} 2020, The
  Astronomical Journal, 161, 24

\bibitem[{{Ginzburg} {et~al.}(2018){Ginzburg}, {Schlichting}, \&
  {Sari}}]{ginzburg2018MNRAS.476..759G}
{Ginzburg}, S., {Schlichting}, H.~E., \& {Sari}, R. 2018, \mnras, 476, 759

\bibitem[{{Goodman} \& {Weare}(2010)}]{2010CAMCS...5...65G}
{Goodman}, J. \& {Weare}, J. 2010, Communications in Applied Mathematics and
  Computational Science, 5, 65

\bibitem[{{G{\"u}nther} {et~al.}(2019){G{\"u}nther}, {Pozuelos}, {Dittmann},
  {Dragomir}, {Kane}, {Daylan}, {Feinstein}, {Huang}, {Morton}, {Bonfanti},
  {Bouma}, {Burt}, {Collins}, {Lissauer}, {Matthews}, {Montet}, {Vanderburg},
  {Wang}, {Winters}, {Ricker}, {Vanderspek}, {Latham}, {Seager}, {Winn},
  {Jenkins}, {Armstrong}, {Barkaoui}, {Batalha}, {Bean}, {Caldwell}, {Ciardi},
  {Collins}, {Crossfield}, {Fausnaugh}, {Furesz}, {Gan}, {Gillon}, {Guerrero},
  {Horne}, {Howell}, {Ireland}, {Isopi}, {Jehin}, {Kielkopf}, {Lepine},
  {Mallia}, {Matson}, {Myers}, {Palle}, {Quinn}, {Relles}, {Rojas-Ayala},
  {Schlieder}, {Sefako}, {Shporer}, {Su{\'a}rez}, {Tan}, {Ting}, {Twicken}, \&
  {Waite}}]{Gunther2019}
{G{\"u}nther}, M.~N., {Pozuelos}, F.~J., {Dittmann}, J.~A., {et~al.} 2019,
  Nature Astronomy, 3, 1099

\bibitem[{{Gupta} \& {Schlichting}(2019)}]{gupta2019MNRAS.487...24G}
{Gupta}, A. \& {Schlichting}, H.~E. 2019, \mnras, 487, 24

\bibitem[{{Hakim} {et~al.}(2018){Hakim}, {Rivoldini}, {Van Hoolst},
  {Cottenier}, {Jaeken}, {Chust}, \& {Steinle-Neumann}}]{Hakim2018}
{Hakim}, K., {Rivoldini}, A., {Van Hoolst}, T., {et~al.} 2018, Icarus, 313, 61

\bibitem[{{Haldemann} {et~al.}(2020){Haldemann}, {Alibert}, {Mordasini}, \&
  {Benz}}]{Haldemann2020}
{Haldemann}, J., {Alibert}, Y., {Mordasini}, C., \& {Benz}, W. 2020, Astronomy
  and Astrophysics, 643, A105

\bibitem[{Haywood {et~al.}(2014)Haywood, Collier~Cameron, Queloz, Barros,
  Deleuil, Fares, Gillon, Lanza, Lovis, Moutou, Pepe, Pollacco, Santerne,
  Ségransan, \& Unruh}]{haywood14}
Haywood, R.~D., Collier~Cameron, A., Queloz, D., {et~al.} 2014, Monthly Notices
  of the Royal Astronomical Society, 443, 2517

\bibitem[{{Hippke} \& {Heller}(2019)}]{HippkeTLS2019}
{Hippke}, M. \& {Heller}, R. 2019, \aap, 623, A39

\bibitem[{{Izidoro} {et~al.}(2022){Izidoro}, {Schlichting}, {Isella},
  {Dasgupta}, {Zimmermann}, \& {Bitsch}}]{izidoro2022ApJ...939L..19I}
{Izidoro}, A., {Schlichting}, H.~E., {Isella}, A., {et~al.} 2022, \apjl, 939,
  L19

\bibitem[{{Jenkins} {et~al.}(2016){Jenkins}, {Twicken}, {McCauliff},
  {Campbell}, {Sanderfer}, {Lung}, {Mansouri-Samani}, {Girouard}, {Tenenbaum},
  {Klaus}, {Smith}, {Caldwell}, {Chacon}, {Henze}, {Heiges}, {Latham},
  {Morgan}, {Swade}, {Rinehart}, \& {Vanderspek}}]{SPOC2016}
{Jenkins}, J.~M., {Twicken}, J.~D., {McCauliff}, S., {et~al.} 2016, in Society
  of Photo-Optical Instrumentation Engineers (SPIE) Conference Series, Vol.
  9913, Software and Cyberinfrastructure for Astronomy IV, ed. G.~{Chiozzi} \&
  J.~C. {Guzman}, 99133E

\bibitem[{{Jin} {et~al.}(2014){Jin}, {Mordasini}, {Parmentier}, {van Boekel},
  {Henning}, \& {Ji}}]{jin2014ApJ...795...65J}
{Jin}, S., {Mordasini}, C., {Parmentier}, V., {et~al.} 2014, \apj, 795, 65

\bibitem[{{Kipping}(2013)}]{kipping2013}
{Kipping}, D.~M. 2013, \mnras, 435, 2152

\bibitem[{{Kochanek} {et~al.}(2017){Kochanek}, {Shappee}, {Stanek}, {Holoien},
  {Thompson}, {Prieto}, {Dong}, {Shields}, {Will}, {Britt}, {Perzanowski}, \&
  {Pojma{\'n}ski}}]{kochanek2017PASP..129j4502K}
{Kochanek}, C.~S., {Shappee}, B.~J., {Stanek}, K.~Z., {et~al.} 2017, \pasp,
  129, 104502

\bibitem[{{Kreidberg}(2015)}]{batman05}
{Kreidberg}, L. 2015, \pasp, 127, 1161

\bibitem[{{Kurucz}(1993)}]{Kurucz-93}
{Kurucz}, R.~L. 1993, {SYNTHE spectrum synthesis programs and line data}

\bibitem[{{Lavie} {et~al.}(2023){Lavie}, {Bouchy}, {Lovis}, {Zapatero Osorio},
  {Deline}, {Barros}, {Figueira}, {Sozzetti}, {Gonz{\'a}lez Hern{\'a}ndez},
  {Lillo-Box}, {Rodrigues}, {Mehner}, {Damasso}, {Adibekyan}, {Alibert},
  {Allende Prieto}, {Cristiani}, {D'Odorico}, {Di Marcantonio}, {Ehrenreich},
  {G{\'e}nova Santos}, {Lo Curto}, {Martins}, {Micela}, {Molaro}, {Nunes},
  {Palle}, {Pepe}, {Poretti}, {Rebolo}, {Santos}, {Sousa}, {Su{\'a}rez
  Mascare{\~n}o}, {Tabrenero}, \& {Udry}}]{lavie2023A&A...673A..69L}
{Lavie}, B., {Bouchy}, F., {Lovis}, C., {et~al.} 2023, \aap, 673, A69

\bibitem[{{Leleu} {et~al.}(2021{\natexlab{a}}){Leleu}, {Alibert}, {Hara},
  {Hooton}, {Wilson}, {Robutel}, {Delisle}, {Laskar}, {Hoyer}, {Lovis},
  {Bryant}, {Ducrot}, {Cabrera}, {Delrez}, {Acton}, {Adibekyan}, {Allart},
  {Allende Prieto}, {Alonso}, {Alves}, {Anderson}, {Angerhausen}, {Anglada
  Escud{\'e}}, {Asquier}, {Barrado}, {Barros}, {Baumjohann}, {Bayliss}, {Beck},
  {Beck}, {Bekkelien}, {Benz}, {Billot}, {Bonfanti}, {Bonfils}, {Bouchy},
  {Bourrier}, {Bou{\'e}}, {Brandeker}, {Broeg}, {Buder}, {Burdanov},
  {Burleigh}, {B{\'a}rczy}, {Cameron}, {Chamberlain}, {Charnoz}, {Cooke},
  {Corral Van Damme}, {Correia}, {Cristiani}, {Damasso}, {Davies}, {Deleuil},
  {Demangeon}, {Demory}, {Di Marcantonio}, {Di Persio}, {Dumusque},
  {Ehrenreich}, {Erikson}, {Figueira}, {Fortier}, {Fossati}, {Fridlund},
  {Futyan}, {Gandolfi}, {Garc{\'\i}a Mu{\~n}oz}, {Garcia}, {Gill}, {Gillen},
  {Gillon}, {Goad}, {Gonz{\'a}lez Hern{\'a}ndez}, {Guedel}, {G{\"u}nther},
  {Haldemann}, {Henderson}, {Heng}, {Hogan}, {Isaak}, {Jehin}, {Jenkins},
  {Jord{\'a}n}, {Kiss}, {Kristiansen}, {Lam}, {Lavie}, {Lecavelier des Etangs},
  {Lendl}, {Lillo-Box}, {Lo Curto}, {Magrin}, {Martins}, {Maxted}, {McCormac},
  {Mehner}, {Micela}, {Molaro}, {Moyano}, {Murray}, {Nascimbeni}, {Nunes},
  {Olofsson}, {Osborn}, {Oshagh}, {Ottensamer}, {Pagano}, {Pall{\'e}},
  {Pedersen}, {Pepe}, {Persson}, {Peter}, {Piotto}, {Polenta}, {Pollacco},
  {Poretti}, {Pozuelos}, {Queloz}, {Ragazzoni}, {Rando}, {Ratti}, {Rauer},
  {Raynard}, {Rebolo}, {Reimers}, {Ribas}, {Santos}, {Scandariato},
  {Schneider}, {Sebastian}, {Sestovic}, {Simon}, {Smith}, {Sousa}, {Sozzetti},
  {Steller}, {Su{\'a}rez Mascare{\~n}o}, {Szab{\'o}}, {S{\'e}gransan},
  {Thomas}, {Thompson}, {Tilbrook}, {Triaud}, {Turner}, {Udry}, {Van Grootel},
  {Venus}, {Verrecchia}, {Vines}, {Walton}, {West}, {Wheatley}, {Wolter}, \&
  {Zapatero Osorio}}]{Leleu2021}
{Leleu}, A., {Alibert}, Y., {Hara}, N.~C., {et~al.} 2021{\natexlab{a}},
  Astronomy and Astrophysics, 649, A26

\bibitem[{{Leleu} {et~al.}(2021{\natexlab{b}}){Leleu}, {Alibert}, {Hara},
  {Hooton}, {Wilson}, {Robutel}, {Delisle}, {Laskar}, {Hoyer}, {Lovis},
  {Bryant}, {Ducrot}, {Cabrera}, {Delrez}, {Acton}, {Adibekyan}, {Allart},
  {Allende Prieto}, {Alonso}, {Alves}, {Anderson}, {Angerhausen}, {Anglada
  Escud{\'e}}, {Asquier}, {Barrado}, {Barros}, {Baumjohann}, {Bayliss}, {Beck},
  {Beck}, {Bekkelien}, {Benz}, {Billot}, {Bonfanti}, {Bonfils}, {Bouchy},
  {Bourrier}, {Bou{\'e}}, {Brandeker}, {Broeg}, {Buder}, {Burdanov},
  {Burleigh}, {B{\'a}rczy}, {Cameron}, {Chamberlain}, {Charnoz}, {Cooke},
  {Corral Van Damme}, {Correia}, {Cristiani}, {Damasso}, {Davies}, {Deleuil},
  {Demangeon}, {Demory}, {Di Marcantonio}, {Di Persio}, {Dumusque},
  {Ehrenreich}, {Erikson}, {Figueira}, {Fortier}, {Fossati}, {Fridlund},
  {Futyan}, {Gandolfi}, {Garc{\'\i}a Mu{\~n}oz}, {Garcia}, {Gill}, {Gillen},
  {Gillon}, {Goad}, {Gonz{\'a}lez Hern{\'a}ndez}, {Guedel}, {G{\"u}nther},
  {Haldemann}, {Henderson}, {Heng}, {Hogan}, {Isaak}, {Jehin}, {Jenkins},
  {Jord{\'a}n}, {Kiss}, {Kristiansen}, {Lam}, {Lavie}, {Lecavelier des Etangs},
  {Lendl}, {Lillo-Box}, {Lo Curto}, {Magrin}, {Martins}, {Maxted}, {McCormac},
  {Mehner}, {Micela}, {Molaro}, {Moyano}, {Murray}, {Nascimbeni}, {Nunes},
  {Olofsson}, {Osborn}, {Oshagh}, {Ottensamer}, {Pagano}, {Pall{\'e}},
  {Pedersen}, {Pepe}, {Persson}, {Peter}, {Piotto}, {Polenta}, {Pollacco},
  {Poretti}, {Pozuelos}, {Queloz}, {Ragazzoni}, {Rando}, {Ratti}, {Rauer},
  {Raynard}, {Rebolo}, {Reimers}, {Ribas}, {Santos}, {Scandariato},
  {Schneider}, {Sebastian}, {Sestovic}, {Simon}, {Smith}, {Sousa}, {Sozzetti},
  {Steller}, {Su{\'a}rez Mascare{\~n}o}, {Szab{\'o}}, {S{\'e}gransan},
  {Thomas}, {Thompson}, {Tilbrook}, {Triaud}, {Turner}, {Udry}, {Van Grootel},
  {Venus}, {Verrecchia}, {Vines}, {Walton}, {West}, {Wheatley}, {Wolter}, \&
  {Zapatero Osorio}}]{2021A&A...649A..26L}
{Leleu}, A., {Alibert}, Y., {Hara}, N.~C., {et~al.} 2021{\natexlab{b}}, \aap,
  649, A26

\bibitem[{{Lillo-Box} {et~al.}(2014){Lillo-Box}, {Barrado}, \&
  {Bouy}}]{lillobox2014A&A...566A.103L}
{Lillo-Box}, J., {Barrado}, D., \& {Bouy}, H. 2014, \aap, 566, A103

\bibitem[{{Lillo-Box} {et~al.}(2020{\natexlab{a}}){Lillo-Box}, {Figueira, P.},
  {Leleu, A.}, {Acu\~na, L.}, {Faria, J. P.}, {Hara, N.}, {Santos, N. C.},
  {Correia, A. C. M.}, {Robutel, P.}, {Deleuil, M.}, {Barrado, D.}, {Sousa,
  S.}, {Bonfils, X.}, {Mousis, O.}, {Almenara, J. M.}, {Astudillo-Defru, N.},
  {Marcq, E.}, {Udry, S.}, {Lovis, C.}, \& {Pepe, F.}}]{lillobox2020}
{Lillo-Box}, J., {Figueira, P.}, {Leleu, A.}, {et~al.} 2020{\natexlab{a}},
  A\&A, 642, A121

\bibitem[{{Lillo-Box} {et~al.}(2023){Lillo-Box}, {Gandolfi}, {Armstrong},
  {Collins}, {Nielsen}, {Luque}, {Korth}, {Sousa}, {Quinn}, {Acu{\~n}a},
  {Howell}, {Morello}, {Hellier}, {Giacalone}, {Hoyer}, {Stassun}, {Palle},
  {Aguichine}, {Mousis}, {Adibekyan}, {Azevedo Silva}, {Barrado}, {Deleuil},
  {Eastman}, {Fukui}, {Hawthorn}, {Irwin}, {Jenkins}, {Latham}, {Muresan},
  {Narita}, {Persson}, {Santerne}, {Santos}, {Savel}, {Osborn}, {Teske},
  {Wheatley}, {Winn}, {Barros}, {Butler}, {Caldwell}, {Charbonneau},
  {Cloutier}, {Crane}, {Demangeon}, {D{\'\i}az}, {Dumusque}, {Esposito},
  {Falk}, {Gill}, {Hojjatpanah}, {Kreidberg}, {Mireles}, {Osborn}, {Ricker},
  {Rodriguez}, {Schwarz}, {Seager}, {Serrano Bell}, {Shectman}, {Shporer},
  {Vezie}, {Wang}, \& {Zhou}}]{2023A&A...669A.109L}
{Lillo-Box}, J., {Gandolfi}, D., {Armstrong}, D.~J., {et~al.} 2023, \aap, 669,
  A109

\bibitem[{{Lillo-Box} {et~al.}(2020{\natexlab{b}}){Lillo-Box}, {Lopez},
  {Santerne}, {Nielsen}, {Barros}, {Deleuil}, {Acu{\~n}a}, {Mousis}, {Sousa},
  {Adibekyan}, {Armstrong}, {Barrado}, {Bayliss}, {Brown}, {Demangeon},
  {Dumusque}, {Figueira}, {Hojjatpanah}, {Osborn}, {Santos}, \&
  {Udry}}]{lillobox2020A&A...640A..48L}
{Lillo-Box}, J., {Lopez}, T.~A., {Santerne}, A., {et~al.} 2020{\natexlab{b}},
  \aap, 640, A48

\bibitem[{{Lopez} \& {Fortney}(2014)}]{Lopez2014}
{Lopez}, E.~D. \& {Fortney}, J.~J. 2014, The Astrophysical Journal, 792, 1

\bibitem[{{Luger} {et~al.}(2019){Luger}, {Agol}, {Foreman-Mackey}, {Fleming},
  {Lustig-Yaeger}, \& {Deitrick}}]{starry2019}
{Luger}, R., {Agol}, E., {Foreman-Mackey}, D., {et~al.} 2019, \aj, 157, 64

\bibitem[{{Luque} \& {Pall{\'e}}(2022)}]{luque2022Sci...377.1211L}
{Luque}, R. \& {Pall{\'e}}, E. 2022, Science, 377, 1211

\bibitem[{{Luque} {et~al.}(2021){Luque}, {Serrano}, {Molaverdikhani}, {Nixon},
  {Livingston}, {Guenther}, {Pall{\'e}}, {Madhusudhan}, {Nowak}, {Korth},
  {Cochran}, {Hirano}, {Chaturvedi}, {Goffo}, {Albrecht}, {Barrag{\'a}n},
  {Brice{\~n}o}, {Cabrera}, {Charbonneau}, {Cloutier}, {Collins}, {Collins},
  {Col{\'o}n}, {Crossfield}, {Csizmadia}, {Dai}, {Deeg}, {Esposito},
  {Fridlund}, {Gandolfi}, {Georgieva}, {Glidden}, {Goeke}, {Grziwa}, {Hatzes},
  {Henze}, {Howell}, {Irwin}, {Jenkins}, {Jensen}, {K{\'a}bath}, {Kidwell},
  {Kielkopf}, {Knudstrup}, {Lam}, {Latham}, {Lissauer}, {Mann}, {Matthews},
  {Mireles}, {Narita}, {Paegert}, {Persson}, {Redfield}, {Ricker}, {Rodler},
  {Schlieder}, {Scott}, {Seager}, {{\v{S}}ubjak}, {Tan}, {Ting}, {Vanderspek},
  {Van Eylen}, {Winn}, \& {Ziegler}}]{2021A&A...645A..41L}
{Luque}, R., {Serrano}, L.~M., {Molaverdikhani}, K., {et~al.} 2021, \aap, 645,
  A41

\bibitem[{{Mandel} \& {Agol}(2002)}]{2002ApJ...580L.171M}
{Mandel}, K. \& {Agol}, E. 2002, \apjl, 580, L171

\bibitem[{{Marboeuf} {et~al.}(2014){Marboeuf}, {Thiabaud}, {Alibert}, {Cabral},
  \& {Benz}}]{Marboeuf2014}
{Marboeuf}, U., {Thiabaud}, A., {Alibert}, Y., {Cabral}, N., \& {Benz}, W.
  2014, Astronomy and Astrophysics, 570, A36

\bibitem[{{Morello} {et~al.}(2023){Morello}, {Parviainen}, {Murgas},
  {Pall{\'e}}, {Oshagh}, {Fukui}, {Hirano}, {Ishikawa}, {Mori}, {Narita},
  {Collins}, {Barkaoui}, {Lewin}, {Cadieux}, {de Leon}, {Soubkiou}, {Abreu
  Garcia}, {Crouzet}, {Esparza-Borges}, {Fern{\'a}ndez Rodr{\'\i}guez},
  {Gal{\'a}n}, {Hori}, {Ikoma}, {Isogai}, {Kagetani}, {Kawauchi}, {Kimura},
  {Kodama}, {Korth}, {Kotani}, {Krishnamurthy}, {Kurita}, {Laza-Ramos},
  {Livingston}, {Luque}, {Madrigal-Aguado}, {Nishiumi}, {Orell-Miquel},
  {Puig-Subir{\`a}}, {S{\'a}nchez-Benavente}, {Stangret}, {Tamura}, {Terada},
  {Watanabe}, {Zou}, {Benkhaldoun}, {Collins}, {Doyon}, {Garcia}, {Ghachoui},
  {Gillon}, {Jehin}, {Pozuelos}, {Schwarz}, \&
  {Timmermans}}]{2023A&A...673A..32M}
{Morello}, G., {Parviainen}, H., {Murgas}, F., {et~al.} 2023, \aap, 673, A32

\bibitem[{{Mori} {et~al.}(2022){Mori}, {Livingston}, {Leon}, {Narita},
  {Hirano}, {Fukui}, {Collins}, {Fujita}, {Hori}, {Ishikawa}, {Kawauchi},
  {Stassun}, {Watanabe}, {Giacalone}, {Gore}, {Schroeder}, {Dressing},
  {Bieryla}, {Jensen}, {Massey}, {Shporer}, {Kuzuhara}, {Charbonneau},
  {Ciardi}, {Doty}, {Esparza-Borges}, {Harakawa}, {Hodapp}, {Ikoma}, {Ikuta},
  {Isogai}, {Jenkins}, {Kagetani}, {Kimura}, {Kodama}, {Kotani},
  {Krishnamurthy}, {Kudo}, {Kurita}, {Kurokawa}, {Kusakabe}, {Latham},
  {McLean}, {Murgas}, {Nishikawa}, {Nishiumi}, {Omiya}, {Osborn}, {Palle},
  {Parviainen}, {Ricker}, {Seager}, {Serizawa}, {Teng}, {Terada}, {Twicken},
  {Ueda}, {Vanderspek}, {Vievard}, {Winn}, {Zou}, \&
  {Tamura}}]{2022AJ....163..298M}
{Mori}, M., {Livingston}, J.~H., {Leon}, J.~d., {et~al.} 2022, \aj, 163, 298

\bibitem[{{Mugrauer} \& {Michel}(2020)}]{mugrauer2020}
{Mugrauer}, M. \& {Michel}, K.-U. 2020, Astronomische Nachrichten, 341, 996

\bibitem[{{Mugrauer} \& {Michel}(2021)}]{mugrauer2021}
{Mugrauer}, M. \& {Michel}, K.-U. 2021, Astronomische Nachrichten, 342, 840

\bibitem[{{Murgas} {et~al.}(2023){Murgas}, {Castro-Gonz{\'a}lez}, {Pall{\'e}},
  {Pozuelos}, {Millholland}, {Foo}, {Korth}, {Marfil}, {Amado}, {Caballero},
  {Christiansen}, {Ciardi}, {Collins}, {Di Sora}, {Fukui}, {Gan}, {Gonzales},
  {Henning}, {Herrero}, {Isopi}, {Jenkins}, {Lillo-Box}, {Lodieu}, {Luque},
  {Mallia}, {Morello}, {Narita}, {Orell-Miquel}, {Parviainen},
  {P{\'e}rez-Torres}, {Quirrenbach}, {Reiners}, {Ribas}, {Safonov}, {Seager},
  {Schwarz}, {Schweitzer}, {Schlecker}, {Strakhov}, {Vanaverbeke}, {Watanabe},
  \& {Winn}}]{2023arXiv230409220M}
{Murgas}, F., {Castro-Gonz{\'a}lez}, A., {Pall{\'e}}, E., {et~al.} 2023, arXiv
  e-prints, arXiv:2304.09220

\bibitem[{{Nissen}(2015)}]{nissen2015A&A...579A..52N}
{Nissen}, P.~E. 2015, \aap, 579, A52

\bibitem[{{Noyes} {et~al.}(1984){Noyes}, {Hartmann}, {Baliunas}, {Duncan}, \&
  {Vaughan}}]{noyes1984ApJ...279..763N}
{Noyes}, R.~W., {Hartmann}, L.~W., {Baliunas}, S.~L., {Duncan}, D.~K., \&
  {Vaughan}, A.~H. 1984, \apj, 279, 763

\bibitem[{{Owen} \& {Campos Estrada}(2020)}]{owen2020}
{Owen}, J.~E. \& {Campos Estrada}, B. 2020, \mnras, 491, 5287

\bibitem[{{Owen} \& {Morton}(2016)}]{Owen2016}
{Owen}, J.~E. \& {Morton}, T.~D. 2016, \apjl, 819, L10

\bibitem[{{Owen} \& {Wu}(2013)}]{Owen2013ApJ...775..105O}
{Owen}, J.~E. \& {Wu}, Y. 2013, \apj, 775, 105

\bibitem[{{Owen} \& {Wu}(2017)}]{owen2017ApJ...847...29O}
{Owen}, J.~E. \& {Wu}, Y. 2017, \apj, 847, 29

\bibitem[{{Pepe} {et~al.}(2021){Pepe}, {Cristiani}, {Rebolo}, {Santos},
  {Dekker}, {Cabral}, {Di Marcantonio}, {Figueira}, {Lo Curto}, {Lovis},
  {Mayor}, {M{\'e}gevand}, {Molaro}, {Riva}, {Zapatero Osorio}, {Amate},
  {Manescau}, {Pasquini}, {Zerbi}, {Adibekyan}, {Abreu}, {Affolter}, {Alibert},
  {Aliverti}, {Allart}, {Allende Prieto}, {{\'A}lvarez}, {Alves}, {Avila},
  {Baldini}, {Bandy}, {Barros}, {Benz}, {Bianco}, {Borsa}, {Bourrier},
  {Bouchy}, {Broeg}, {Calderone}, {Cirami}, {Coelho}, {Conconi}, {Coretti},
  {Cumani}, {Cupani}, {D'Odorico}, {Damasso}, {Deiries}, {Delabre},
  {Demangeon}, {Dumusque}, {Ehrenreich}, {Faria}, {Fragoso}, {Genolet},
  {Genoni}, {G{\'e}nova Santos}, {Gonz{\'a}lez Hern{\'a}ndez}, {Hughes},
  {Iwert}, {Kerber}, {Knudstrup}, {Landoni}, {Lavie}, {Lillo-Box}, {Lizon},
  {Maire}, {Martins}, {Mehner}, {Micela}, {Modigliani}, {Monteiro}, {Monteiro},
  {Moschetti}, {Murphy}, {Nunes}, {Oggioni}, {Oliveira}, {Oshagh}, {Pall{\'e}},
  {Pariani}, {Poretti}, {Rasilla}, {Rebord{\~a}o}, {Redaelli}, {Santana
  Tschudi}, {Santin}, {Santos}, {S{\'e}gransan}, {Schmidt}, {Segovia},
  {Sosnowska}, {Sozzetti}, {Sousa}, {Span{\`o}}, {Su{\'a}rez Mascare{\~n}o},
  {Tabernero}, {Tenegi}, {Udry}, \& {Zanutta}}]{pepe21}
{Pepe}, F., {Cristiani}, S., {Rebolo}, R., {et~al.} 2021, \aap, 645, A96

\bibitem[{{Pepper} {et~al.}(2017){Pepper}, {Gillen}, {Parviainen},
  {Hillenbrand}, {Cody}, {Aigrain}, {Stauffer}, {Vrba}, {David}, {Lillo-Box},
  {Stassun}, {Conroy}, {Pope}, \& {Barrado}}]{2017AJ....153..177P}
{Pepper}, J., {Gillen}, E., {Parviainen}, H., {et~al.} 2017, \aj, 153, 177

\bibitem[{{Piaulet} {et~al.}(2023){Piaulet}, {Benneke}, {Almenara}, {Dragomir},
  {Knutson}, {Thorngren}, {Peterson}, {Crossfield}, {Kempton}, {Kubyshkina},
  {Howard}, {Angus}, {Isaacson}, {Weiss}, {Beichman}, {Fortney}, {Fossati},
  {Lammer}, {McCullough}, {Morley}, \& {Wong}}]{piaulet2023NatAs...7..206P}
{Piaulet}, C., {Benneke}, B., {Almenara}, J.~M., {et~al.} 2023, Nature
  Astronomy, 7, 206

\bibitem[{{Ricker} {et~al.}(2016){Ricker}, {Vanderspek}, {Winn}, {Seager},
  {Berta-Thompson}, {Levine}, {Villasenor}, {Latham}, {Charbonneau}, {Holman},
  {Johnson}, {Sasselov}, {Szentgyorgyi}, {Torres}, {Bakos}, {Brown},
  {Christensen-Dalsgaard}, {Kjeldsen}, {Clampin}, {Rinehart}, {Deming}, {Doty},
  {Dunham}, {Ida}, {Kawai}, {Sato}, {Jenkins}, {Lissauer}, {Jernigan},
  {Kaltenegger}, {Laughlin}, {Lin}, {McCullough}, {Narita}, {Pepper},
  {Stassun}, \& {Udry}}]{ricker2016}
{Ricker}, G.~R., {Vanderspek}, R., {Winn}, J., {et~al.} 2016, in Society of
  Photo-Optical Instrumentation Engineers (SPIE) Conference Series, Vol. 9904,
  Space Telescopes and Instrumentation 2016: Optical, Infrared, and Millimeter
  Wave, ed. H.~A. {MacEwen}, G.~G. {Fazio}, M.~{Lystrup}, N.~{Batalha},
  N.~{Siegler}, \& E.~C. {Tong}, 99042B

\bibitem[{{Robin} {et~al.}(2003){Robin}, {Reyl{\'e}}, {Derri{\`e}re}, \&
  {Picaud}}]{2003A&A...409..523R}
{Robin}, A.~C., {Reyl{\'e}}, C., {Derri{\`e}re}, S., \& {Picaud}, S. 2003,
  \aap, 409, 523

\bibitem[{{Rodrigues} {et~al.}(2017){Rodrigues}, {Bossini}, {Miglio},
  {Girardi}, {Montalb{\'a}n}, {Noels}, {Trabucchi}, {Coelho}, \&
  {Marigo}}]{PARAM3}
{Rodrigues}, T.~S., {Bossini}, D., {Miglio}, A., {et~al.} 2017, MNRAS, 467,
  1433

\bibitem[{{Rodrigues} {et~al.}(2014){Rodrigues}, {Girardi}, {Miglio},
  {Bossini}, {Bovy}, {Epstein}, {Pinsonneault}, {Stello}, {Zasowski}, {Allende
  Prieto}, {Chaplin}, {Hekker}, {Johnson}, {M{\'e}sz{\'a}ros}, {Mosser},
  {Anders}, {Basu}, {Beers}, {Chiappini}, {da Costa}, {Elsworth},
  {Garc{\'\i}a}, {Garc{\'\i}a P{\'e}rez}, {Hearty}, {Maia}, {Majewski},
  {Mathur}, {Montalb{\'a}n}, {Nidever}, {Santiago}, {Schultheis}, {Serenelli},
  \& {Shetrone}}]{PARAM2}
{Rodrigues}, T.~S., {Girardi}, L., {Miglio}, A., {et~al.} 2014, MNRAS, 445,
  2758

\bibitem[{{Rogers} {et~al.}(2023){Rogers}, {Jan{\'o} Mu{\~n}oz}, {Owen}, \&
  {Makinen}}]{Rogers2021b}
{Rogers}, J.~G., {Jan{\'o} Mu{\~n}oz}, C., {Owen}, J.~E., \& {Makinen}, T.~L.
  2023, \mnras, 519, 6028

\bibitem[{Salvatier {et~al.}(2016)Salvatier, Wiecki, \&
  Fonnesbeck}]{exoplanet:pymc3}
Salvatier, J., Wiecki, T.~V., \& Fonnesbeck, C. 2016, PeerJ Computer Science,
  2, e55

\bibitem[{{Santos} {et~al.}(2013){Santos}, {Sousa}, {Mortier}, {Neves},
  {Adibekyan}, {Tsantaki}, {Delgado Mena}, {Bonfils}, {Israelian}, {Mayor}, \&
  {Udry}}]{Santos-13}
{Santos}, N.~C., {Sousa}, S.~G., {Mortier}, A., {et~al.} 2013, \aap, 556, A150

\bibitem[{{Schlieder} {et~al.}(2021){Schlieder}, {Gonzales}, {Ciardi}, {Patel},
  {Crossfield}, {Crepp}, {Dressing}, {Barclay}, \&
  {Howard}}]{schlieder2021FrASS...8...63S}
{Schlieder}, J.~E., {Gonzales}, E.~J., {Ciardi}, D.~R., {et~al.} 2021,
  Frontiers in Astronomy and Space Sciences, 8, 63

\bibitem[{{Shappee} {et~al.}(2014){Shappee}, {Prieto}, {Grupe}, {Kochanek},
  {Stanek}, {De Rosa}, {Mathur}, {Zu}, {Peterson}, {Pogge}, {Komossa}, {Im},
  {Jencson}, {Holoien}, {Basu}, {Beacom}, {Szczygie{\l}}, {Brimacombe},
  {Adams}, {Campillay}, {Choi}, {Contreras}, {Dietrich}, {Dubberley},
  {Elphick}, {Foale}, {Giustini}, {Gonzalez}, {Hawkins}, {Howell}, {Hsiao},
  {Koss}, {Leighly}, {Morrell}, {Mudd}, {Mullins}, {Nugent}, {Parrent},
  {Phillips}, {Pojmanski}, {Rosing}, {Ross}, {Sand}, {Terndrup}, {Valenti},
  {Walker}, \& {Yoon}}]{shappee2014ApJ...788...48S}
{Shappee}, B.~J., {Prieto}, J.~L., {Grupe}, D., {et~al.} 2014, \apj, 788, 48

\bibitem[{{Siverd} {et~al.}(2012){Siverd}, {Beatty}, {Pepper}, {Eastman},
  {Collins}, {Bieryla}, {Latham}, {Buchhave}, {Jensen}, {Crepp}, {Street},
  {Stassun}, {Gaudi}, {Berlind}, {Calkins}, {DePoy}, {Esquerdo}, {Fulton},
  {F{\H{u}}r{\'e}sz}, {Geary}, {Gould}, {Hebb}, {Kielkopf}, {Marshall},
  {Pogge}, {Stanek}, {Stefanik}, {Szentgyorgyi}, {Trueblood}, {Trueblood},
  {Stutz}, \& {van Saders}}]{Siverd12}
{Siverd}, R.~J., {Beatty}, T.~G., {Pepper}, J., {et~al.} 2012, \apj, 761, 123

\bibitem[{{Skrutskie} {et~al.}(2006){Skrutskie}, {Cutri}, {Stiening},
  {Weinberg}, {Schneider}, {Carpenter}, {Beichman}, {Capps}, {Chester},
  {Elias}, {Huchra}, {Liebert}, {Lonsdale}, {Monet}, {Price}, {Seitzer},
  {Jarrett}, {Kirkpatrick}, {Gizis}, {Howard}, {Evans}, {Fowler}, {Fullmer},
  {Hurt}, {Light}, {Kopan}, {Marsh}, {McCallon}, {Tam}, {Van Dyk}, \&
  {Wheelock}}]{Skrutskie06}
{Skrutskie}, M.~F., {Cutri}, R.~M., {Stiening}, R., {et~al.} 2006, \aj, 131,
  1163

\bibitem[{{Smith} {et~al.}(2012){Smith}, {Stumpe}, {Van Cleve}, {Jenkins},
  {Barclay}, {Fanelli}, {Girouard}, {Kolodziejczak}, {McCauliff}, {Morris}, \&
  {Twicken}}]{PDCSAP2012II}
{Smith}, J.~C., {Stumpe}, M.~C., {Van Cleve}, J.~E., {et~al.} 2012, \pasp, 124,
  1000

\bibitem[{{Sneden}(1973)}]{Sneden-73}
{Sneden}, C.~A. 1973, PhD thesis, The University of Texas at Austin

\bibitem[{{Sotin} {et~al.}(2007){Sotin}, {Grasset}, \& {Mocquet}}]{Sotin2007}
{Sotin}, C., {Grasset}, O., \& {Mocquet}, A. 2007, Icarus, 191, 337

\bibitem[{{Soto} {et~al.}(2021){Soto}, {Anglada-Escud{\'e}}, {Dreizler},
  {Molaverdikhani}, {Kemmer}, {Rodr{\'\i}guez-L{\'o}pez}, {Lillo-Box},
  {Pall{\'e}}, {Espinoza}, {Caballero}, {Quirrenbach}, {Ribas}, {Reiners},
  {Narita}, {Hirano}, {Amado}, {B{\'e}jar}, {Bluhm}, {Burke}, {Caldwell},
  {Charbonneau}, {Cloutier}, {Collins}, {Cort{\'e}s-Contreras}, {Girardin},
  {Guerra}, {Harakawa}, {Hatzes}, {Irwin}, {Jenkins}, {Jensen}, {Kawauchi},
  {Kotani}, {Kudo}, {Kunimoto}, {Kuzuhara}, {Latham}, {Montes}, {Morales},
  {Mori}, {Nelson}, {Omiya}, {Pedraz}, {Passegger}, {Rackham}, {Rudat},
  {Schlieder}, {Sch{\"o}fer}, {Schweitzer}, {Selezneva}, {Stockdale}, {Tamura},
  {Trifonov}, {Vanderspek}, \& {Watanabe}}]{2021A&A...649A.144S}
{Soto}, M.~G., {Anglada-Escud{\'e}}, G., {Dreizler}, S., {et~al.} 2021, \aap,
  649, A144

\bibitem[{{Sousa}(2014)}]{Sousa-14}
{Sousa}, S.~G. 2014, in Determination of Atmospheric Parameters of B, A, F and
  G Type Stars, 297--310

\bibitem[{{Sousa} {et~al.}(2021){Sousa}, {Adibekyan}, {Delgado-Mena}, {Santos},
  {Rojas-Ayala}, {Soares}, {Legoinha}, {Ulmer-Moll}, {Camacho}, {Barros},
  {Demangeon}, {Hoyer}, {Israelian}, {Mortier}, {Tsantaki}, \&
  {Monteiro}}]{Sousa-21}
{Sousa}, S.~G., {Adibekyan}, V., {Delgado-Mena}, E., {et~al.} 2021, \aap, 656,
  A53

\bibitem[{{Sousa} {et~al.}(2015){Sousa}, {Santos}, {Adibekyan}, {Delgado-Mena},
  \& {Israelian}}]{Sousa-15}
{Sousa}, S.~G., {Santos}, N.~C., {Adibekyan}, V., {Delgado-Mena}, E., \&
  {Israelian}, G. 2015, \aap, 577, A67

\bibitem[{{Sousa} {et~al.}(2007){Sousa}, {Santos}, {Israelian}, {Mayor}, \&
  {Monteiro}}]{Sousa-07}
{Sousa}, S.~G., {Santos}, N.~C., {Israelian}, G., {Mayor}, M., \& {Monteiro},
  M.~J.~P.~F.~G. 2007, A\&A, 469, 783

\bibitem[{{Sousa} {et~al.}(2008){Sousa}, {Santos}, {Mayor}, {Udry},
  {Casagrande}, {Israelian}, {Pepe}, {Queloz}, \& {Monteiro}}]{Sousa-08}
{Sousa}, S.~G., {Santos}, N.~C., {Mayor}, M., {et~al.} 2008, \aap, 487, 373

\bibitem[{{Southworth}(2011)}]{2011MNRAS.417.2166S}
{Southworth}, J. 2011, \mnras, 417, 2166

\bibitem[{{Sozzetti} {et~al.}(2021){Sozzetti}, {Damasso}, {Bonomo}, {Alibert},
  {Sousa}, {Adibekyan}, {Zapatero Osorio}, {Gonz{\'a}lez Hern{\'a}ndez},
  {Barros}, {Lillo-Box}, {Stassun}, {Winn}, {Cristiani}, {Pepe}, {Rebolo},
  {Santos}, {Allart}, {Barclay}, {Bouchy}, {Cabral}, {Ciardi}, {Di
  Marcantonio}, {D'Odorico}, {Ehrenreich}, {Fasnaugh}, {Figueira}, {Haldemann},
  {Jenkins}, {Latham}, {Lavie}, {Lo Curto}, {Lovis}, {Martins}, {M{\'e}gevand},
  {Mehner}, {Micela}, {Molaro}, {Nunes}, {Oshagh}, {Otegi}, {Pall{\'e}},
  {Poretti}, {Ricker}, {Rodriguez}, {Seager}, {Su{\'a}rez Mascare{\~n}o},
  {Twicken}, \& {Udry}}]{sozzetti2021}
{Sozzetti}, A., {Damasso}, M., {Bonomo}, A.~S., {et~al.} 2021, \aap, 648, A75

\bibitem[{{Sozzetti} {et~al.}(2007){Sozzetti}, {Torres}, {Charbonneau},
  {Latham}, {Holman}, {Winn}, {Laird}, \&
  {O'Donovan}}]{sozzetti2007ApJ...664.1190S}
{Sozzetti}, A., {Torres}, G., {Charbonneau}, D., {et~al.} 2007, \apj, 664, 1190

\bibitem[{{Stef{\'a}nsson} {et~al.}(2020){Stef{\'a}nsson}, {Kopparapu}, {Lin},
  {Mahadevan}, {Ca{\~n}as}, {Kanodia}, {Ninan}, {Cochran}, {Endl}, {Hebb},
  {Wisniewski}, {Gupta}, {Everett}, {Bender}, {Diddams}, {Ford}, {Fredrick},
  {Halverson}, {Hearty}, {Levi}, {Maney}, {Metcalf}, {Monson}, {Ramsey},
  {Robertson}, {Roy}, {Schwab}, {Terrien}, \&
  {Wright}}]{stefansson2020AJ....160..259S}
{Stef{\'a}nsson}, G., {Kopparapu}, R., {Lin}, A., {et~al.} 2020, \aj, 160, 259

\bibitem[{{Stumpe} {et~al.}(2014){Stumpe}, {Smith}, {Catanzarite}, {Van Cleve},
  {Jenkins}, {Twicken}, \& {Girouard}}]{PDCSAP2014}
{Stumpe}, M.~C., {Smith}, J.~C., {Catanzarite}, J.~H., {et~al.} 2014, \pasp,
  126, 100

\bibitem[{{Stumpe} {et~al.}(2012){Stumpe}, {Smith}, {Van Cleve}, {Twicken},
  {Barclay}, {Fanelli}, {Girouard}, {Jenkins}, {Kolodziejczak}, {McCauliff}, \&
  {Morris}}]{PDCSAP2012}
{Stumpe}, M.~C., {Smith}, J.~C., {Van Cleve}, J.~E., {et~al.} 2012, \pasp, 124,
  985

\bibitem[{{Teyssandier} {et~al.}(2015){Teyssandier}, {Owen}, {Adams}, \&
  {Quillen}}]{Teyssandier2015}
{Teyssandier}, J., {Owen}, J.~E., {Adams}, F.~C., \& {Quillen}, A.~C. 2015,
  \mnras, 452, 1743

\bibitem[{{Thiabaud} {et~al.}(2014){Thiabaud}, {Marboeuf}, {Alibert}, {Cabral},
  {Leya}, \& {Mezger}}]{Thiabaud2014}
{Thiabaud}, A., {Marboeuf}, U., {Alibert}, Y., {et~al.} 2014, Astronomy and
  Astrophysics, 562, A27

\bibitem[{{Tokovinin}(2018)}]{tokovinin2018PASP..130c5002T}
{Tokovinin}, A. 2018, \pasp, 130, 035002

\bibitem[{{Van Eylen} {et~al.}(2021{\natexlab{a}}){Van Eylen},
  {Astudillo-Defru}, {Bonfils}, {Livingston}, {Hirano}, {Luque}, {Lam},
  {Justesen}, {Winn}, {Gandolfi}, {Nowak}, {Palle}, {Albrecht}, {Dai}, {Campos
  Estrada}, {Owen}, {Foreman-Mackey}, {Fridlund}, {Korth}, {Mathur},
  {Forveille}, {Mikal-Evans}, {Osborne}, {Ho}, {Almenara}, {Artigau},
  {Barrag{\'a}n}, {Barros}, {Bouchy}, {Cabrera}, {Caldwell}, {Charbonneau},
  {Chaturvedi}, {Cochran}, {Csizmadia}, {Damasso}, {Delfosse}, {De Medeiros},
  {D{\'\i}az}, {Doyon}, {Esposito}, {F{\H{u}}r{\'e}sz}, {Figueira},
  {Georgieva}, {Goffo}, {Grziwa}, {Guenther}, {Hatzes}, {Jenkins}, {Kabath},
  {Knudstrup}, {Latham}, {Lavie}, {Lovis}, {Mennickent}, {Mullally}, {Murgas},
  {Narita}, {Pepe}, {Persson}, {Redfield}, {Ricker}, {Santos}, {Seager},
  {Serrano}, {Smith}, {Su{\'a}rez Mascare{\~n}o}, {Subjak}, {Twicken}, {Udry},
  {Vanderspek}, \& {Zapatero Osorio}}]{vaneylen2021MNRAS.507.2154V}
{Van Eylen}, V., {Astudillo-Defru}, N., {Bonfils}, X., {et~al.}
  2021{\natexlab{a}}, \mnras, 507, 2154

\bibitem[{{Van Eylen} {et~al.}(2021{\natexlab{b}}){Van Eylen},
  {Astudillo-Defru}, {Bonfils}, {Livingston}, {Hirano}, {Luque}, {Lam},
  {Justesen}, {Winn}, {Gandolfi}, {Nowak}, {Palle}, {Albrecht}, {Dai}, {Campos
  Estrada}, {Owen}, {Foreman-Mackey}, {Fridlund}, {Korth}, {Mathur},
  {Forveille}, {Mikal-Evans}, {Osborne}, {Ho}, {Almenara}, {Artigau},
  {Barrag{\'a}n}, {Barros}, {Bouchy}, {Cabrera}, {Caldwell}, {Charbonneau},
  {Chaturvedi}, {Cochran}, {Csizmadia}, {Damasso}, {Delfosse}, {De Medeiros},
  {D{\'\i}az}, {Doyon}, {Esposito}, {F{\H{u}}r{\'e}sz}, {Figueira},
  {Georgieva}, {Goffo}, {Grziwa}, {Guenther}, {Hatzes}, {Jenkins}, {Kabath},
  {Knudstrup}, {Latham}, {Lavie}, {Lovis}, {Mennickent}, {Mullally}, {Murgas},
  {Narita}, {Pepe}, {Persson}, {Redfield}, {Ricker}, {Santos}, {Seager},
  {Serrano}, {Smith}, {Su{\'a}rez Mascare{\~n}o}, {Subjak}, {Twicken}, {Udry},
  {Vanderspek}, \& {Zapatero Osorio}}]{VanEylen2021}
{Van Eylen}, V., {Astudillo-Defru}, N., {Bonfils}, X., {et~al.}
  2021{\natexlab{b}}, \mnras, 507, 2154

\bibitem[{{Wizinowich} {et~al.}(2000){Wizinowich}, {Acton}, {Shelton},
  {Stomski}, {Gathright}, {Ho}, {Lupton}, {Tsubota}, {Lai}, {Max}, {Brase},
  {An}, {Avicola}, {Olivier}, {Gavel}, {Macintosh}, {Ghez}, \&
  {Larkin}}]{wizinowich2000}
{Wizinowich}, P., {Acton}, D.~S., {Shelton}, C., {et~al.} 2000, \pasp, 112, 315

\bibitem[{{Zacharias} {et~al.}(2012){Zacharias}, {Finch}, {Girard}, {Henden},
  {Bartlett}, {Monet}, \& {Zacharias}}]{2012yCat.1322....0Z}
{Zacharias}, N., {Finch}, C.~T., {Girard}, T.~M., {et~al.} 2012, VizieR Online
  Data Catalog, I/322A

\bibitem[{{Zechmeister} {et~al.}(2019){Zechmeister}, {Dreizler}, {Ribas},
  {Reiners}, {Caballero}, {Bauer}, {B{\'e}jar}, {Gonz{\'a}lez-Cuesta},
  {Herrero}, {Lalitha}, {L{\'o}pez-Gonz{\'a}lez}, {Luque}, {Morales},
  {Pall{\'e}}, {Rodr{\'\i}guez}, {Rodr{\'\i}guez L{\'o}pez}, {Tal-Or},
  {Anglada-Escud{\'e}}, {Quirrenbach}, {Amado}, {Abril}, {Aceituno},
  {Aceituno}, {Alonso-Floriano}, {Ammler-von Eiff}, {Antona Jim{\'e}nez},
  {Anwand-Heerwart}, {Arroyo-Torres}, {Azzaro}, {Baroch}, {Barrado},
  {Becerril}, {Ben{\'\i}tez}, {Berdi{\~n}as}, {Bergond}, {Bluhm},
  {Brinkm{\"o}ller}, {del Burgo}, {Calvo Ortega}, {Cano}, {Cardona
  Guill{\'e}n}, {Carro}, {C{\'a}rdenas V{\'a}zquez}, {Casal},
  {Casasayas-Barris}, {Casanova}, {Chaturvedi}, {Cifuentes}, {Claret},
  {Colom{\'e}}, {Cort{\'e}s-Contreras}, {Czesla}, {D{\'\i}ez-Alonso}, {Dorda},
  {Fern{\'a}ndez}, {Fern{\'a}ndez-Mart{\'\i}n}, {Fuhrmeister}, {Fukui},
  {Galad{\'\i}-Enr{\'\i}quez}, {Gallardo Cava}, {Garcia de la Fuente},
  {Garcia-Piquer}, {Garc{\'\i}a Vargas}, {Gesa}, {G{\'o}ngora Rueda},
  {Gonz{\'a}lez-{\'A}lvarez}, {Gonz{\'a}lez Hern{\'a}ndez},
  {Gonz{\'a}lez-Peinado}, {Gr{\"o}zinger}, {Gu{\`a}rdia}, {Guijarro}, {de
  Guindos}, {Hatzes}, {Hauschildt}, {Hedrosa}, {Helmling}, {Henning},
  {Hermelo}, {Hern{\'a}ndez Arabi}, {Hern{\'a}ndez Casta{\~n}o}, {Hern{\'a}ndez
  Otero}, {Hintz}, {Huke}, {Huber}, {Jeffers}, {Johnson}, {de Juan},
  {Kaminski}, {Kemmer}, {Kim}, {Klahr}, {Klein}, {Kl{\"u}ter}, {Klutsch},
  {Kossakowski}, {K{\"u}rster}, {Labarga}, {Lafarga}, {Llamas}, {Lamp{\'o}n},
  {Lara}, {Launhardt}, {L{\'a}zaro}, {Lodieu}, {L{\'o}pez del Fresno},
  {L{\'o}pez-Puertas}, {L{\'o}pez Salas}, {L{\'o}pez-Santiago}, {Mag{\'a}n
  Madinabeitia}, {Mall}, {Mancini}, {Mandel}, {Marfil}, {Mar{\'\i}n Molina},
  {Maroto Fern{\'a}ndez}, {Mart{\'\i}n}, {Mart{\'\i}n-Fern{\'a}ndez},
  {Mart{\'\i}n-Ruiz}, {Marvin}, {Mirabet}, {Monta{\~n}{\'e}s-Rodr{\'\i}guez},
  {Montes}, {Moreno-Raya}, {Nagel}, {Naranjo}, {Narita}, {Nortmann}, {Nowak},
  {Ofir}, {Oshagh}, {Panduro}, {Parviainen}, {Pascual}, {Passegger}, {Pavlov},
  {Pedraz}, {P{\'e}rez-Calpena}, {P{\'e}rez Medialdea}, {Perger}, {Perryman},
  {Rabaza}, {Ram{\'o}n Ballesta}, {Rebolo}, {Redondo}, {Reffert}, {Reinhardt},
  {Rhode}, {Rix}, {Rodler}, {Rodr{\'\i}guez Trinidad}, {Rosich}, {Sadegi},
  {S{\'a}nchez-Blanco}, {S{\'a}nchez Carrasco}, {S{\'a}nchez-L{\'o}pez},
  {Sanz-Forcada}, {Sarkis}, {Sarmiento}, {Sch{\"a}fer}, {Schmitt},
  {Sch{\"o}fer}, {Schweitzer}, {Seifert}, {Shulyak}, {Solano}, {Sota}, {Stahl},
  {Stock}, {Strachan}, {Stuber}, {St{\"u}rmer}, {Su{\'a}rez}, {Tabernero},
  {Tala Pinto}, {Trifonov}, {Veredas}, {Vico Linares}, {Vilardell}, {Wagner},
  {Wolthoff}, {Xu}, {Yan}, \& {Zapatero Osorio}}]{zech2019}
{Zechmeister}, M., {Dreizler}, S., {Ribas}, I., {et~al.} 2019, \aap, 627, A49

\bibitem[{{Zechmeister} {et~al.}(2009){Zechmeister}, {K\"urster}, \&
  {Endl}}]{zech2009}
{Zechmeister}, M., {K\"urster}, M., \& {Endl}, M. 2009, A\&A, 505, 859

\bibitem[{{Zeng} {et~al.}(2019){Zeng}, {Jacobsen}, {Sasselov}, {Petaev},
  {Vanderburg}, {Lopez-Morales}, {Perez-Mercader}, {Mattsson}, {Li}, {Heising},
  {Bonomo}, {Damasso}, {Berger}, {Cao}, {Levi}, \&
  {Wordsworth}}]{zeng2019PNAS..116.9723Z}
{Zeng}, L., {Jacobsen}, S.~B., {Sasselov}, D.~D., {et~al.} 2019, Proceedings of
  the National Academy of Science, 116, 9723

\bibitem[{{Ziegler} {et~al.}(2020){Ziegler}, {Tokovinin}, {Brice{\~n}o},
  {Mang}, {Law}, \& {Mann}}]{ziegler2020AJ....159...19Z}
{Ziegler}, C., {Tokovinin}, A., {Brice{\~n}o}, C., {et~al.} 2020, \aj, 159, 19

\end{thebibliography}

\begin{appendix}

\section{Dataset}

\begin{small}
\longtab[1]{
\begin{longtable}{ccccccccc}
\caption{ESPRESSO RVs and activity diagnostics.}          
\label{table:dataespresso} \\     
\hline\hline      
\noalign{\smallskip}
Time & RV & $\sigma_{\rm RV}$ & CCF FWHM & $\sigma_{\rm FWHM}$ & CCF BIS & $\sigma_{\rm BIS}$ & $\log R^{\prime}_{\rm HK}$ & $\sigma_{\log R^{\prime}_{\rm HK}}$ \\    
(BJD$-2\,450\,000$) & \ms & \ms & (\ms) & (\ms) & (\ms) & (\ms) & (dex) & (dex)\\ 
\noalign{\smallskip}
\hline\hline
\endfirsthead
\caption{Continued.}\\
\hline\hline      
\noalign{\smallskip}
Time & RV & $\sigma_{\rm RV}$ & CCF FWHM & $\sigma_{\rm FWHM}$ & CCF BIS & $\sigma_{\rm BIS}$  & $\log R^{\prime}_{\rm HK}$ & $\sigma_{\log R^{\prime}_{\rm HK}}$\\    
(BJD$-2\,450\,000$) & \ms & \ms & (\ms) & (\ms) & (\ms) & (\ms) & (dex) & (dex)\\ 
\noalign{\smallskip}
\hline\hline
\endhead
\hline
\endfoot
\hline  
\endlastfoot
\noalign{\smallskip}
8765.880033 & 81722.91 & 0.61 & 6668.53 & 1.22 &  -68.02 & 1.22 & -5.983 &  0.014 \\
8792.789313 & 81730.59 & 0.27 & 6672.29 & 0.53 &  -67.64 & 0.53 & -5.178 &  0.001 \\
8799.679952 & 81734.23 & 0.54 & 6681.47 & 1.08 &  -60.22 & 1.08 & -5.502 &  0.004 \\
8803.799572 & 81726.85 & 0.26 & 6675.84 & 0.53 &  -67.25 & 0.53 & -5.192 &  0.001 \\
8806.664396 & 81732.28 & 0.34 & 6675.23 & 0.67 &  -68.05 & 0.67 & -5.263 &  0.001 \\
8814.627483 & 81731.51 & 0.29 & 6671.59 & 0.59 &  -66.70 & 0.59 & -5.250 &  0.001 \\
8823.809857 & 81734.16 & 0.29 & 6673.61 & 0.59 &  -64.45 & 0.59 & -5.224 &  0.001 \\
8850.752661 & 81730.02 & 0.33 & 6668.18 & 0.65 &  -66.91 & 0.65 & -5.178 &  0.001 \\
8853.763918 & 81727.91 & 0.30 & 6671.10 & 0.60 &  -70.75 & 0.60 & -5.247 &  0.001 \\
8858.667699 & 81726.97 & 0.27 & 6672.31 & 0.55 &  -66.85 & 0.55 & -5.245 &  0.001 \\
8879.662595 & 81729.70 & 0.26 & 6667.42 & 0.52 &  -69.56 & 0.52 & -5.214 &  0.001 \\
8910.636691 & 81726.05 & 0.27 & 6666.97 & 0.53 &  -69.43 & 0.53 & -5.146 &  0.001 \\
8924.501697\tablefootmark{a} & 81723.16 & 0.32 & 6667.98 & 0.63 &  -66.17 & 0.63 & -5.138 &  0.001 \\
9210.702206 & 81728.43 & 0.27 & 6674.93 & 0.55 &  -67.75 & 0.55 & -5.147 &  0.001 \\
9271.610597 & 81725.24 & 0.31 & 6679.71 & 0.62 &  -69.80 & 0.62 & -5.286 &  0.001 \\
9292.636622 & 81726.03 & 0.32 & 6678.49 & 0.63 &  -67.02 & 0.63 & -5.339 &  0.001 \\
9297.543406 & 81723.08 & 0.25 & 6674.31 & 0.50 &  -65.01 & 0.50 & -5.222 &  0.000 \\
9308.586832 & 81728.94 & 0.38 & 6676.01 & 0.77 &  -66.62 & 0.77 & -5.384 &  0.002 \\
9311.524799 & 81727.62 & 0.37 & 6680.99 & 0.74 &  -66.85 & 0.74 & -5.288 &  0.001 \\
9314.503978 & 81731.62 & 0.41 & 6678.74 & 0.81 &  -65.49 & 0.81 & -5.342 &  0.002 \\
9316.538824 & 81733.79 & 0.49 & 6673.21 & 0.99 &  -65.67 & 0.99 & -5.456 &  0.003 \\
9319.526315 & 81734.20 & 0.30 & 6674.49 & 0.60 &  -65.43 & 0.60 & -5.250 &  0.001 \\
9322.537378 & 81724.98 & 0.37 & 6672.55 & 0.73 &  -68.36 & 0.73 & -5.306 &  0.001 \\
9323.462970 & 81723.33 & 0.39 & 6676.22 & 0.79 &  -65.95 & 0.79 & -5.342 &  0.001 \\
9325.500282 & 81727.66 & 0.45 & 6670.68 & 0.91 &  -66.57 & 0.91 & -5.339 &  0.002 \\
9333.486134 & 81736.81 & 0.43 & 6682.57 & 0.85 &  -67.13 & 0.85 & -5.312 &  0.002 \\
9459.864986 & 81729.04 & 0.33 & 6695.25 & 0.67 &  -63.62 & 0.67 & -5.200 &  0.001 \\
9461.893582 & 81733.35 & 0.30 & 6691.33 & 0.60 &  -59.97 & 0.60 & -5.143 &  0.001 \\
9464.820578 & 81732.78 & 0.42 & 6685.37 & 0.84 &  -60.58 & 0.84 & -5.300 &  0.002 \\
9466.877164 & 81732.89 & 0.26 & 6679.74 & 0.52 &  -61.62 & 0.52 & -5.214 &  0.001 \\
9471.873632 & 81727.38 & 0.31 & 6676.67 & 0.63 &  -67.83 & 0.63 & -5.242 &  0.001 \\
9491.819141 & 81730.93 & 0.33 & 6679.05 & 0.65 &  -68.48 & 0.65 & -5.193 &  0.001 \\
9493.867756 & 81736.29 & 0.50 & 6684.76 & 1.00 &  -67.38 & 1.00 & -5.372 &  0.002 \\
9496.759000 & 81732.75 & 0.54 & 6689.72 & 1.09 &  -62.76 & 1.09 & -5.431 &  0.003 \\
9509.702920 & 81726.17 & 0.36 & 6680.48 & 0.72 &  -67.23 & 0.72 & -5.180 &  0.001 \\
9514.757558 & 81725.60 & 0.40 & 6670.68 & 0.81 &  -65.90 & 0.81 & -5.222 &  0.001 \\
9516.846022 & 81725.83 & 0.26 & 6674.79 & 0.52 &  -68.41 & 0.52 & -5.177 &  0.001 \\
9522.788552 & 81726.88 & 0.23 & 6677.84 & 0.46 &  -66.42 & 0.46 & -5.163 &  0.000 \\
9529.766884 & 81723.33 & 0.29 & 6676.66 & 0.58 &  -66.40 & 0.58 & -5.198 &  0.001 \\
9531.783334 & 81733.24 & 0.31 & 6676.43 & 0.62 &  -67.01 & 0.62 & -5.214 &  0.001 \\
9533.799604 & 81730.25 & 0.22 & 6681.51 & 0.45 &  -66.46 & 0.45 & -5.114 &  0.000 \\
9536.660100 & 81729.13 & 0.35 & 6682.03 & 0.70 &  -64.53 & 0.70 & -5.210 &  0.001 \\
9538.670826 & 81730.69 & 0.55 & 6684.61 & 1.10 &  -63.55 & 1.10 & -5.219 &  0.002 \\
9563.655314 & 81733.25 & 0.27 & 6673.60 & 0.55 &  -64.86 & 0.55 & -5.193 &  0.001 \\
9575.673661 & 81729.39 & 0.34 & 6675.02 & 0.67 &  -65.30 & 0.67 & -5.130 &  0.001 \\
9577.726829 & 81734.23 & 0.27 & 6673.65 & 0.54 &  -66.10 & 0.54 & -5.198 &  0.001 \\
9579.606638 & 81723.26 & 0.23 & 6676.50 & 0.45 &  -66.75 & 0.45 & -5.203 &  0.000 \\
9580.636695 & 81724.01 & 0.42 & 6672.95 & 0.83 &  -66.96 & 0.83 & -5.293 &  0.001 \\
9582.596259 & 81724.62 & 0.29 & 6673.58 & 0.57 &  -64.55 & 0.57 & -5.164 &  0.001 \\
9584.574904 & 81728.62 & 0.26 & 6671.79 & 0.53 &  -66.65 & 0.53 & -5.187 &  0.001 \\
9586.545678 & 81724.82 & 0.27 & 6674.32 & 0.54 &  -65.77 & 0.54 & -5.168 &  0.001 \\
9588.550479 & 81730.38 & 0.35 & 6672.88 & 0.70 &  -67.02 & 0.70 & -5.163 &  0.001 \\
9590.692903 & 81727.38 & 0.35 & 6670.57 & 0.70 &  -67.06 & 0.70 & -5.159 &  0.001 \\
9592.642227 & 81725.67 & 0.28 & 6671.58 & 0.56 &  -67.36 & 0.56 & -5.219 &  0.001 \\
9595.600121 & 81728.65 & 0.23 & 6672.27 & 0.47 &  -69.01 & 0.47 & -5.217 &  0.000 \\
9597.551898 & 81725.48 & 0.29 & 6669.92 & 0.57 &  -67.87 & 0.57 & -5.245 &  0.001 \\
9600.662966 & 81727.07 & 0.33 & 6670.65 & 0.66 &  -69.33 & 0.66 & -5.285 &  0.001 \\
9603.598412 & 81730.60 & 0.49 & 6669.48 & 0.98 &  -67.40 & 0.98 & -5.342 &  0.002 \\
9606.624832 & 81727.23 & 0.31 & 6675.20 & 0.63 &  -67.78 & 0.63 & -5.247 &  0.001 \\
9607.625044 & 81724.58 & 0.28 & 6674.22 & 0.56 &  -68.30 & 0.56 & -5.196 &  0.001 \\
9611.565663 & 81725.85 & 0.25 & 6675.23 & 0.49 &  -64.67 & 0.49 & -5.134 &  0.000 \\
9612.549153 & 81728.05 & 0.37 & 6679.25 & 0.73 &  -66.70 & 0.73 & -5.124 &  0.001 \\
9613.652209 & 81731.31 & 0.34 & 6676.01 & 0.67 &  -66.49 & 0.67 & -5.174 &  0.001 \\
9614.596492 & 81729.53 & 0.33 & 6676.97 & 0.67 &  -65.57 & 0.67 & -5.214 &  0.001 \\
9615.596521 & 81732.92 & 0.25 & 6677.61 & 0.50 &  -64.44 & 0.50 & -5.191 &  0.000 \\
9616.547147 & 81734.14 & 0.29 & 6673.56 & 0.57 &  -65.61 & 0.57 & -5.207 &  0.001 \\
9617.624635 & 81729.65 & 0.27 & 6672.39 & 0.54 &  -65.12 & 0.54 & -5.206 &  0.001 \\
9618.582003 & 81725.46 & 0.30 & 6676.58 & 0.60 &  -65.28 & 0.60 & -5.237 &  0.001 \\
9619.547977 & 81727.04 & 0.36 & 6674.64 & 0.71 &  -64.90 & 0.71 & -5.261 &  0.001 \\
9620.603519 & 81728.22 & 0.29 & 6668.16 & 0.57 &  -66.24 & 0.57 & -5.135 &  0.001 \\
9622.700868 & 81724.29 & 0.64 & 6679.09 & 1.28 &  -70.72 & 1.28 & -5.137 &  0.002 \\
9623.540034 & 81727.00 & 0.28 & 6671.53 & 0.56 &  -67.64 & 0.56 & -5.176 &  0.001 \\
9624.581316 & 81724.43 & 0.24 & 6671.47 & 0.49 &  -67.22 & 0.49 & -5.191 &  0.000 \\
9625.569889 & 81721.72 & 0.24 & 6671.20 & 0.48 &  -68.67 & 0.48 & -5.161 &  0.000 \\
9626.613695 & 81725.81 & 0.38 & 6668.95 & 0.76 &  -68.68 & 0.76 & -5.192 &  0.001 \\
9627.536012 & 81729.76 & 0.33 & 6670.44 & 0.65 &  -67.82 & 0.65 & -5.172 &  0.001 \\
9635.549347 & 81727.38 & 0.27 & 6675.11 & 0.54 &  -65.97 & 0.54 & -5.186 &  0.001 \\
9648.520045 & 81730.23 & 0.43 & 6671.02 & 0.86 &  -68.06 & 0.86 & -5.425 &  0.002 \\
9650.528840 & 81722.07 & 0.40 & 6672.41 & 0.80 &  -66.46 & 0.80 & -5.366 &  0.002 \\
9652.549356 & 81728.62 & 0.33 & 6674.25 & 0.67 &  -67.48 & 0.67 & -5.324 &  0.001 \\
9658.552623 & 81730.69 & 0.42 & 6668.66 & 0.83 &  -65.78 & 0.83 & -5.547 &  0.003 \\
9660.588322 & 81729.33 & 0.34 & 6667.72 & 0.68 &  -66.48 & 0.68 & -5.172 &  0.001 \\
9662.577262 & 81727.48 & 0.30 & 6665.08 & 0.60 &  -67.01 & 0.60 & -5.237 &  0.001 \\
\hline                                   
\end{longtable}
\tablefoot{
    \tablefoottext{a}{This represents the last epoch of the data sub-sample labelled as ESP19. The sub-sample ESP21 starts from the following epoch. }}
}
\end{small}

\newpage

\section{Additional plots} \label{app:plots}

\begin{figure}
    \centering
    \includegraphics[width=\linewidth]{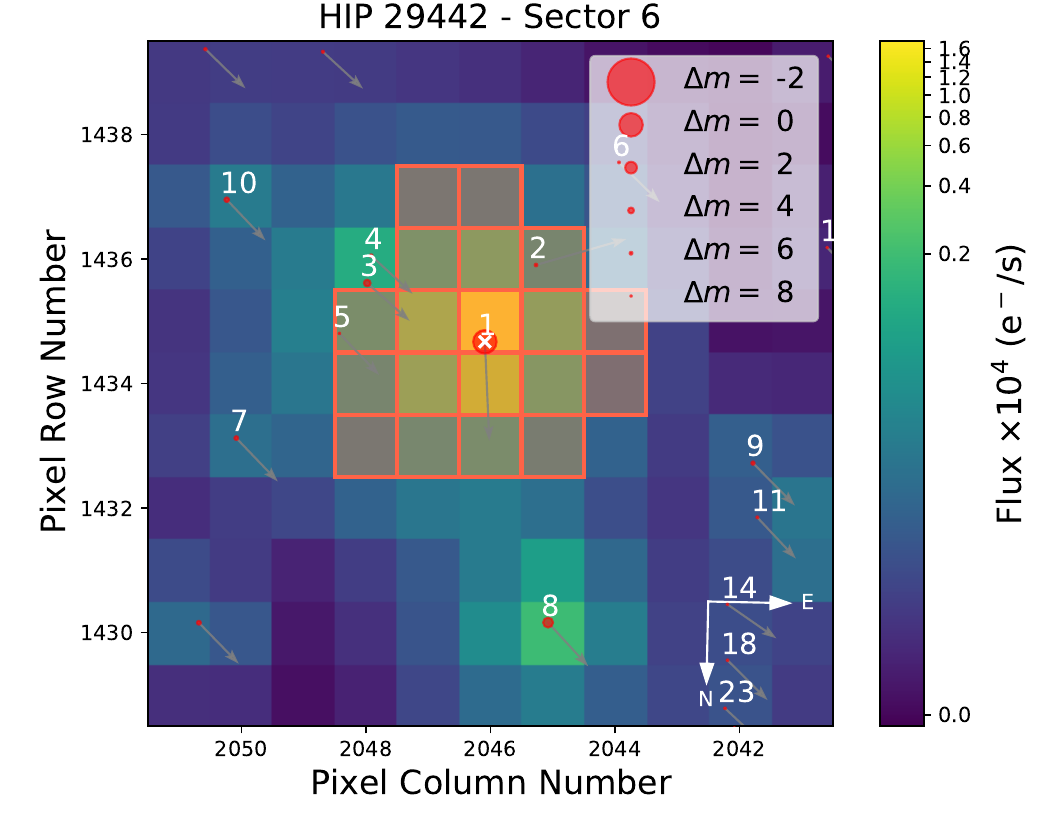}
    \includegraphics[width=\linewidth]{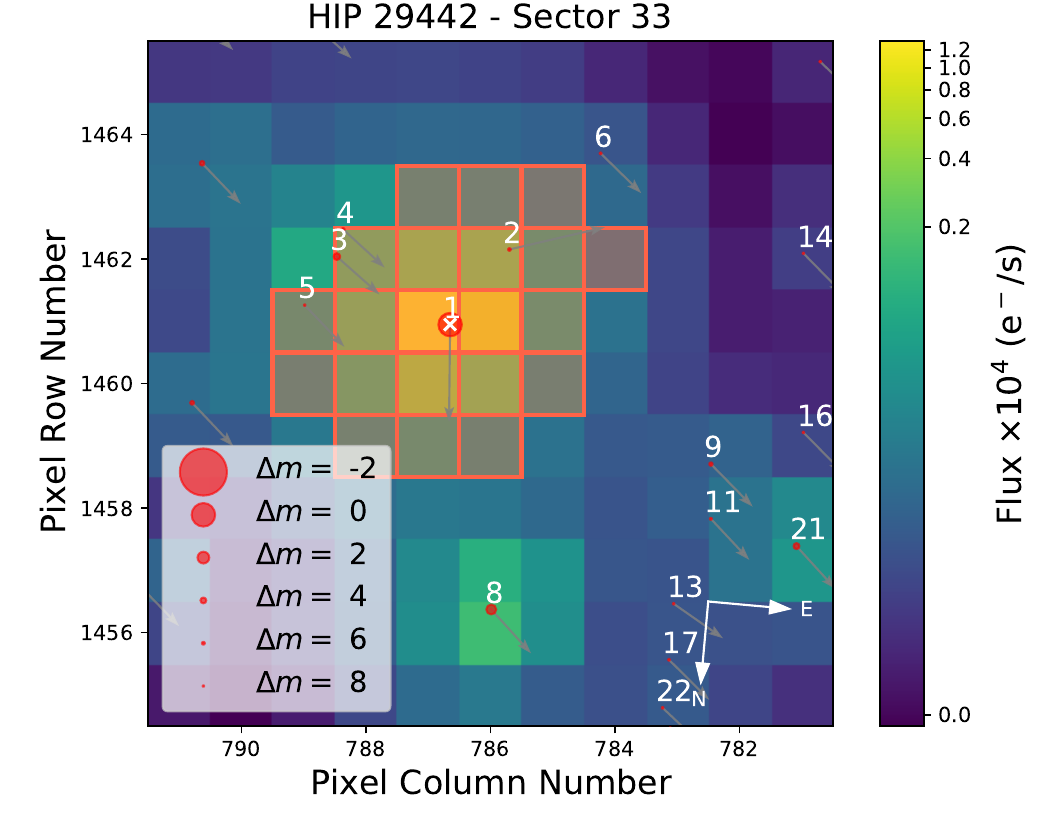}
    \caption{TESS Target Pixel Files for sectors 6 and 33 obtained with \textsc{tpfplotter} (\citealt{tpfplotter2020}; the code is publicly available at \url{www.github.com/jlillo/tpfplotter}). Orange squares identify the aperture masks used to extract the light curve. Sources cross-matched with the Gaia DR3 catalog are indicated by red dots, whose size is scaled with their relative magnitude compared to that of HIP\,29442 (TOI-469). Proper motions are indicated by arrows. The pixel scale is 21$\arcsec$ pixel$^{-1}$.}
    \label{fig:tpfplotter}
\end{figure}

\begin{figure}
    \centering
    \includegraphics[width=0.4\textwidth]{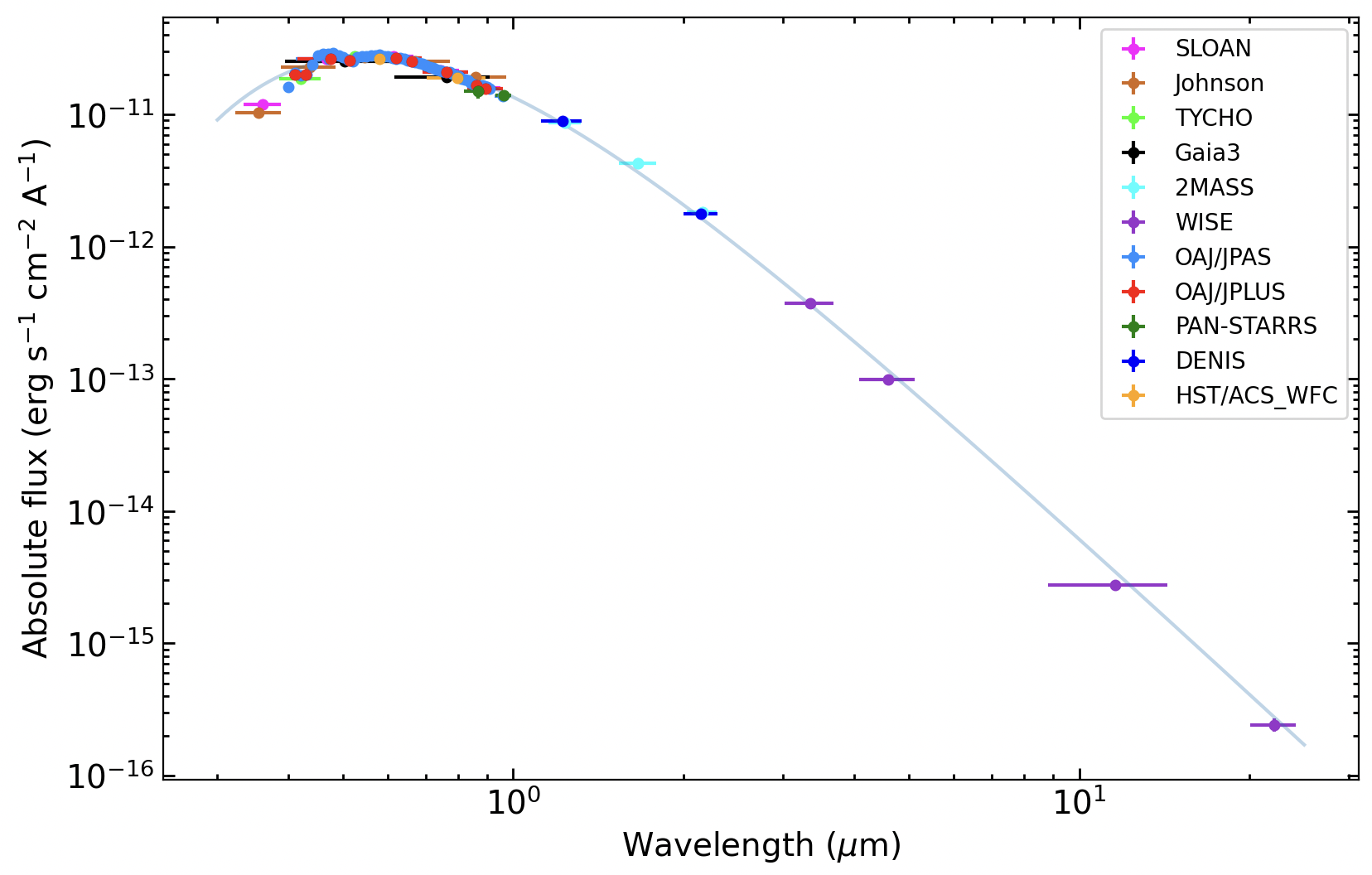}
    \caption{Spectral energy distribution in the range $\sim$0.3-25 $\mu$m, used for deriving the bolometric luminosity and the stellar radius through the Stefan-Boltzmann equation. The adopted stellar distance is the trigonometric parallax from the Gaia EDR3.}
    \label{fig:sed}
\end{figure}

\begin{figure}
    \centering
    \includegraphics[width=0.5\textwidth]{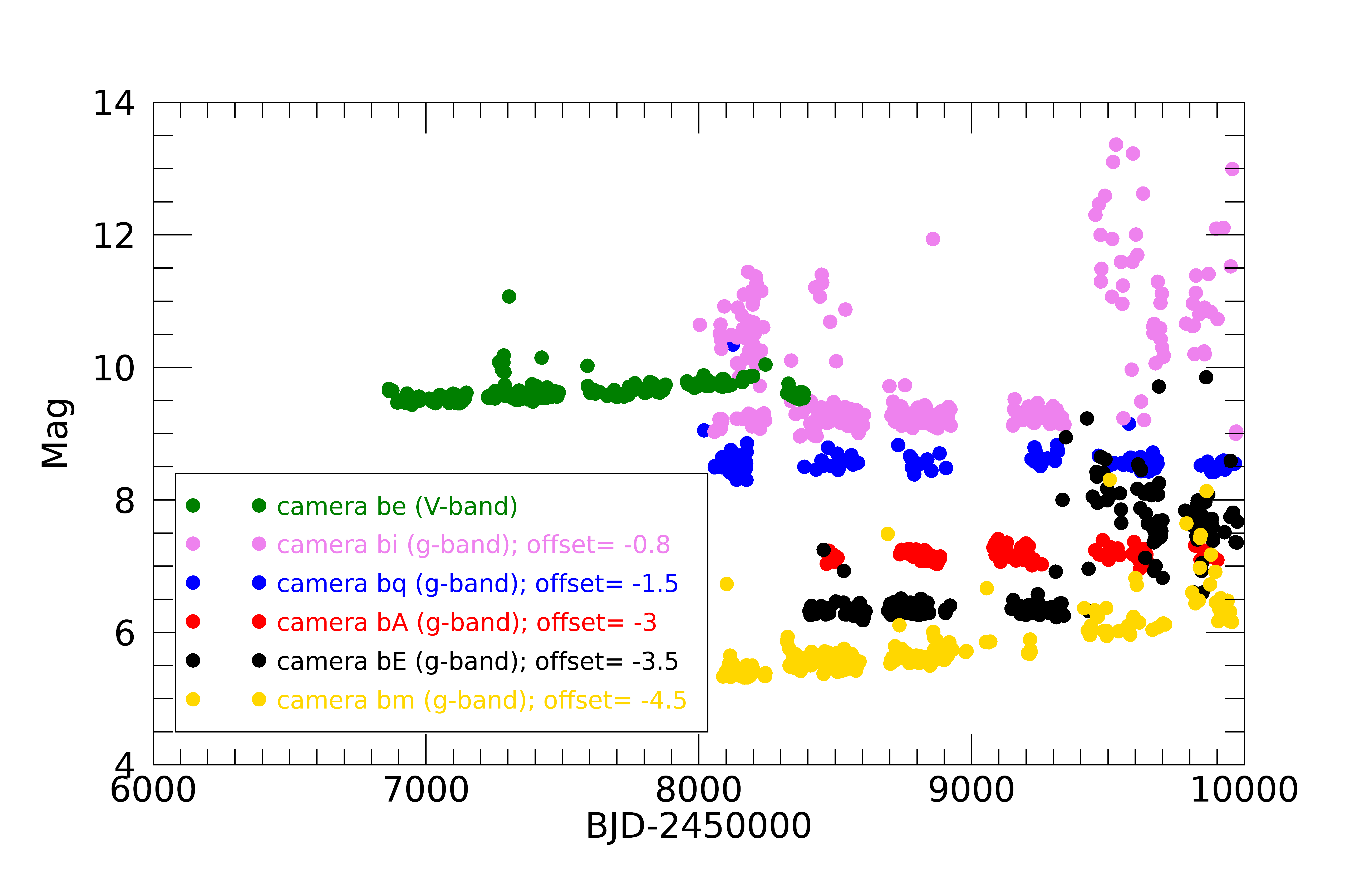}
    \caption{ASAS-SN photometry of HIP\,29442. For a better visualisation, offsets have been applied to the original magnitudes, as indicated in the legend.}
    \label{fig:asas}
\end{figure}

\begin{figure}
    \centering
    \includegraphics[width=0.4\textwidth]{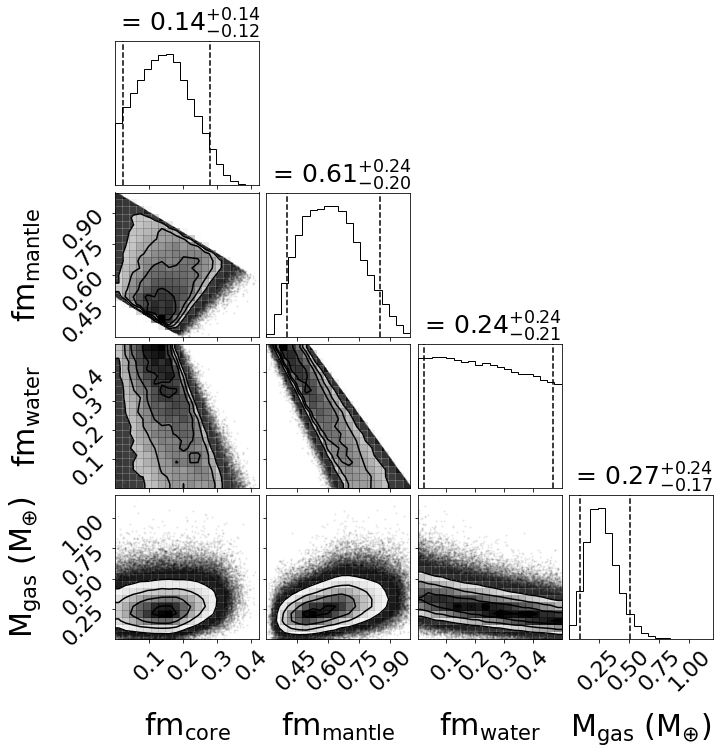}
    \includegraphics[width=0.4\textwidth]{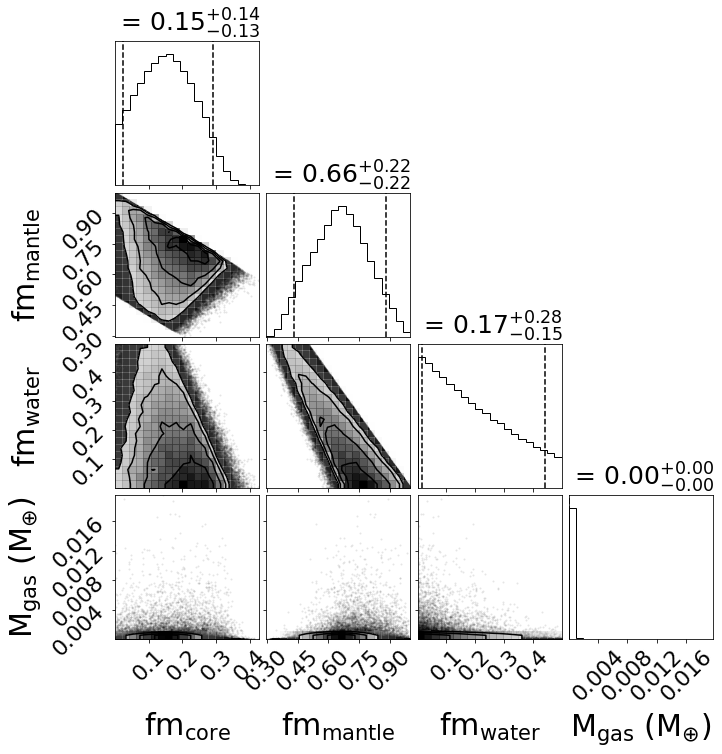}
    \includegraphics[width=0.4\textwidth]{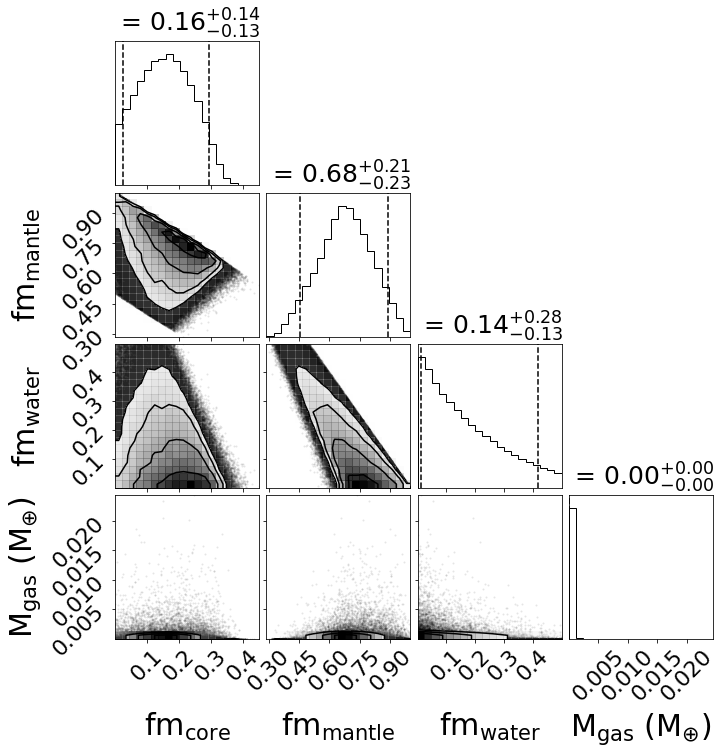}
    \caption{Corner plots showing the posteriors of the free parameters adopted for the planetary internal structure modelling described in Sect. \ref{sec:internalstructure}. From top to bottom, the plots refer to HIP\,29442\,$b$,  HIP\,29442\,$c$, and  HIP\,29442\,$d$. }
    \label{fig:corner_internal_structure}
\end{figure}

\end{appendix}
\end{document}